\documentclass[aps,twocolumn,amsfonts,amssymb,amsmath,floatfix,tightenlines,superscriptaddress,longbibliography]{revtex4-1}

\usepackage{amsfonts,amssymb}
\usepackage{amsmath,times}
\usepackage{graphicx}
\usepackage[colorlinks=true,linkcolor=blue,anchorcolor=red,citecolor=blue,urlcolor=blue]{hyperref}
\usepackage{color}
\usepackage{cleveref}
\usepackage[lofdepth,lotdepth,caption=false]{subfig}
\usepackage[normalem]{ulem}

\renewcommand{\vec}[1]{\mathbf{#1}}

\newcommand{\Psivar}{\Psi_{\mathrm{var}}}

\newcommand{\Otri}{\Omega}
\newcommand{\veck}{\vec{k}}

\newcommand{\asbg}{a_{\mathrm{bg}}}
\newcommand{\ammbg}{a_{\mathrm{mm}}^{\mathrm{bg}}}

\newcommand{\ammeff}{a_{\mathrm{mm}}^{\mathrm{eff}}}
\newcommand{\amm}{a_{\mathrm{mm}}}

\begin{document}

\title{Tunable Molecular Interactions Near an Atomic Feshbach Resonance: Stability and Collapse of a Molecular Bose-Einstein Condensate}

\author{Zhiqiang Wang}
\email[]{wzhiqustc@ustc.edu.cn}
\affiliation{Hefei National Research Center for Physical Sciences at the Microscale and School of Physical Sciences, University of Science and Technology of China,  Hefei, Anhui 230026, China}
\affiliation{Shanghai Research Center for Quantum Science and CAS Center for Excellence in Quantum Information and Quantum Physics, University of  Science and Technology of China, Shanghai 201315, China}
\affiliation{Hefei National Laboratory, University of  Science and Technology of China, Hefei 230088, China}
\affiliation{Department of Physics and James Franck Institute, University of Chicago, Chicago, Illinois 60637, USA}
\author{Ke Wang}
\affiliation{Department of Physics and James Franck Institute, University of Chicago, Chicago, Illinois 60637, USA}
\author{Zhendong Zhang}
\affiliation{E. L. Ginzton Laboratory and Department of Applied Physics, Stanford University, Stanford, CA 94305, USA}
\affiliation{Department of Physics and Hong Kong Institute of Quantum Science and Technology, The University of Hong Kong, Hong Kong, China}
\author{Qijin Chen}
\affiliation{Hefei National Research Center for Physical Sciences at the Microscale and School of Physical Sciences, University of Science and Technology of China,  Hefei, Anhui 230026, China}
\affiliation{Shanghai Research Center for Quantum Science and CAS Center for Excellence in Quantum Information and Quantum Physics, University of  Science and Technology of China, Shanghai 201315, China}
\affiliation{Hefei National Laboratory, University of  Science and Technology of China, Hefei 230088, China}
\author{Cheng Chin}
\affiliation{Department of Physics and James Franck Institute, University of Chicago, Chicago, Illinois 60637, USA}
\affiliation{Enrico Fermi Institute, University of Chicago, Chicago, Illinois 60637, USA}
\author{K. Levin}
\affiliation{Department of Physics and James Franck Institute, University of Chicago, Chicago, Illinois 60637, USA}
\date{\today}

\begin{abstract}
Understanding and controlling interactions of ultracold molecules is a cornerstone of
quantum chemistry.
While the laboratory creation of degenerate molecular gases comprised of bosonic
atoms has unlocked powerful new platforms
for quantum simulation, progress is limited by the absence of a robust theoretical framework for 
characterizing inter-molecular
interactions. 
This is in stark contrast to the situation for Fermi gases.
In this Letter, we present such a framework providing 
universal expressions for these molecular scattering lengths as functions of
experimentally measurable quantities. 
Our discoveries are crucial for understanding molecular condensate formation. Calculations of 
the compressibility reveal that a sign change
in such molecular scattering lengths is directly correlated with the instability of these condensates.
These results offer fresh insight with broad applications for
atomic, molecular, and condensed matter physics, as well as quantum
chemistry.
\end{abstract}

\maketitle

\textbf{Introduction ---}
Molecular condensates of bosonic atoms represent a key frontier in physics, offering a path to discover new states of matter and quantum phase transitions
\cite{Radzihovsky2004,Romans2004,Radzihovsky2008,Radzihovsky2009}.
The use of Feshbach resonances
~\cite{Chin2010}, 
particularly through magnetoassociation, has revolutionized our ability to create stable diatomic molecules from ultracold atomic pairs. This has allowed for the preparation of equilibrium molecular condensates
\cite{Zhang2021,Zhang2023}.

Despite this progress, bosonic systems face significant challenges. Condensates become unstable near a resonance, 
as they experience particle loss and heating due to three-body recombination
~\cite{Fedichev1996,Bedaque2000,Weber2003,Fletcher2013}. 
Feshbach interactions can add to this condensate destabilization when they introduce an attractive force 
between molecules.
Such instabilities contrast sharply with two-component Fermi gases, where the Pauli exclusion principle
~\cite{Petrov2004,Petrov2005}
allows for the robust formation of stable molecular condensates. The stability in fermionic systems has enabled an extensive exploration of the Bardeen-Cooper-Schrieffer (BCS)
to Bose-Einstein condensation (BEC) crossover
~\cite{Zwerger2012,Regal2004,Zwierlein2004,Chen2005}, 
a level of understanding that has yet to be replicated with bosons.
Here, BCS-BEC crossover refers to a superfluid undergoing a smooth evolution, with increasing interaction strength, from
large, overlapping Cooper pairs to tightly bound diatomic molecules.

\begin{figure}[tp]
\begin{center}
\includegraphics[width=0.95\linewidth,clip,trim=0 5pt 0 0]
{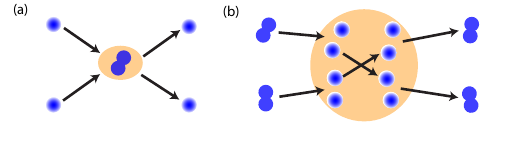}
\caption{ 
Scattering due to atomic Feshbach coupling:
(a) Leading-order scattering process contributing to the resonant term in the atomic scattering length, $a_s$ [Eq.~\eqref{eq:asdef}]. Two incoming atoms (single blue circles) temporarily combine into a molecule (two attached circles) via Feshbach coupling, and then dissociate into two outgoing free atoms. The orange region represents virtual processes.
(b) Corresponding leading-order contribution to the molecular scattering length, $\amm$, near the resonance. Here, two molecules approach each other and temporarily break up into four free atoms, which then propagate, interact, and recombine into two molecules. Similar to panel (a), this process universally depends on the Feshbach coupling but is of higher order in the coupling strength.
}
\label{fig:Sketch}
\end{center}
\end{figure}

In this Letter, we address this critical issue of bosonic molecular condensate instability by examining the molecule-molecule scattering length. 
Our approach is inspired by atomic condensates, where the two-body $s$-wave scattering length is typically described by the following equation:

\begin{align}
a_s  & =  \asbg  \bigg(1 - \frac{\Delta B}{B-B_0} \bigg),  \label{eq:asdef}
\end{align}
Here, $B_0$ is the resonance value of the magnetic field $B$, $\Delta B$ is the resonance width, and $\asbg$ is the atomic $s$-wave background scattering length.
The leading-order contribution to $a_s$ due to the Feshbach coupling can be visualized as in Fig.~\ref{fig:Sketch}(a)
\footnote{While higher-order contributions, involving repeated processes from
Fig.~\ref{fig:Sketch}(a) renormalize $B_0$ they don't change the resonant structure of
$a_s$.}.

It is well known that atomic condensates collapse when $a_s$ becomes negative. This typically happens when the
magnetic field $B$ is above the resonance, specifically within the range
$B_0 <B<B_0+\Delta B$
(assuming a positive atomic background scattering length,
$\asbg>0$). 
For concreteness, here we assume $\Delta B>0$, so that the bare molecules' energy drops below the atom continuum when the magnetic field is below resonance ($B<B_0$)
~\footnote{The formulas derived here can be easily generalized to the case of negative $\Delta B$}.
What has not been widely appreciated is that Feshbach coupling can also modify the scattering properties of molecules, which can lead to the instability of a molecular Bose-Einstein condensate. Understanding this important effect is central to our investigation.

In this Letter, we demonstrate how Feshbach resonances dramatically alter the intermolecular scattering 
length and consequently the behavior of bosonic molecular condensates, mirroring their profound effect in atomic systems. Crucially, this 
tunability can drive molecular condensates into instability when this scattering length (called $\amm$ at zero
and $\ammeff$ at finite density) becomes negative. Notable is a rather pronounced
density dependence found in $\ammeff$ for narrow resonances,
which arises from many-body effects and introduces additional complexity; only in the 
limit of very wide Feshbach resonances does $\ammeff$ simplify, converging with the two-body scattering length $\amm$.

The accurate determination of both the fundamental molecular scattering length ($\amm$) and the density-dependent effective scattering length ($\ammeff$) 
presents a significant theoretical challenge. As Figure \ref{fig:Sketch}(b) illustrates, even a leading-order intermolecular scattering process 
can be considerably more complex as it involves higher order in Feshbach coupling strength than what is seen in simpler atomic systems 
[Figure \ref{fig:Sketch}(a)]. Additionally, we also need to precisely account for background interactions between atoms and molecules.

\begin{figure}[tp]
\begin{center}
\includegraphics[width=0.96\linewidth,clip,trim=0 0pt 0 0]
 {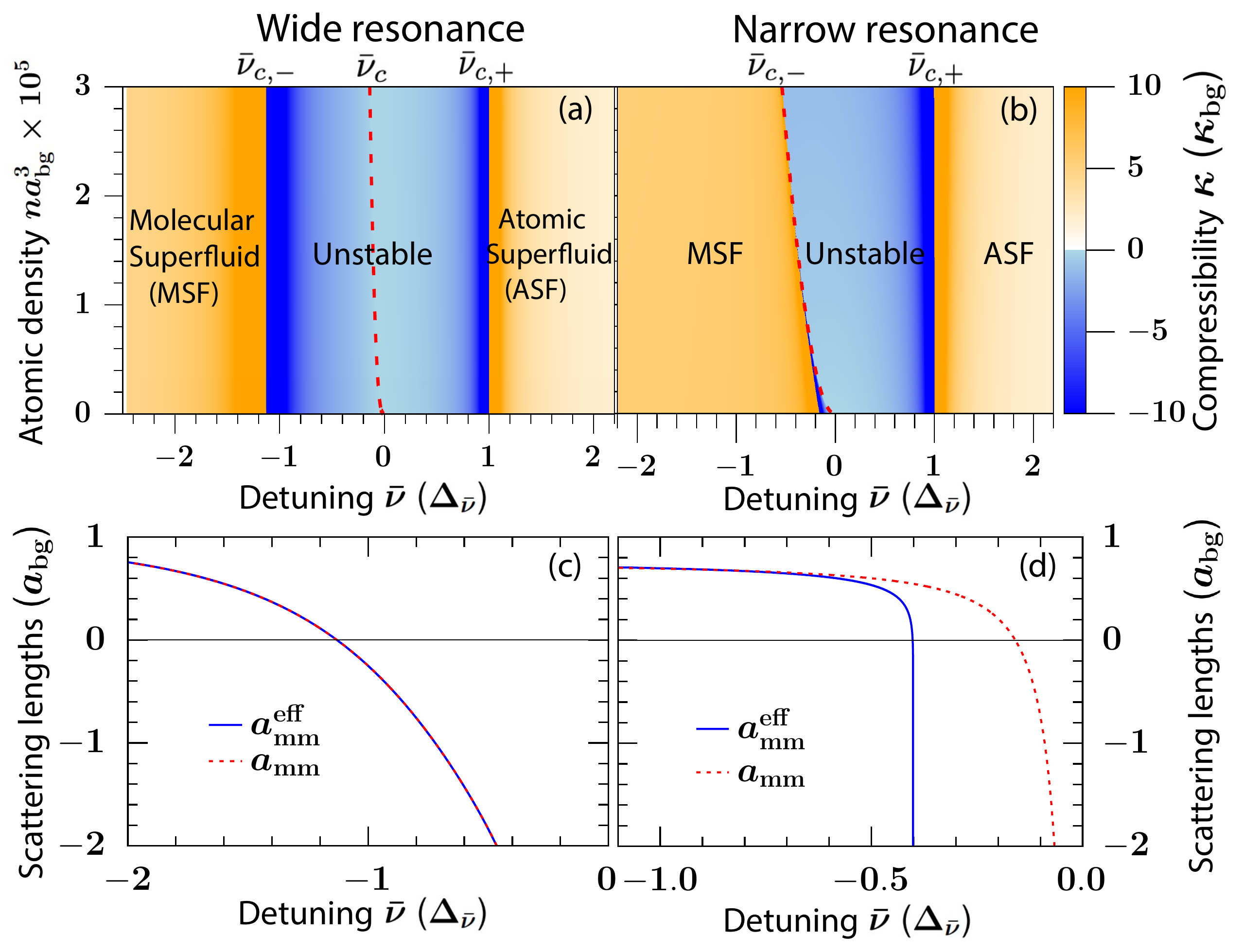}
\caption{ 
Phase diagrams and molecular scattering lengths.
Ground-state stability phase diagrams for (a) wide and (b) narrow resonances, showing compressibility $\kappa$ [normalized by 
$\kappa_{\mathrm{bg}}=m_1/(4\pi \hbar^2 a_{\mathrm{bg}})$] 
as a function of atom number density $n$ and of detuning 
$\bar{\nu} =\Delta \mu_m (B-B_0)$
(normalized by resonance width 
$\Delta_{\bar{\nu}}=\Delta \mu_m \Delta B$).  
Orange indicates stable regions ($\kappa>0$), and blue indicates unstable regions. 
The atomic condensate is present only in the atomic superfluid (ASF) phase, while the molecular condensate is present in 
both ASF and molecular superfluid (MSF) phases. Red dashed lines denote the quantum critical point 
$\bar{\nu}_c(n)$
separating ASF from MSF, while
$\bar{\nu}_{c,-}$ and $\bar{\nu}_{c,+}$
mark boundaries between unstable and stable regions.
(c),(d) Corresponding two-body molecular scattering length 
$\amm$, and its many-body analogue $\ammeff(n)$
for $n \asbg^3=1.68\times 10^{-5}$.
Parameters for the narrow resonance [panels (b) and (d)] are from the $^{133}$Cs resonance 
at $B_0 = 19.849$ G
\cite{Zhang2021,Zhang2023,Wang2024}. For the wide resonance [panels (a) and (c)], 
$\Delta_{\bar{\nu}}$
is increased by $10^2$ times relative to panels (b) and (d).
}
\label{fig:Fig1}
\end{center}
\end{figure}

\textbf{Summary of main results ---}
Our theoretical framework employs a two-channel variational wavefunction
treatment to systematically analyze the system's compressibility and, from it, deduce the crucial scattering lengths. 
We begin by presenting general insights into the zero-temperature stability phase diagrams presented in Figure~\ref{fig:Fig1},
which are determined through numerical calculations of the compressibility $\kappa= d n/ d \mu$.

Figures~\ref{fig:Fig1}(a,b) show the zero-temperature stability phase diagrams for both a wide and narrow
resonance, in terms of density $n$ on the vertical axis and detuning $\bar{\nu}$ on the horizontal axis. 
Both panels indicate a stable molecular superfluid (MSF) phase at large negative detuning,
a stable atomic superfluid (ASF) phase at large positive detuning, and an unstable region in between,
where the compressibility becomes negative $\kappa <0$.
The boundaries of this unstable region are marked by the detunings
$\bar{\nu}_{c,+}$ and $\bar{\nu}_{c,-}$. 
In Fig.~\ref{fig:Fig1}(c,d), we plot both our calculated two-body scattering length, $\amm$,
and its many body analogue $\ammeff(n)$.

Note a striking similarity between panels (a) and (b) of Fig.~\ref{fig:Fig1} on the atomic side. For both narrow and wide resonances, the phase boundary of the unstable region is essentially dictated by the sign change of the two-body atomic scattering length, $a_s$.  This means that in both scenarios, the upper boundary detuning of the unstable regime, 
$\bar{\nu}_{c,+}$ remains nearly independent of density.

In contrast, the behavior of the compressibility on the molecular side of the phase diagrams in Figs.~\ref{fig:Fig1}(a,b) reveals a significant difference between wide and narrow resonances. We summarize our key observations:

 (1) \textit{Wide resonance case with density-independent stability}. In the case of wide resonances, the entire unstable region shows very little dependence on density. This is a significant finding because it means we can roughly characterize the molecular condensate's stability using just a two-body scattering length, similarly to how we treat atomic condensates. Figure~\ref{fig:Fig1}(c) confirms this, showing that $\ammeff(n)$ is nearly equivalent to $\amm$, and importantly, $\amm$ itself changes sign at the instability onset detuning, 
$\bar{\nu}_{c,-}$, 
in Fig.~\ref{fig:Fig1}(a).

(2) \textit{Narrow resonance case with strong density dependence}. By contrast, for narrow resonances, the lower boundary detuning, 
$\bar{\nu}_{c,-}$, 
of the unstable region in Fig.~\ref{fig:Fig1}(b) is highly dependent on density. Here, it is the density-dependent many-body scattering length, $\ammeff$, not the two-body $\amm$, 
that changes sign at this instability boundary, as clearly shown in Fig.~\ref{fig:Fig1}(d).

(3) \textit{Proximity to quantum critical point (QCP)}. What is particularly interesting is that for narrow resonances, at moderate to high densities, the boundary detuning 
$\bar{\nu}_{c,-}(n)$
in Fig.~\ref{fig:Fig1}(b) nearly aligns with the detuning 
$\bar{\nu}_c(n)$. 
This detuning $\bar{\nu}_c(n)$ is associated with a QCP that separates the MSF from the ASF phase. 
This strong correlation suggests that close proximity to the QCP is a crucial factor in the instability of these 
narrow resonance molecular superfluids.

To quantitatively confirm the observations above,
we present three limiting forms for the scattering lengths $\amm$ and $\ammeff$.
These formulas are particularly relevant in the interesting detuning regions where a stable MSF exists and
where the compressibility changes its sign~\cite{supplement}.
It is important to note that they are applicable only for the magnetic field below the resonance $B<B_0$.

For a wide resonance at large negative detuning, we find that the scattering lengths $\ammeff$ and $\amm$ are given by
\begin{subequations} \label{eq:ammwide}
\begin{align}
\ammeff(n) & \approx
 \amm   = \ammbg \bigg[ 1 - \bigg( \frac{\Delta_B^{\mathrm{wide}}}{B_0 -B} \bigg)^2  \bigg]  \label{eq:ammwide1} \\
 \text{where } \qquad
\Delta_B^{\mathrm{wide}}   & = \frac{\pi}{ \sqrt{6}} \sqrt{ \frac{ \hbar^2 }{ m_1 (\bar{a} -\asbg)^2  \Delta \mu_m } \Delta B  }  \label{eq:ammwide2}
\end{align}
\end{subequations}
is an effective molecular resonance width. Here,
$\ammbg$, which is chosen to be positive~\footnote{In this Letter, we consider only the case where both $\asbg$ and $\ammbg$ are positive.},
$m_1$, and $\Delta \mu_m$ are the molecular $s$-wave background scattering length, the atomic mass, and the magnetic
moment difference between the open and closed channels, respectively.
As in the literature~\cite{Lange2009,Chin2010},
$\bar{a}$ in Eq.~\eqref{eq:ammwide2} can be taken as $0.96 R_{\mathrm{vdW}}$, where $R_{\mathrm{vdW}}$ is the van der Waals length.

For the narrow resonance case at zero density and with detuning somewhat
away from unitarity, we can define a two-body molecular scattering length
\begin{subequations} \label{eq:ammnarrow}
\begin{align}
\amm & = \ammbg \bigg[ 1-  \bigg( \frac{ \Delta_B^{\text{narrow}} }{B_0-B} \bigg)^{ \frac{3}{2} }   \bigg]  \label{eq:ammnarrow1}  \\
\text{where }  \quad
\Delta_B^{\text{narrow}}   &  =  \bigg( \frac{\asbg}{\ammbg} \bigg)^{ \frac{2}{3} }  \bigg( \frac{ m_1 \asbg^2 \Delta \mu_m }{\hbar^2}   \bigg)^{ \frac{1}{3} }  \;  (\Delta B)^{ \frac{4}{3} }\label{eq:ammnarrow2}
\end{align}
\end{subequations}
is the counterpart molecular resonance width.
The scattering process underlying the resonant term in Eq.~\eqref{eq:ammnarrow1}
\footnote{Note that the whole contribution of the resonance term in Eq.~\eqref{eq:ammnarrow1} does not depend on $\ammbg$, since the overall prefactor
of $\ammbg$ is cancelled out by a factor of $1/\ammbg$ from $(\Delta_B^{\mathrm{narrow}})^{2}$.}
is illustrated in Fig.~\ref{fig:Sketch}(b), which
involves the exchange of two bosonic atom constituents between molecules.
Importantly, the negative sign in front of this term is fundamentally connected to
Bose statistics.
It can be shown that for a \textit{fermionic} Feshbach resonance, $\amm$ takes a similar form
to Eq.~\eqref{eq:ammnarrow1}, but with the negative sign in front of the resonant term replaced by a \textit{positive} sign (see Refs.~\cite{Gurarie2007,Levinsen2011}~\footnote{On the other hand, in the fermionic studies of Refs.~\cite{Gurarie2007,Levinsen2011}, the effect of the molecular background interaction $\ammbg$ on $\amm$ was not considered. For additional discussions, see Ref.~\cite{supplement}.}).

Finally we present an expression for
the many-body molecular scattering length 
$\ammeff(n)$ in the case of a narrow resonance at finite density,
\begin{subequations} \label{eq:ammfiniten}
\begin{align}
\ammeff(n) & \approx \ammbg \bigg[  1 -  \bigg( \frac{      \Delta_B^{  \text{narrow}  } } { B_0 -B } \bigg)^{    \frac{3}{2} }   f(B,n) \bigg] \label{eq:ammfiniten1} \\
\text{with } \; 
f(B,n) & = 1 + \frac{2 \sqrt{2}}{\pi } \ln \bigg( \frac{ \bar{\nu}^2 }{ \bar{\nu}^2 -  \bar{\nu}_c^2(n) } \bigg),  \label{eq:ammfiniten2}
\end{align}
\end{subequations}
where $\bar{\nu}=\Delta \mu_m (B-B_0)$ is the detuning and $\bar{\nu}_c(n)\approx -2 \bar{\alpha} \sqrt{n}$
is its corresponding QCP value, with $\bar{\alpha}$ representing the renormalized Feshbach coupling strength defined in Eq.~\eqref{eq:rstar}.
Equation~\eqref{eq:ammfiniten} is applicable for the detuning less than the QCP value, $\bar{\nu} < \bar{\nu}_c<0$. 
At zero density, the QCP coincides with the resonance $\bar{\nu}_c=0$ and Eq.~\eqref{eq:ammfiniten1} reduces to Eq.~\eqref{eq:ammnarrow} for the two-body scattering length $\amm$.

Qualitatively, one can understand why many-body physics plays an important role in the molecular condensate stability,
as the main fluctuations that destabilize the MSF phase in Fig.~\ref{fig:Fig1} are atomic Bogoliubov excitations arising from
the presence of the ground-state molecular condensate.
These excitations exhibit a significant energy gap and behave like free atoms when the detuning is large and negative, 
as is relevant in a wide resonance.  However,
they become soft with a phonon-like, low-energy dispersion as the detuning
approaches the QCP. 
For a sufficiently narrow resonance at a given density, it is this softness of the Bogoliubov excitations, rather than proximity 
to the two-body resonance point, that drives the instability.

These many-body effects are manifested in the logarithmic term in $f(B,n)$ 
appearing in Eq.~\ref{eq:ammfiniten}. 
This logarithmic dependence has the important consequence that
the boundary detuning $\bar{\nu}_{c,-}(n)$ for the narrow resonance
is generally exponentially close to
the QCP~\cite{supplement}.
Importantly,
this narrow resonance case presents an interesting opportunity as it provides experimental 
access to the associated QCP physics from the stable MSF side.

\textbf{Variational analysis ---}
To arrive at all of these results, we use a 
variational ground state analysis that incorporates
both atomic and molecular condensates, as well as depletion and atomic Cooper-pairing contributions.
Some aspects of these stability issues due to the presence of Cooper pairing of bosons have been discussed in prior work~\cite{Evans1969,Nozieres1982,Jeon2002,Koetsier2009,Lee2010,Basu2008}. There, the emphasis
was primarily restricted to single-channel models.

We adopt a two-channel model Hamiltonian that includes both inter-channel Feshbach coupling and
intra-channel background interactions: 
\begin{align} \label{eq:Hamiltonian}
\hat{H} & = 
\sum_{\sigma=1}^2  \sum_{\veck}  h_{\sigma \veck} a_{\sigma \veck}^\dagger a_{ \sigma \veck}   
 - \frac{\alpha}{\sqrt{V}} \sum_{\veck_1,\veck_2}\big( a^\dagger_{1 \veck_1 } a^\dagger_{1 \veck_2 } a_{ 2, \veck_1+\veck_2} \nonumber \\
 & + h. c.\big)
+ \sum_{\sigma=1}^2  \sum_{\veck_{i}}   \frac{g_\sigma}{2 V} 
a^\dagger_{ \sigma \veck_1 } a^\dagger_{ \sigma \veck_2 } a_{  \sigma \veck_3} a_{ \sigma , \veck_1 +\veck_2 -\veck_3 }. 
\end{align}
Here, $a_{\sigma \veck}$ is an annihilation operator for the open-channel atoms (with $\sigma=1$) or closed-channel molecules ($\sigma=2$).
The summation $V^{-1}\sum_{\veck}$, with $V$ the volume, represents an integral over the momentum $\veck$ with a cutoff $\Lambda$, 
which is needed to regularize ultraviolet divergences in various $\veck$ dependent integrals.
Finally, we assume three-dimensional isotropy and ignore trap effects. 

In Eq.~\eqref{eq:Hamiltonian}, we define kinetic energy contributions
$h_{1 \veck}=(\hbar\veck)^2/2 m_1 -\mu$ and $h_{2 \veck}=(\hbar \veck)^2/2 m_2 -(2\mu -\nu)$ with $m_2=2 m_1$,
$\mu$ the chemical potential, and $\nu$ the \textit{bare} molecule detuning.
The parameter $g_\sigma>0$ corresponds to a \textit{repulsive} intra-channel density-density  interaction, and $\alpha$ represents the \textit{bare} 
Feshbach coupling between the two channels. 
The parameters $\{\nu,\alpha, g_1,g_2\}$ depend on the cutoff $\Lambda$ and are related~\cite{Wang2024,supplement} to experimental observables
by $ \nu  = \bar{\nu}  + \sqrt{2} \beta \alpha \bar{\alpha}$, 
$ \alpha  = \bar{\alpha} \Gamma / \sqrt{2}$, $ g_1    = \bar{g}_1 \Gamma$,
and $ g_2    = \bar{g}_2 /(1- (2/\pi)  \Lambda \ammbg)$
with
$\bar{\alpha}=\sqrt{4\pi \hbar^2 \asbg \Delta \mu_m \Delta B/m_1}$,
$ \bar{g}_1  = 4\pi \hbar^2 \asbg / m_1>0$,  $ \bar{g}_2  = 4\pi \hbar^2 \ammbg / m_2 >0 $,
$\beta  =  m_1 \Lambda/ (2\pi^2 \hbar^2) $, and
$\Gamma  =1/(1-\beta \bar{g}_1)$. 
These relations are chosen to reproduce the atomic scattering length
$a_s$ in Eq.~\eqref{eq:asdef} in the two-atom scattering limit. 
Finally, the momentum cutoff $\Lambda$ is related to the length scale $\bar{a}$ in Eqs.~\eqref{eq:ammwide},\eqref{eq:ammnarrow} and \eqref{eq:ammfiniten} by $\Lambda=\pi/(2\bar{a})$. 

From the renormalized Feshbach coupling parameter $\bar{\alpha}$ one can define a characteristic length scale $r_*$~\cite{Gurarie2007,Levinsen2011,Chin2010,Ho2012}
as
\begin{align}
r_*& \equiv  4\pi \hbar^4/m_1^2 \bar{\alpha}^2 = \hbar^2 /  ( \asbg m_1 \Delta \mu_m \Delta B),    \label{eq:rstar}
\end{align}
which allows us to classify the resonance width quantitatively. 
If the ratio $w_{\mathrm{res}} \equiv \asbg/r_*$ is much smaller than unity $w_{\mathrm{res}}\ll 1$, the resonance is viewed as narrow; otherwise, it is classified as wide.  
For the narrow and wide resonances used in this Letter, we choose $w_{\mathrm{res}} \approx 0.006$~\cite{Wang2024,supplement}
and $w_{\mathrm{res}} \sim 1$, respectively
~\footnote{We note that using an even wider resonance does not change our conclusions. We use the moderately wide resonance to make our numerical calculation easier.}.

To address the ground-state stability at zero temperature
we adopt the following many-body variational wavefunction as an approximation to the true ground state of $\hat{H}$,
\begin{align} \label{eq:Psivar}
|\Psivar \rangle & =\mathcal{N}^{-1}  e^{ \sum_\sigma \Psi_{\sigma 0} \sqrt{V}  a^\dagger_{\sigma 0}   + \sum_{\veck}^\prime \sum_{\sigma} \chi_{\sigma \veck}\; a^\dagger_{\sigma \veck} a^\dagger_{\sigma -\veck} } \vert 0 \rangle,
\end{align}
where $\{\Psi_{\sigma 0}, \chi_{\sigma \veck}\}$ are the variational parameters and $\mathcal{N} $ is the normalization factor. 
In the exponent, the $\veck-$sum is over half of $\veck-$space, and the prime in the $\veck$-summation implies the origin $\veck=0$ is excluded. 
The vacuum $|0\rangle$ satisfies $a_{\sigma\veck}|0\rangle=0$ for all annihilation operators $a_{\sigma\veck}$. 
In the spirit of generalized Bogoliubov theory, this variational wavefunction includes only pair-wise correlations between atoms or between molecules
in the exponent.
We emphasize that this approximation is adequate for our focus on the detuning regime that is not too close to the resonance\cite{Wang2025}. 
In Ref.\cite{supplement}, we provide concrete estimates showing that neglecting higher-order correlations results in less than $2\%$ 
uncertainty in the phase boundaries of the phase diagram in Fig.~\ref{fig:Fig1}, 
as well as in the scattering length formulas given in Eqs.~\eqref{eq:ammwide} \textendash \eqref{eq:ammfiniten}. 

The trial ground state energy associated with $|\Psivar \rangle$ is then
\begin{align}   \label{eq:Etri}
 &  \Otri [\Psi_{10}, \Psi_{20}, \chi_{1\veck},\chi_{2\veck}] 
 =  \langle \Psivar | \hat{H} | \Psivar\rangle,   
\end{align}
which is a functional of the parameters $\{\Psi_{10}, \Psi_{20}, \chi_{1\veck},\chi_{2\veck}\}$.
Here, $\Psi_{\sigma 0}=\langle a_{\sigma 0} \rangle/ \sqrt{V}$ indicates that $\Psi_{10}$ ($\Psi_{20}$) also represents the amplitude of the atomic (molecular) condensate. 
Minimizing $\Otri$ with respect to $\{ \Psi_{10}^*, \Psi_{20}^*, \chi_{1\veck}^*, \chi_{2\veck}^* \}$ yields
a set of saddle-point equations~\cite{Wang2024}.
Those derived from the derivative $\partial \Omega/\partial \chi_{\sigma \veck}^*$  can be recast into the form of the BCS-like gap equation by introducing
the Cooper-like pairing order parameter
$ \Delta_\sigma   \equiv  g_\sigma  V^{-1}\sum_{\veck \ne 0}   \langle a_{\sigma \veck}  a_{\sigma, -\veck} \rangle$.
Using the pairing order parameter $\Delta_\sigma$ one can rewrite the ground state energy
$\Otri$ in Eq.~\eqref{eq:Etri} as a function of only five unknowns: $\Otri= \Otri[\Psi_{10},\Psi_{20},\Delta_{1},\Delta_2,\mu]$, 
whose first-order derivative with respect to the five parameters
lead to four saddle-point equations plus one total particle number density constraint: $ n = (|\Psi_{10}|^2 + n_1) + 2 (|\Psi_{20}|^2 + n_2) $ with 
 $ n_\sigma  = V^{-1} \sum_{\veck \ne 0} \langle a_{\sigma \veck}^\dagger a_{\sigma \veck} \rangle$.

\begin{figure}[tp]
\begin{center}
\includegraphics[width=0.96\linewidth,clip,trim=0 0pt 0 0]{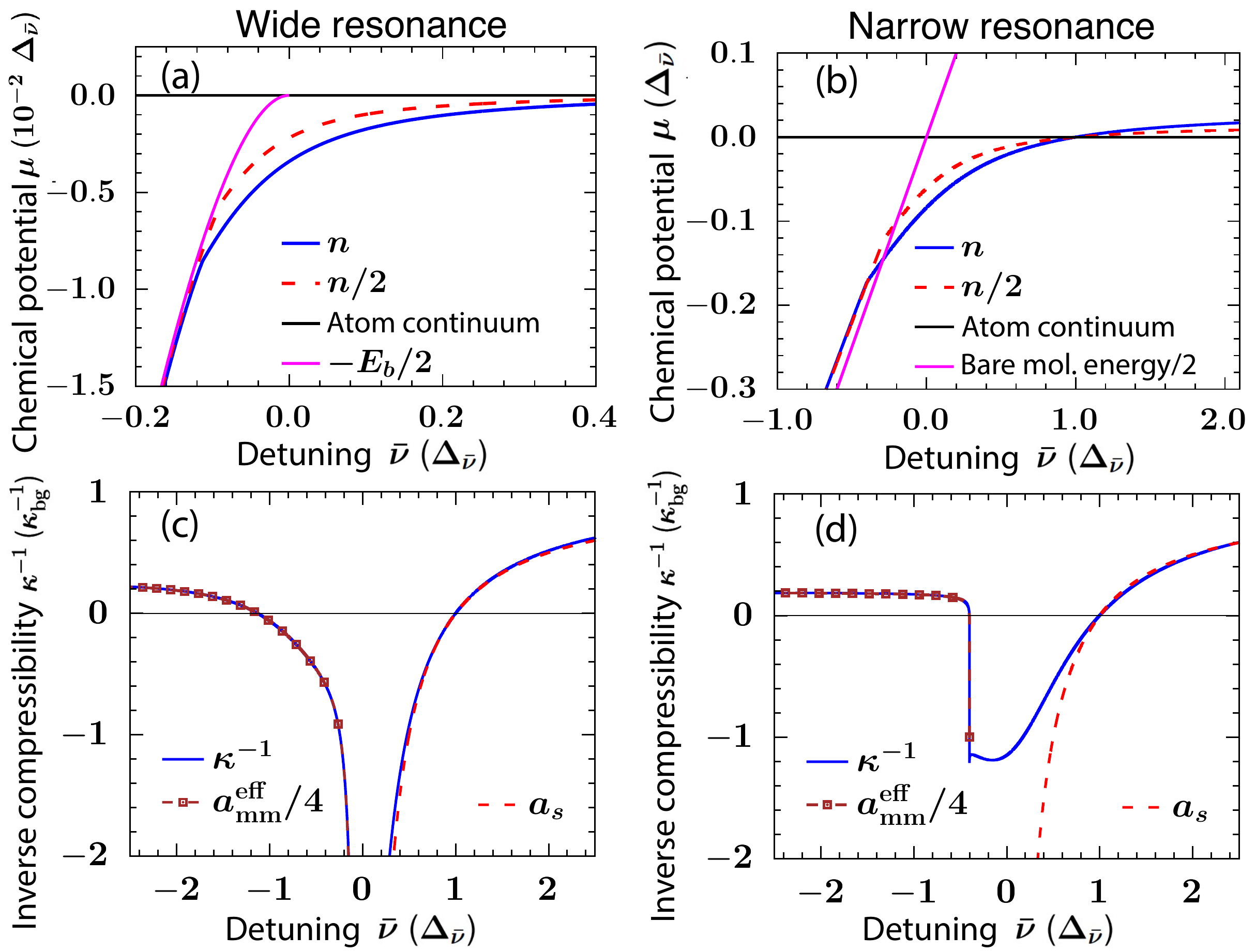}
\caption{
Chemical potential, inverse compressibility and molecular scattering lengths.
(a,b) Chemical potential $\mu$ (blue solid and red dashed lines) vs detuning
$\bar{\nu}$ at densities $n$ and $n/2$, with $n\asbg^3 =1.68\times 10^{-5}$
[same as Figs.~\ref{fig:Fig1}(c) and \ref{fig:Fig1}(d)]. The ``atom continuum" threshold is at $\mu=0$. In panel (a),
``$-E_b$" represents the dressed molecular energy; in panel (b), it is replaced by its bare value, 
$\bar{\nu}$. (c), (d) Comparison of the inverse compressibility (blue solid lines) at 
$n\asbg^3 =1.68\times 10^{-5}$ with the atomic scattering length $a_s$
(red dashed lines) and the many-body molecular scattering length 
$\ammeff$ (brown dashed lines with squares), defined in Eq.~\eqref{eq:ammeffdef}. 
Both $\ammeff$ and $a_s$ are plotted in units of $\asbg$.
}
\label{fig:Fig4}
\end{center}
\end{figure}

We now numerically solve the five equations. Figures~\ref{fig:Fig4}(a) and \ref{fig:Fig4}(b) plot
the calculated chemical potentials near resonance at two different densities
for both the wide and narrow resonances.
This provides useful insight into the anomalous negative sign of the compressibility
and the related condensate instability. 
From both figures we observe that the chemical potential $\mu$, which represents the average energy per atom,
falls below the two energy levels corresponding to
the bare atomic continuum threshold on the atomic side and the two-body molecular energy$/2$
on the molecular side. 
Furthermore, as the density $n$ decreases,  in the near-resonance regime the chemical potential $\mu$ approaches the
two energy levels from below.
Consequently, the inverse compressibility $\kappa^{-1}=d\mu/dn$ must be \textit{negative}
near and on both sides of the resonance
\footnote{In contrast to the near-resonance regime in Figs.~\ref{fig:Fig4}(a,b), the chemical potential $\mu$ rises above one of the two
energy levels when the detuning is far off the resonance.
This is true for both the wide- and narrow-resonance cases (see Ref.~\cite{supplement} for additional details). 
It implies that, in these far-off-resonance regimes, the chemical potential
must approach these levels from above as the density vanishes ($n\rightarrow 0$), consistent with the fact
that the compressibility is positive in those detuning regimes. }. 
The numerical results that were shown previously in Figs.~\ref{fig:Fig1}(a) and \ref{fig:Fig1}(b)
support this analysis; there we directly evaluated the derivative $d\mu/dn$ numerically, 
from which we constructed the ground-state stability phase diagrams.

\textbf{Many-body effective scattering length --- }
Armed with this understanding we now derive
the many-body \textit{effective} scattering length $\ammeff$ for the dressed molecules
in terms of an effective interaction $g_{2}^{\mathrm{eff}}$:
\begin{subequations} \label{eq:ammeffdef}
\begin{align}
  \ammeff(n)    & =  \frac{m_2}{4\pi \hbar^2} g_{2}^{\mathrm{eff}}(n)   \label{eq:ammeff} \\
\text{with } \quad
g_{2}^{\mathrm{eff}}
 & = \frac{ (\mathcal{Z}^{\mathrm{eff}})^2 }{2} 
\frac{\partial^4 \Otri[\Psi_{10}, \Psi_{20}, \Delta_1, \Delta_2, \mu] /V}{\partial \Psi_{20}^2 \partial (\Psi_{20}^*)^2 }.   \label{eq:g2eff}
\end{align} 
\end{subequations}
Here, $\mathcal{Z}^{\mathrm{eff}}$
is the fraction of the dressed-molecular condensate in the closed channel, defined by $ \mathcal{Z}^{\mathrm{eff}}= |\Psi_{20}|^2 / (|\Psi_{20}|^2 + n_1/2)$
\footnote{The factor $(\mathcal{Z}^{\mathrm{eff}})^2$ in Eq.~\eqref{eq:g2eff} is needed to reconcile two facts:
(1) the many-body scattering length $\ammeff$, as well as the two-body one $\amm$, is defined for a \textit{dressed} molecule~\cite{supplement}, which is an admixture of closed-channel molecules with open-channel atom pairs,
and (2) the derivative in Eq.~\eqref{eq:g2eff} is taken with respect to the \textit{bare} closed-channel molecular condensate amplitude $\Psi_{20}$.
Although this factor is essentially equal to $1$ for a narrow resonance, 
it becomes important for wide resonances, particularly in deriving Eq.~\eqref{eq:ammwide}~\cite{supplement} in the zero density limit. }. 
The quartic derivative in Eq.~\eqref{eq:g2eff} can be carried out approximately~\cite{supplement}
\footnote{When performing derivatives in Eq.~\eqref{eq:g2eff} we set $\{\Psi_{10},\Delta_1,\Delta_2\}$ at their saddle-point values. 
They are viewed as functions of $\{\Psi_{20},\mu\}$ only.
In arriving at Eq.~\eqref{eq:g2effL3} we have also neglected terms 
that are higher order in the ratios $\{g_1 n/E_{1\veck=0}, g_2 n/E_{1\veck=0},
\alpha \sqrt{n}/E_{1\veck=0} \}$. 
},
leading to
\begin{equation}
\frac{ g_{2}^{\mathrm{eff}} } { (\mathcal{Z}^{\mathrm{eff}})^2 }
   \approx  \widetilde{\bar{g}}_2  - \widetilde{\bar{\alpha}}^4 \bigg[
\frac{1}{V}  \sum_\veck \frac{1}{2 E_{1\veck}^3} - 2 g_1  \bigg( \frac{1}{V} \sum_\veck  \frac{ \widetilde{\epsilon}_{1\veck}}{ 2 E_{1\veck}^3 } \bigg)^2 \bigg] ,\label{eq:g2effL3}
\end{equation}
where $E_{1 \veck}   = \sqrt{  \widetilde{\epsilon}_{1 \veck}^2  - |\widetilde{\Delta}_1|^2 } $
is the atomic Bogoliubov quasiparticle energy,
with $ \widetilde{\epsilon}_{1 \veck}  =  h_{1 \veck}  +2 g_1 (|\Psi_{1 0}|^2  + n_1)$
and $ \widetilde{\Delta}_{1}    =   \Delta_1 + g_1 \Psi_{1 0}^2 -  2 \alpha \Psi_{20}$.
In Eq.~\eqref{eq:g2effL3}, $\widetilde{\bar{\alpha}}$ and $ \widetilde{\bar{g}}_2$ are two interaction parameters, related to
the Feshbach coupling $\bar{\alpha}$ and molecule-molecule interaction $\bar{g}_2$ by
$ \widetilde{\bar{\alpha}}  =\sqrt{2}\alpha/(1+g_1 V^{-1} \sum_\veck  1/2 E_{1\veck})$
and $ \widetilde{\bar{g}}_2  =g_2/(1+g_2 V^{-1} \sum_\veck 1/2 E_{2\veck})$. 
It is important to note that Eq.~\eqref{eq:g2effL3} is applicable as long as the detuning
$\bar{\nu} $ is smaller than and not too close to its QCP value $\bar{\nu}_c$
as we have already set $\Psi_{10}=0$. 
Evaluating Eqs.~\eqref{eq:ammeffdef} and \eqref{eq:g2effL3}
leads to the $\ammeff$ plots in Figs.~\ref{fig:Fig1}(c) and \ref{fig:Fig1}(d). 

The overall minus sign associated with the term $\sum_\veck 1/2 E_{1\veck}^3$ in Eq.~\eqref{eq:g2effL3}
should be noted.
This term arises from contributions related to the scattering process shown in Fig.~\ref{fig:Sketch}(b). 
It should be clear that the presence of $E_{1\veck}$ in the denominator,
reflects the fact that the excitations involved in the intermediate scattering state are atomic Bogoliubov quasiparticles.

In general, the $\veck$ integral in Eq.~\eqref{eq:g2effL3} can not be done analytically.
However, for the narrow resonance case and detuning near the QCP, Eqs.~\eqref{eq:g2effL3} and \eqref{eq:ammeffdef} can be further simplified
to yield the simple analytical expression for the many-body scattering length $\ammeff$~\cite{supplement} presented in Eq.~\eqref{eq:ammfiniten}.
In Figs.~\ref{fig:Fig4}(c) and \ref{fig:Fig4}(d), we numerically evaluate $\ammeff$ for a generic detuning in both the narrow and wide resonance cases,
and compare it with the numerically calculated compressibility inverse $\kappa^{-1}$
~\footnote{At large negative detuning $\bar{\nu}$, the many-body scattering length $\ammeff$ approaches its two-body background counterpart $\ammbg$.}. 
Notably, the two, $\ammeff$ and $\kappa^{-1}$, show rather precise agreement, for both the narrow and wide resonances at detunings away from
the immediate vicinity of the QCP
\footnote{We note that the more extended unstable regime in Fig.~\ref{fig:Fig1}(a) for a wide resonance
is associated with the presence of a sizable number of atom Cooper pairs~\cite{supplement}, 
which is also a consequence of the larger Feshbach coupling strength $\bar{\alpha}$. 
}. 

All of this allows us to understand why the compressibility in Fig.~\ref{fig:Fig1}(a) and \ref{fig:Fig1}(b)
behaves so differently when comparing the behavior of
wide and narrow resonances.
For a very narrow resonance, because of the small Feshbach coupling $\bar{\alpha}$, the factor $\widetilde{\bar{\alpha}}^4 $
in Eq.~\eqref{eq:g2effL3}, which is proportional to $\bar{\alpha}^4$,
is very small. Consequently the effective scattering length $\ammeff$ becomes negative only when the atomic Bogoliubov quasiparticle energy gap
(contained in $E_{1\veck=0}$) is sufficiently small. This insures that
the detuning at which $\ammeff$ changes sign is sufficiently close to the QCP.
In contrast, for a wide resonance this occurs when the gap $E_{1\veck=0}$ is still large, which corresponds to a detuning well away from the QCP. 

\textbf{Zero density limit ---}The expressions for the two-body scattering length $\amm$, presented
in Eqs.~\eqref{eq:ammwide} and \eqref{eq:ammnarrow}, were obtained from the zero-density limit of $\ammeff$ in Eq.~\eqref{eq:ammeffdef},
which leads to
\begin{align} \label{eq:ammunify}
 \amm =  \ammbg  - \frac{\ammbg }{k_b^4 r_* \bar{a}^3}  A_1 - \frac{1}{k_b^6 r_*^2 \bar{a}^3} A_2 +  \frac{\asbg}{k_b^8 r_*^2 \bar{a}^6} A_3,
\end{align}
where $r_*$ was defined in Eq.~\eqref{eq:rstar} and $k_b=\sqrt{m E_b}/\hbar$ is the detuning-dependent momentum corresponding to
the molecular binding energy $E_b$. The three dimensionless and positive coefficients
$\{A_1,A_2,A_3\}$ contain sub-leading dependences on $1/k_b$~\cite{supplement}, and their expressions are given in the Supplemental Material~\cite{supplement}. 
To arrive at Eqs.~\eqref{eq:ammwide} and \eqref{eq:ammnarrow}, we retain the background contribution $\ammbg$ in Eq.~\eqref{eq:ammunify} along with the $A_1$ and $A_2$ terms,
which are dominant for the wide and narrow resonances, respectively, near the boundary detuning $\bar{\nu}_{c,-}$ in Fig.~\ref{fig:Fig1}.

\textbf{Conclusions--}
In this Letter we have conducted an in-depth study on the stability of molecular condensates near Feshbach resonances in bosonic atoms, establishing the relationship to the inter-molecular scattering lengths.
In contrast with the Fermi gases, there has, thus far, been very little theoretical work on
characterizing this property for those molecules comprised of bosonic atoms. 
Such calculations can not naturally build on past work for the Fermi systems~\cite{Levinsen2006} 
as in the Bose
problem in the more immediate vicinity of resonance one has to contend with Efimov and tetramer bound states
as well as other background contributions.
With the caveat that, following experiment
~\cite{Zhang2021,Zhang2023},
we are not too close to resonance (where instability is
guaranteed) we have shown how just as Feshbach resonances allow tuning of the scattering
lengths of atoms, they also modify the scattering length of molecules constituted from these atoms.

Our predictions are ripe for experimental validation.
The concrete expressions for the molecular scattering lengths we derived make it possible to address
a number of issues that were previously inaccessible; these include, among others, the equilibrium equation of state
in the dilute molecular gas regime, the expansion dynamics, the collective modes,
and the molecule-molecule scattering induced relaxation~\cite{Petrov2005,Levinsen2006,Levinsen2011,
Chin2005,Chen2023}.

\textbf{Acknowledgment ---} We thank P. Julienne and B. Zhao for helpful discussions.
Z. W. and Q. C. were supported by the Innovation Program for Quantum Science and Technology (Grant No. 2021ZD0301904). 
This work was also supported by the National Science Foundation under Grant No. PHY-2409612 and No. PHY-2103542,
and by the Air Force Office of Scientific Research under Award No. FA9550-21-1-0447.
Z. Z. acknowledges the Bloch Postdoctoral Fellowship.
We also acknowledge the University of Chicago's Research Computing Center for their support of this work.


\bibliography{References}

\end{document}


\title{
\large{\textbf{Supplementary Materials for: \\
Tunable Molecular Interactions Near an Atomic Feshbach Resonance: Stability and Collapse of a Molecular Bose-Einstein Condensate
}
}
}

\author{Zhiqiang Wang}
\email[]{wzhiqustc@ustc.edu.cn}
\affiliation{Hefei National Research Center for Physical Sciences at the Microscale and School of Physical Sciences, University of Science and Technology of China,  Hefei, Anhui 230026, China}
\affiliation{Shanghai Research Center for Quantum Science and CAS Center for Excellence in Quantum Information and Quantum Physics, University of  Science and Technology of China, Shanghai 201315, China}
\affiliation{Hefei National Laboratory, University of  Science and Technology of China, Hefei 230088, China}
\affiliation{Department of Physics and James Franck Institute, University of Chicago, Chicago, Illinois 60637, USA}
\author{Ke Wang}
\affiliation{Department of Physics and James Franck Institute, University of Chicago, Chicago, Illinois 60637, USA}
\author{Zhendong Zhang}
\affiliation{E. L. Ginzton Laboratory and Department of Applied Physics, Stanford University, Stanford, CA 94305, USA}
\affiliation{Department of Physics and Hong Kong Institute of Quantum Science and Technology,
The University of Hong Kong, Hong Kong, China}
\author{Qijin Chen}
\affiliation{Hefei National Research Center for Physical Sciences at the Microscale and School of Physical Sciences, University of Science and Technology of China,  Hefei, Anhui 230026, China}
\affiliation{Shanghai Research Center for Quantum Science and CAS Center for Excellence in Quantum Information and Quantum Physics, University of  Science and Technology of China, Shanghai 201315, China}
\affiliation{Hefei National Laboratory, University of  Science and Technology of China, Hefei 230088, China}
\author{Cheng Chin}
\affiliation{Department of Physics and James Franck Institute, University of Chicago, Chicago, Illinois 60637, USA}
\affiliation{Enrico Fermi Institute, University of Chicago, Chicago, Illinois 60637, USA}
\author{K. Levin}
\affiliation{Department of Physics and James Franck Institute, University of Chicago, Chicago, Illinois 60637, USA}
\date{\today}

\maketitle

\beginsupplement

\tableofcontents

 \section{Hamiltonian and choices of interaction parameters}
 
All our analysis starts with the following two-channel Hamiltonian $H$,
\begin{align} \label{eq:Hamiltonian}
\hat{H} & = 
\sum_{\sigma=1}^2  \sum_{\veck}  h_{\sigma \veck} a_{\sigma \veck}^\dagger a_{ \sigma \veck}   
+ \sum_{\sigma=1}^2  \frac{1}{V}\sum_{\veck_1,\veck_2,\veck_3}\frac{g_\sigma}{2}  
a^\dagger_{ \sigma \veck_1 } a^\dagger_{ \sigma \veck_2 } a_{  \sigma \veck_3} a_{ \sigma , \veck_1 +\veck_2 -\veck_3 }  
 - \frac{\alpha}{\sqrt{V}} \sum_{\veck_1,\veck_2}\big( a^\dagger_{1 \veck_1 } a^\dagger_{1 \veck_2 } a_{ 2, \veck_1+\veck_2} + h. c.\big). 
\end{align}
In this Hamiltonian,  $a_{\sigma \veck}$ is an annihilation operator for the open-channel atoms (with $\sigma=1$) or closed-channel molecules ($\sigma=2$).
In the first term of $H$ above, $h_{1\veck}  = \veck^2/2m_1 -\mu$ and $h_{2\veck}  =  \veck^2/2 m_2 - (2 \mu - \nu)$ are the kinetic energy contributions for the atoms and molecules,
respectively, with $m_1$ and $m_2$ the corresponding atomic and  molecular mass,
$\mu$ the chemical potential, and $\nu$ the \textit{bare} closed-channel molecule detuning. 
The second term $g_\sigma$ is an intra-channel density-density interaction, which we consider repulsive. 
The third term is the \textit{bare} Feshbach coupling with $\alpha$ the corresponding coupling strength. 
The summation $V^{-1}\sum_{\veck}$, with $V$ the volume, represents a three dimensional integral over the momentum $\veck$, $\int^\Lambda d \veck/(2\pi)^3$, 
where $\Lambda$ is a cutoff that is needed to regularize ultraviolet divergences in various $\veck$ integrals because of our use of contact interactions in the Hamiltonian.
In our theory, we assume three-dimensional isotropy and ignore trap effects.

\subsection{Regularization}
\label{sec:parameter}

The relation between the interaction parameters $\{g_1, g_2, \alpha\}$, as well as the detuning $\nu$, and experimentally measured Feshbach resonance parameters
can be summarized by
\begin{subequations} \label{eq:regularization}
\begin{align}
\alpha  & = \frac{\bar{\alpha} }{\sqrt{2}} \frac{1}{1-\asbg/\bar{a}} ,    \label{eq:alpharelation} \\
\nu  & = \bar{\nu} + \frac{m_1 }{4\pi \hbar^2 \bar{a}}    \frac{ \bar{\alpha}^2}{1-\asbg/\bar{a}} , \label{eq:nurelation} \\
g_1  & =  \bar{g}_1  \frac{1}{1-\asbg/\bar{a}} ,  \label{eq:g1}\\
g_2 & = \bar{g}_2 \frac{1}{1-\ammbg/\bar{a}} , \label{eq:g2} \\
\Lambda   & = \pi/ (2\bar{a}),  \label{eq:abar}
\end{align}
\end{subequations}
where
\begin{subequations} \label{eq:renormquant}
\begin{align}
\bar{\nu}   & = \Delta \mu_m (B- B_0), \\
\bar{\alpha}^2 & =  4\pi \hbar^2 \asbg \Delta \mu_m \Delta B/ m_1 , \\
\bar{g}_1 &  = 4\pi \hbar^2 \asbg / m_1, \\
\bar{g}_2 &  = 4\pi \hbar^2 \ammbg / m_2. 
\end{align}
\end{subequations}
Because the relations in Eq.~\eqref{eq:regularization} involve the momentum cutoff $\Lambda$, which is used to regularize ultraviolet divergences in integrals over the momentum,
we also call these relations regularization conditions. 
In the above equations, we have denoted quantities that are directly related to experimental observables, such as $\bar{g}_1$ and $\bar{g}_2$,  by a bar atop. 
We also define $\asbg$ and $\ammbg$ as the atomic and molecular background scattering lengths, respectively,
$B_0$ as the experimental resonance point, $\Delta B$ as the corresponding width of the atomic Feshbach resonance,
and $\Delta \mu_m$ as the magnetic moment difference between a closed-channel molecule and a pair of atoms in the open channel. 
The length scale $\bar{a}$ is defined from the momentum cutoff $\Lambda$ in our contact interaction calculation by $\bar{a}=\pi/(2\Lambda)$. 
When making connections to calculations with a more realistic van der Waals potential, 
one can take $\bar{a} \approx 0.96 R_{\text{vdW}} $~\cite{Lange2009}, where $R_{\text{vdW}} $ is the van der Waals length.

Detailed derivations of the relations in Eq.~\eqref{eq:regularization} can be found in Ref.~\cite{Wang2024}. 
These relations are chosen such that in the two-atom scattering limit, the atomic $s$-wave scattering length $a_s$
takes the following well-known form, 
\begin{align}
a_s &  = \asbg ( 1- \frac{\Delta B}{B-B_0} ), \label{eq:asdef}
\end{align}
which can be also rewritten as
\begin{align}
a_s &  = \asbg - \frac{m_1}{4\pi \hbar^2} \frac{\bar{\alpha}^2}{\bar{\nu}}
\end{align}
using Eq.~\eqref{eq:renormquant}. 

\subsection{Choice of interaction parameters}
\label{sec:parameters}

Our numerical studies in the main text involve both narrow and wide resonances. To be concrete, for the narrow resonance, we use the experimental values of the parameters $\{\asbg, \ammbg, m_1, \Delta \mu_m, \Delta B \}$ for the resonance of \Cs~at magnetic field $B=19.849$G from Refs.~\cite{Zhang2021,Zhang2023} as input. 
We only consider cases where both the atomic background scattering length $\asbg$ and the molecular background scattering length $\ammbg$ are positive, as found in the \Cs~experiments~\cite{Zhang2021,Zhang2023}.
The positive $\asbg$ and $\ammbg$ allow for a stable atomic superfluid phase and molecular superfluid phase
to exist in the phase diagrams we explore. 
Combining these experimental parameters with the regularizations in Eq.~\eqref{eq:regularization} and choosing $\Lambda=\pi/5/\asbg$, which is equivalent to $\bar{a}=(5/2) \asbg$ for our numerical calculation, we can then uniquely determine the interaction parameters $\{g_1, g_2,\alpha \}$,
as well as the bare detuning $\nu$ for a given physical detuning $\bar{\nu}$. 
Details about our chosen parameters for the narrow-resonance simulation are shown in Table~\ref{tab:parameters}. 

\begin{table}[htp]
    \center
    \caption{Parameters used in our numerical simulations for the narrow resonance in this paper. The values of the bare interaction parameters $\{ \alpha, g_1, g_2 \}$
    are determined from the experimental value of the parameters
    $\{\asbg, \ammbg, \Delta \mu_m , \Delta B \}$ for the resonance of \Cs~at magnetic field $B=19.849$G~\cite{Zhang2021,Zhang2023}, using the relations in Eq.~\eqref{eq:regularization}.  }
    \begin{tabular}{ | c |  c |  c | c |   }
       \hline
       \hline
           ~~ $\asbg$~~               &     ~~ $\ammbg$ ~~                       &    ~~ $\Delta \mu_m$ ~~                                &     ~~ $\Delta B$  ~~             \\
        \hline
            ~~  163 $a_{\text{B}}$  ~~        &    ~~  220 $a_{\text{B}}$ ~~  &    ~~  0.57 $\mu_{\text{B}}$ ~~              &     ~~  8.3 mG  ~~          \\
        \hline
        \hline
           ~~ $\Lambda$~~               &     ~~ $\alpha$ ~~                       &    ~~ $g_1$ ~~                                &     ~~ $g_2$  ~~             \\
        \hline
            ~~ $\pi/5$ $\asbg^{-1}$ ~~        &    ~~ 0.34 $\hbar^2/(m_1 \sqrt{\asbg})$ ~~  &    ~~  $20.9$ $ \hbar^2\asbg/m_1$ ~~              &     ~~   $18.4$ $\hbar^2 \asbg/m_1$  ~~          \\
        \hline
        \hline
          \end{tabular}
    \label{tab:parameters}
\end{table}

\subsubsection{Narrow versus wide}
\label{sec:narrowide}

In the paper we draw attention to the distinction between wide and narrow resonances.
The resonance of \Cs~at $B=19.849$G is considered very narrow. There are two different
standards for how to quantify this, associated with a two-body ~\cite{Chin2010} and a many-body classification 
scheme~\cite{Ho2012}.

In the two-body scheme, one can define the following length-scale parameter $r_*$~\cite{Gurarie2007,Levinsen2011,Chin2010,Ho2012}
from the Feshbach coupling $\bar{\alpha}$,
\begin{align}
r_* & \equiv \frac{4\pi \hbar^4}{m_1^2 \bar{\alpha}^2} =  \frac{\hbar^2}{m_1 \asbg \Delta \mu_m \Delta B},  \label{eq:rstar}
\end{align}
to classify the resonance width.
Using $r_*$ and $\asbg$ we can then define a dimensionless resonance width parameter~\cite{Ho2012} $\mathit{w}_{\mathrm{res}}=\asbg/r_*$.
 If this width parameter is much smaller than unity, $w_{\mathrm{res}}=\asbg/r_*\ll 1$,  the resonance is called narrow; otherwise, it is wide. Using the parameters for \Cs~from experiments (see Table~\ref{tab:parameters}) we find that $\mathit{w}_{\mathrm{res}} \approx 0.006 \ll 1$. 
 
In the many-body scheme, another dimensionless resonance width parameter, $x\equiv 1/(k_n r_*)$, is defined, where $k_n=(6\pi^2 n)^{1/3}$.
This parameter $x$ depends on the atom number density. For the density used in Refs.~\cite{Zhang2021,Zhang2023}, which will be our primary focus, we estimate~\cite{Wang2024} that $x\approx 0.07 \ll 1$. 
In either scheme, the resonance of \Cs~at $B=19.849$G can be 
established to be very narrow.

For our wide-resonance simulation, instead of extracting parameters from actual experiments for a particular atomic superfluid, 
we simply increase the resonance width $\Delta_{\bar{\nu}} = \Delta \mu_m \Delta B$ from our narrow-resonance value by $10^2$ times
while leaving all other parameters the same as in Table~\ref{tab:parameters}. 
This results in a width parameter of $\mathit{w}_{\mathrm{res}} =0.6 \sim 1$ and $x=7 \gg 1$.
In the two-body classification scheme, this resonance is only moderately wide since the dimensionless parameter $\mathit{w}$ is of order unity  $\mathit{w}_{\mathrm{res}} \sim 1$, 
which we adopt so that our numerical calculations can be made easier.
We emphasize that considering an even wider resonance does not change our conclusions in the paper. 

\section{Expression of the two-body molecular scattering length $\amm$}
\label{sec:ammexpression}

In this section, we provide additional details about the general formula of the two-body molecular scattering length $\amm$ presented in Eq. (11) of the main text.
For the convenience of the following discussions, we have reproduced the equation here:
\begin{align} \label{eq:ammunify}
a_{\mathrm{mm}}   =  \ammbg  - \frac{\ammbg}{k_b^4 r_* \bar{a}^3}  A_1  - \frac{1}{k_b^6 r_*^2 \bar{a}^3} A_2 +  \frac{\asbg}{k_b^8 r_*^2 \bar{a}^6} A_3. 
\end{align}
In this equation, the length scale parameter $r_*$ is related to the width of the Feshbach resonance by Eq.~\eqref{eq:rstar}. 
The dependence on the detuning $\bar{\nu}=\Delta \mu_m (B-B_0)$ enters the scattering length $\amm$ in Eq.~\eqref{eq:ammunify} through 
the binding momentum $k_b=\sqrt{m_1 E_b}/\hbar$, where $E_b$ is the molecular binding energy. 
Expressions for the momentum $k_b$ will be discussed in the subsequent Section~\ref{sec:kbdiscussion}.  

Eq.~\eqref{eq:ammunify} is written in such a way as to explicitly show only the leading order dependences on $1/k_b$ for each term.
The coefficients $A_1,A_2$, and $A_3$ are \textit{dimensionless} parameters that include subleading order dependences on $1/k_b$, as well as additional dependence 
on other relevant length scale parameters of the problem. As $k_b$ approaches infinity (i. e. when $\bar{\nu} \rightarrow -\infty$), $A_1 , A_2$ and $A_3$ converge to finite constants. 

Detailed derivations of Eq.~\eqref{eq:ammunify} will be presented in Sections~\ref{sec:zeronlimit}.
In the following discussion, we first focus on the general expressions of the coefficients $\{A_1 ,A_2,A_3\}$ 
and how we derive the various limiting forms of the two-body scattering length $\amm$ from Eq.~\eqref{eq:ammunify},
as presented in Eqs. (2) and (3) of the main text. 

\subsection{Detailed expressions for the coefficients $A_1,A_2$ and $A_3$ in Eq.~\eqref{eq:ammunify} of $\amm$}
The expressions for the coefficients $\{A_1,A_2,A_3\}$ in Eq.~\eqref{eq:ammunify} are given by
\begin{subequations}
 \begin{align}
 A_1 & =  (k_b \bar{a})^2  \, C_2  \frac{ (2 k_b r_*)^2 (1-  C_1 \asbg/\bar{a} )^2  + C_2 r_*/\bar{a} }{\big[ 2 k_b r_* (1-  C_1 \asbg/\bar{a} )^2  +  C_2/ (k_b \bar{a})\big]^2},    \label{eq:A}\\
 A_2 & = (k_b \bar{a})^2 \, C_3  \frac{ (2 k_b r_*)^2 }{\big[  2 k_b r_* (1- C_1 \asbg / \bar{a} )^2   + C_2/ (k_b \bar{a}) \big]^2 },  \label{eq:B} \\
 A_3  & =4 \big( (k_b \bar{a})^2 \, C_2 \big)^2   \frac{ (2 k_b r_*)^2 }{\big[ 2 k_b r_* (1-  C_1 \asbg/\bar{a} )^2  +  C_2/ (k_b \bar{a})\big]^2}.   \label{eq:C}
 \end{align}
 \end{subequations}
Here, $C_1, C_2$, and  $C_3$ are three \textit{dimensionless} functions of $k_b \bar{a}$, i. e.
\begin{subequations} \label{eq:C123}
 \begin{align}
 C_1 & = \bigg(  y \tan^{-1} \frac{1}{ y}  \bigg)  \bigg\vert_{y=(2/\pi) k_b \bar{a}},  \label{eq:C1} \\
 C_2 & = \bigg( y \tan^{-1} \frac{1}{ y }  -   \frac{y^2 }{1+ y^2}  \bigg) \bigg\vert_{y=(2/\pi) k_b \bar{a}},    \label{eq:C2} \\
 C_3 &  =   \bigg( y \tan^{-1} \frac{1}{y} + \frac{y^2(1-y^2)}{(1+y^2)^2} \bigg) \bigg\vert_{y=(2/\pi) k_b \bar{a}},   \label{eq:C3} 
 \end{align}
 \end{subequations}

\subsection{Relation between the binding momentum $k_b$ and the detuning $\bar{\nu}$}
\label{sec:kbdiscussion}

In general, the binding momentum $k_b=\sqrt{m_1 E_b}/\hbar$ in Eq.~\eqref{eq:ammunify} depends on $\bar{\nu}=\Delta \mu_m (B-B_0)$ in a complicated way.
To find out this dependence, one needs to solve the two-atom scattering problem in the presence of both the Feshbach coupling $\alpha$
and the atom background interaction $g_1$.
By deriving the molecule propagator as a function of energy,
one can identify the pole of the propagator and from the pole condition determine the bound-state energy $E_b$ (or equivalently $k_b$) for the dressed molecule,
which consists of components from both the closed-channel and the open channels. 
The derivation can be done by following previous studies on the BCS-BEC crossover
in a \textit{fermionic} Feshbach resonance~\cite{Ohashi2002,Marcelis2004,Falco2007}. 
In the end, finding the poles is equivalent to solving the following transcendental equation of $k_b$, 
\begin{subequations}  \label{eq:root}
\begin{align}
\big(k_b^2 + \frac{m_1 \bar{\nu}}{\hbar^2}  \big)  + \bigg(  \frac{1}{r_* \bar{a}}   - \big(k_b^2 + \frac{m_1 \bar{\nu}}{\hbar^2} \big)  \frac{\asbg} {\bar{a}}\bigg) C_1(k_b \bar{a})  =0,  \\
\text{with  }  C_1(k_b \bar{a}) = \bigg(  y \tan^{-1} \frac{1}{ y}  \bigg)  \bigg\vert_{y=(2/\pi) k_b \bar{a}}.
\end{align}
\end{subequations}
We note that, in the limit of $\bar{a} \rightarrow 0$ while keeping all other parameters finite, Eq.~\eqref{eq:root} reduces to a cubic equation in the momentum $k_b$ (see also Eq.~\eqref{eq:root2} below).
This cubic equation has previously appeared in Ref.~\cite{Falco2007}, but in the context of a fermionic resonance study with an attractive atomic background interaction ($\asbg<0$).

In general, an analytical and closed-form solution to Eq.~\eqref{eq:root} is impossible. 
However, an approximated solution can be obtained by considering three different asymptotic detuning regimes. 
\begin{itemize}
\item \textbf{Large negative detuning regime I : } 
\begin{align}
\text{i. e.  }  \quad (-\bar{\nu})  \gg \hbar^2 / m_1\bar{a}^2 \quad  \& \&  \quad   (-\bar{\nu})  \gg \hbar^2 / (m_1 r_* \bar{a}).
\end{align}
This regime exists for both narrow and wide resonances. 
In this regime, the binding momentum $k_b \sim \sqrt{m_1 (-\bar{\nu})}/\hbar$ dominates over the inverse of all other length scales. 
In particular, we have the relation $k_b \bar{a} \gg 1$, which leads to $C_1(k_b \bar{a}) \approx 1$ in Eq.~\eqref{eq:root}. Under these conditions, one finds from Eq.~\eqref{eq:root} that 
\begin{align}
 E_b  = \frac{\hbar^2 k_b^2}{m_1} \approx -\bar{\nu}  + \frac{\hbar^2}{m_1 r_*} \frac{1}{\asbg -\bar{a}} \approx -\bar{\nu}. 
\end{align}

\item \textbf{Small negative detuning regime II : }  
\begin{align}
\text{i. e. } \quad  E_b \ll  (-\bar{\nu}) \ll \frac{\hbar^2}{m_1 \bar{a}^2}. 
\end{align}
This is the universal regime where the atom background scattering length $a_s$ dominates over all other length scales. 
For this regime, the binding momentum is small $k_b \bar{a} \ll 1$ so that Eq.~\eqref{eq:root} can be rewritten as
\begin{align}
\big(k_b^2 + \frac{m_1 \bar{\nu} }{\hbar^2}  \big)  + \bigg(  \frac{1}{r_*}   - \big( k_b^2 +  \frac{ m_1 \bar{\nu} }{\hbar^2} \big)  \asbg \bigg)  k_b \approx 0.  \label{eq:root2}
\end{align}
This is a cubic equation in $k_b$. It can be further simplified if we assume that the detuning is close to the resonance enough such that $E_b\ll (- \bar{\nu})$. Then the factor $k_b^2 + m_1 \bar{\nu}/\hbar^2$ in Eq.~\eqref{eq:root2} can be replaced by $m_1 \bar{\nu}/\hbar^2$, leading to
\begin{subequations}
\begin{align}
k_b  &  \approx \big( \asbg - \frac{\hbar^2}{r_* m_1 \bar{\nu}} \big)^{-1}= \frac{1}{a_s},  \label{eq:kbasII} \\
\Rightarrow  E_b &  =\frac{\hbar^2 k_b^2}{ m_1} \approx \frac{\hbar^2}{m_1 a_s^2}. 
\end{align}
\end{subequations}
To obtain the first line above we have used the the expression of the atomic scattering length $a_s$ from Eq.~\eqref{eq:asdef} and the definition of the length scale $r_*$ from Eq.~\eqref{eq:rstar}.

\item \textbf{Intermediate negative detuning regime III: } 
\begin{align}
\text{i. e. } \quad   \frac{\hbar^2}{m_1 r_*^2} \ll  (- \bar{\nu} ) \ll \frac{\hbar^2}{m_1 \asbg^2} \sim \frac{\hbar^2}{m_1 \bar{a}^2}.
\end{align} 
This regime exists only for a narrow resonance since it requires $r_* \gg 1/\bar{a}\sim 1/ |\asbg|$. 
We consider this detuning regime explicitly because it is the one where the molecular scattering length $\amm$ changes its sign for a narrow resonance,
which is most relevant to our stability discussions. 
For this regime, one can show that
\begin{align}
 E_b  \approx -\bar{\nu}. 
\end{align}
\end{itemize}

To summarize, for a narrow resonance one can use the relation $E_b \approx -\bar{\nu}$ for nearly all detunings, except when the detuning is very close
to the resonance $(-\bar{\nu}) \lesssim \hbar^2/m_1 (r_*)^2$, a range that is negligible if the resonance is sufficiently narrow. On the other hand, for a wide resonance, 
the relation $E_b \approx - \bar{\nu}$ holds at large negative detuning
while it is replaced by a different one $E_b= \hbar^2 / m_1 a_s^2$ when the detuning is small.

\subsection{ Asymptotic formulae for the molecular scattering length $\amm$}
\label{sec:amm123}

We first emphasize that all the formulae for the scattering lengths $\amm$ and $\ammeff$ derived below are applicable only for negative detuning $\bar{\nu}<0$, i. e. $B<B_0$. 

\subsubsection{Large negative detuning regime I for both narrow and wide resonances} 

In this regime, the terms with the coefficients $A_2$ and $A_3$ in Eq.~\eqref{eq:ammunify} can be dropped as the binding momentum $k_b$ is large and these terms are higher order in $1/k_b$.
Given that $k_b \approx \sqrt{m_1 (-\bar{\nu})}/\hbar =  \sqrt{m_1 \Delta \mu_m (B_0-B)}/\hbar$, Eq.~\eqref{eq:ammunify} then becomes
\begin{subequations}
\begin{align} 
a_{\mathrm{mm}}  &  \approx  \ammbg  - \frac{\ammbg}{k_b^4 r_* \bar{a}^3}  A_1 = \ammbg \bigg[ 1 - \bigg( \frac{\Delta_B^{\text{wide}}}{B_0-B} \bigg)^2  \bigg],
 \label{eq:ammwide}
 \end{align}
with
\begin{align}
\Delta_B^{\text{wide}}   & =  \frac{\pi}{\sqrt{6}} \sqrt{ \frac{\hbar^2}{ m_1 (\bar{a} - \asbg)^2 \Delta \mu_m} }  \sqrt{\Delta B}
\end{align}
\end{subequations}
an effective molecular resonance width. 
To arrive at this expression we have used the definition of $r^*$, i. e. $r_*=\hbar^2/(m_1 \asbg \Delta \mu_m \Delta B)$ 
and also the following approximated expression for the coefficient $A_1$ (see Eq.~\eqref{eq:A}
for its full expression),
\begin{align}
A_1  & \approx  \frac{\pi^2}{6} \frac{1}{(1-\asbg/\bar{a})^2}
\end{align}
which is obtained from Eq.~\eqref{eq:A} by using 
$ C_1   =  1- \frac{1}{3 y^2} + \mathcal{O}( \frac{1}{y^4} )$, 
$ C_2   =   \frac{2}{3 y^2} + \mathcal{O}( \frac{1}{y^4} )$, 
$ C_3   =  \frac{8}{3 y^2} + \mathcal{O}( \frac{1}{y^4} )$ in 
Eqs.~\eqref{eq:C1}, \eqref{eq:C2},\eqref{eq:C3}, respectively.

\subsubsection{Small negative detuning regime II for a wide resonance} 

In this regime, the binding momentum $k_b$ satisfies $k_b r_* \ll \{ k_b \asbg, k_b \bar{a} \} \ll 1$ so that in Eqs.~\eqref{eq:C1}, \eqref{eq:C2},\eqref{eq:C3}, 
$C_1  \approx  C_2  \approx C_3  \approx k_b \bar{a} $.
Then from Eqs.~\eqref{eq:A},\eqref{eq:B}, and \eqref{eq:C}, we obtain the approximate results $A_1 \approx  k_b^4 r_* \bar{a}^3$, $A_2 \approx   4 k_b^5 {r_*}^2 \bar{a}^3$,
and $A_3  \approx   16 k_b^8 {r_*}^2 \bar{a}^6 $. 
Substituting these results into Eq.~\eqref{eq:ammunify} leads to
\begin{subequations} 
\begin{align} 
a_{\mathrm{mm}}   & =  \ammbg  - \frac{\ammbg}{k_b^4 r_* \bar{a}^3}  A_1 - \frac{1}{k_b^6 r_*^2 \bar{a}^3} A_2 +  \frac{\asbg}{k_b^8 r_*^2 \bar{a}^6} A_3    \\
   & \approx \ammbg -\ammbg - \frac{4}{k_b} + 16 \asbg  \\
   & = 12  \asbg \bigg[ 1 - \frac{ \Delta B/3}{B_0-B|} \bigg].    \label{eq:ammwide2}
\end{align}
\end{subequations}
To arrive at the last expression we have used the approximate relation $k_b \approx 1/a_s$ from Eq.~\eqref{eq:kbasII} in Section~\ref{sec:kbdiscussion}
as well as Eq.~\eqref{eq:asdef} for the atomic scattering length $a_s$. 

\subsubsection{Intermediate negative detuning regime III for a narrow resonance}
This is the detuning regime where $\amm$ changes its sign for a narrow resonance.
To be specific, we consider the intermediate detuning regime of $\bar{\nu}$ such that
$\{1/\bar{a} \sim 1/ |\asbg| \} \gg \sqrt{m_1 (-\bar{\nu})}/\hbar \gg 1/\sqrt{r_* |\ammbg|} \gg 1/r_*$.
Then one can show that the $A_1$ and $A_3$ terms in Eq.~\eqref{eq:ammunify} are subdominant than the $A_2$ term by using $k_b \approx \sqrt{m_1 (-\bar{\nu})}/\hbar$,
$ A_1 \approx  (  k_b \bar{a} )^3 $, $A_2  \approx  (  k_b \bar{a} )^3$ and $ A_3  \approx   (  k_b \bar{a} )^6 $ in Eqs.~\eqref{eq:A},\eqref{eq:B}, and \eqref{eq:C},
for which we have used $ C_1  \approx   C_2  \approx   C_3  \approx k_b \bar{a}$ from Eqs.~\eqref{eq:C1}, \eqref{eq:C2},\eqref{eq:C3}.
Dropping the $A_1$ and $A_3$ terms in Eq.~\eqref{eq:ammunify} and using the definition of $r_*$ from Eq.~\eqref{eq:rstar} lead to
\begin{subequations} \label{eq:ammnarrow}
\begin{align}
\amm  & \approx \ammbg \bigg[ 1-  \bigg( \frac{ \Delta_B^{\text{narrow}} }{  B_0-B} \bigg)^{\frac{3}{2}}   \bigg],
\end{align}
where
\begin{align}
 \Delta_B^{\text{narrow}}    & =  \bigg( \frac{\asbg}{\ammbg} \bigg)^{ \frac{2}{3}}  \bigg( \frac{ m_1 \asbg^2 \Delta \mu_m }{\hbar^2}   \bigg)^{\frac{1}{3}}  \;  (\Delta B)^{\frac{4}{3}} 
 \label{eq:DeltaBn}
\end{align}
\end{subequations}
is the counterpart molecular resonance width. 

\subsection{Diagrammatic representation of the terms in Eq.~\eqref{eq:ammunify} for the two-body molecular scattering length $\amm$}
\label{sec:diagrams}

We emphasize that the general expression of the molecular scattering length $\amm$ in Eq.~\eqref{eq:ammunify}
is not perturbative in the inverse of the binding momentum $1/k_b$ or in any of the interaction constants $\{ \alpha, g_1, g_2\}$
in the Hamiltonian. Instead, it contains contributions that are of infinite order in these parameters.
Equation~\eqref{eq:ammunify} was obtained via a self-consistent variational wavefunction calculation of the ground state energy and its quartic derivative,
evaluated in the zero-particle density limit. 

Nevertheless, we can give a Feynman diagram interpretation for each of the $\{A_1, A_2, A_3\}$ terms in 
Eq.~\eqref{eq:ammunify}, when the additional dependencies of the coefficients $\{A_1, A_2 , A_3\}$
on the interaction parameters as well as on the momentum $k_b$ can be ignored. 
These diagrams should be understood as the leading order contributions in the inverse of the binding momentum in a sum of infinite diagrams. 
In the following, we provide the corresponding diagrammatic interpretations  
and also explain how the power dependences of each of the three terms in Eq.~\eqref{eq:ammunify} on the inverse binding momentum $1/k_b$ arise. 

\subsubsection{Leading order diagram for the $A_1$ term}

\begin{figure}[htp]
\begin{center}
\includegraphics[width=0.9\linewidth]
 {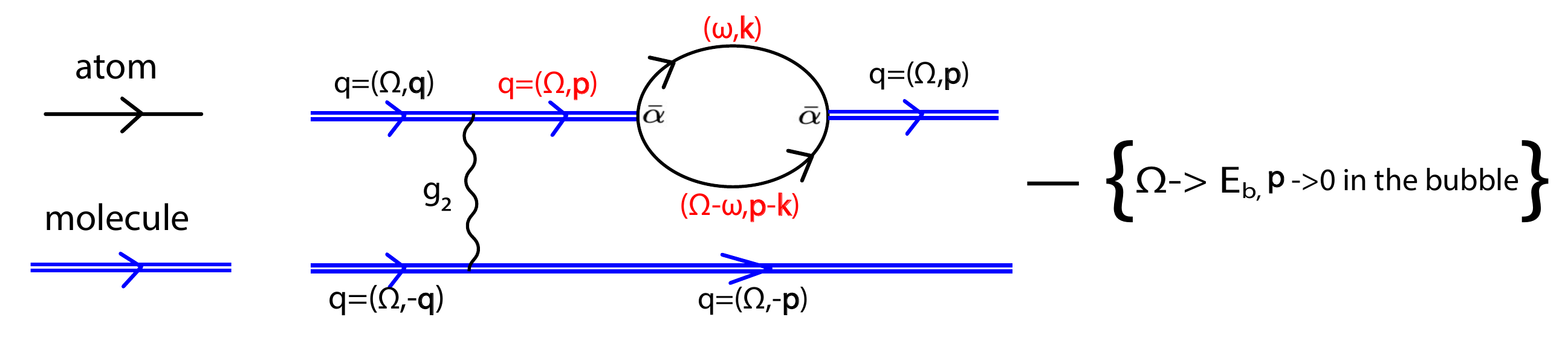}
\caption{One of the leading order diagrammatic contributions to the $A_1$ term in Eq.~\eqref{eq:ammunify}. (We suppress $\hbar$ in the labels of this figure, as well as in the subsequent Feynman diagrams.) 
There are also additional diagrams that contribute at the same order, differing in the placement of the particle-particle bubble within the diagram. 
In the diagram shown, $\bar{\alpha}$ is the Feshbach coupling strength while $g_2$ represents the molecular background interaction.
The four external molecular legs, whose frequency-momentum is indicated in black, do not contribute to the scattering amplitude.
The subtraction, $- \{ \hbar \Omega \rightarrow E_b, \vecp \rightarrow 0 \quad \text{in the bubble} \}$, is needed to ensure a finite result for the diagram
in the limit of $\{\hbar \Omega \rightarrow E_b,  |\vecq|=|\vecp| \rightarrow 0 \}$.
The term being subtracted is singular in this limit and it contributes to the renormalization of the $E_b$ value. 
To calculate the scattering amplitude contribution from the above diagram to $\amm$,
one must average the amplitude over the scattering angles and then consider the low-energy limit for both the incoming and outgoing molecules~\cite{Levinsen2006}.
From the above diagram we see that this contribution involves a combination of the molecule background interaction $g_2$
and a product of two Feshbach coupling constant, $\bar{\alpha}^2$,
leading to the $g_2 \bar{\alpha}^2$ factor in Eq.~\eqref{eq:diagramAl1}.
Additionally, it is evident from the diagram that the $A_1$ term in Eq.~\eqref{eq:ammunify} originates from a molecular self-energy effect. 
}
\label{fig:diagramA}
\end{center}
\end{figure}

One of the leading order diagrams for the $A_1$ term in Eq.~\eqref{eq:ammunify} can be represented by Fig.~\ref{fig:diagramA}, from which
one can deduce the associated factor $ \ammbg/ (k_b^4 r_* \bar{a}^3)$ as follows,
\begin{subequations} \label{eq:diagramA}
\begin{align}
\text{ Fig.~\ref{fig:diagramA}  } & \sim   g_2 \bar{\alpha}^2  \lim_{|\vecq|=|\vecp| \rightarrow 0, \hbar \Omega \rightarrow E_b} 
\bigg \langle G_2(\Omega,\vecp)   \bigg\{ 
i \int \frac{d\omega}{2\pi} \int^\Lambda \frac{d \veck}{(2\pi)^3}  
G_1(\omega, \veck) G_1( \Omega -\omega, \vecp -\veck)   \nonumber \\
& 
-  i \int \frac{d\omega}{2\pi} \int^\Lambda \frac{d \veck}{(2\pi)^3}  
G_1(\omega, \veck) G_1( E_b/\hbar -\omega, \vec{0} -\veck)    \bigg\}
 \bigg \rangle_{\hat{\vecp},\hat{\vecq}}                            \label{eq:diagramAl1} \\
& \propto  \frac{\ammbg}{r_*}   \lim_{ \hbar \Omega \rightarrow E_b}  
\frac{1}{ \hbar \Omega +i \delta - E_b}  i \int \frac{d\omega}{2\pi} \bigg\{ \frac{1}{ \hbar \omega + i\delta} \frac{1}{\hbar \Omega- \hbar \omega + i \delta} 
-  \frac{1}{\hbar \omega + i\delta} \frac{1}{E_b - \hbar \omega + i \delta} \bigg\}   \int^\Lambda  \frac{d \veck}{(2\pi)^3}  \label{eq:diagramAl2} \\
& \propto  - \frac{\ammbg}{k_b^4 r_* \bar{a}^3}.  \label{eq:diagramAl3}
\end{align}
\end{subequations}
In Eq.~\eqref{eq:diagramAl1}, $G_1(\omega,\veck) = (\hbar \omega + i \delta - \hbar^2 \veck^2/2m_1)^{-1}$ 
and $G_2(\Omega,\vecp) = (\hbar \Omega + i\delta - \hbar^2 \vecp^2/2m_2  -\nu )^{-1}$ 
represent the retarded non-interacting atom and molecule Green's functions in vacuum, respectively.
The limit $ \lim_{|\vecq|=|\vecp| \rightarrow 0, \hbar \Omega \rightarrow E_b}  \langle  \cdots \rangle_{\hat{\vecp},\hat{\vecq}}$ arises
because we are calculating the scattering amplitude contribution from Fig.~\ref{fig:diagramA} to the $s$-wave scattering length $\amm$.
For this calculation, it is necessary to average the amplitude in Fig.~\ref{fig:diagramA}, over the scattering angle between $\vecq$ and $\vecp$ 
and also to consider the on-shell condition~\cite{Levinsen2006} along with the low-energy limit
for the incoming and outgoing molecules, i. e. $\hbar \Omega=E_b+\hbar^2 \vecp^2/2m_2\rightarrow E_b$ and $|\vecp|=|\vecq| \rightarrow 0$. 
The subtraction in the curly brace in Eq.~\eqref{eq:diagramAl1} comes from the curly brace in Fig.~\ref{fig:diagramA}, which is needed to obtain
a finite result for $\amm$ in the considered low-energy limit. The term subtracted is singular in this limit, and it contributes to the renormalization of the molecular binding
energy $E_b$. 

From Eq.~\eqref{eq:diagramAl1} to \eqref{eq:diagramAl2} we have used the relations $g_2 \propto \ammbg$, $\bar{\alpha}^2 \propto 1/r_*$
and also that in the asymptotic limit of large detuning $|\bar{\nu}|$, the momentum dependences in all Green's functions can be neglected.

The overall minus sign in Eq.~\eqref{eq:diagramAl3} originates from the fact that the diagram in Fig.~\ref{fig:diagramA} represents a molecular self-energy
effect. More specifically, it comes from the reduction of the closed-channel fraction $\mathcal{Z}$ of a dressed molecule
(see Eq.~\eqref{eq:ammABC2} of Section~\ref{sec:zeronlimit} below)
\footnote{The leading order deviation of $\mathcal{Z}$ from $1$ is $\sim
- \partial \Sigma_2(\Omega,\vecp)/\partial \Omega \vert_{\Omega=E_b, \vecp=0}$~\cite{Falco2005} where $\Sigma_2$ is the molecular self energy.
The subtraction in Fig.~\ref{fig:diagramA} comes from the fact that $\mathcal{Z}$ is related to the derivative of $\Sigma_2$,
not the self energy $\Sigma_2$ itself. 
}. 
This reduction of $\mathcal{Z}$ is due to Feshbach coupling to the open channel, which decreases the closed-channel content of the dressed molecules. 
Consequently, this diminishes
the contribution of the closed-channel background interaction $g_2$ to $\amm$, leading to the overall minus sign
in front of the $A_1$ term in Eq.~\eqref{eq:ammunify}.

\subsubsection{Leading order diagram for the $A_2$ term}
\begin{figure}[htp]
\begin{center}
\includegraphics[width=0.85\linewidth]
 {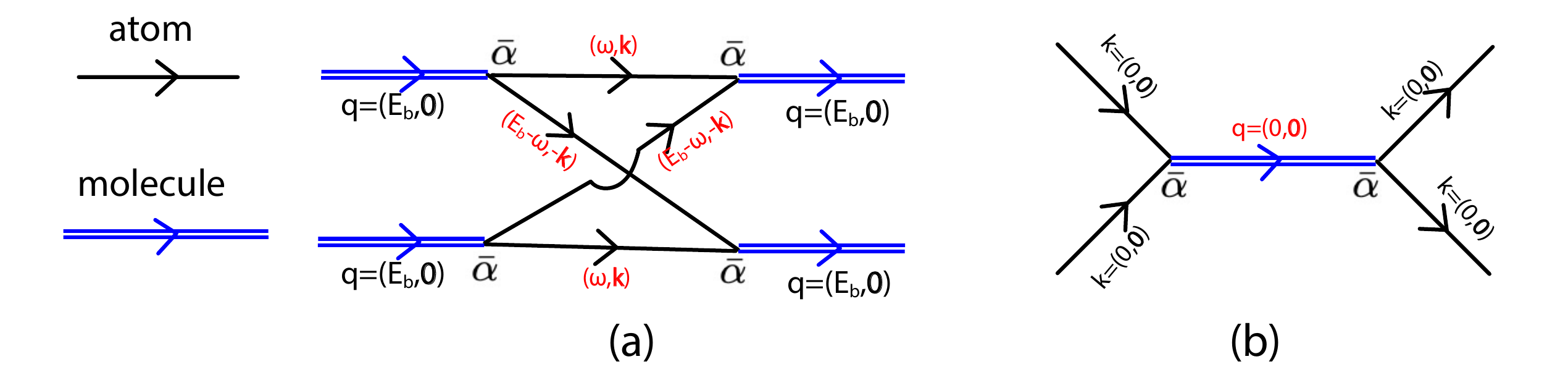}
\caption{
(a) Leading-order contribution for the $A_2$ term in Eq.~\eqref{eq:ammunify}. It corresponds to the schematic diagram in Fig. 1(b) of the main text. 
In this scattering process, two molecules come close together, temporarily break up into four free atoms,
and then recombine into another two molecules. The net effect is one exchange of two atoms.
(b) The corresponding leading order scattering process that contributes to the atom scattering length $a_s$.
This diagram is lower order in the Feshbach coupling $\bar{\alpha}$ compared to the one shown in (a). Its contribution to the resonant term in $a_s$
is given by $\bar{\alpha}^2 G_2(\Omega=0,\vecq=0) =\frac{\bar{\alpha}^2}{-\nu}$, which differs from the resonant term in Eq.~\eqref{eq:asdef} by the denominator. 
This difference arises from the higher-order atomic scattering process. 
}
\label{fig:DiagramC}
\end{center}
\end{figure}

The leading-order diagram for the $A_2$ term in Eq.~\eqref{eq:ammunify} is shown in Fig.~\ref{fig:DiagramC}(a),
whose leading dependence on the inverse binding momentum $1/k_b$ can be estimated as follows,
\begin{subequations}
\begin{align}
\text{ Fig.~\ref{fig:DiagramC}(a) } & \sim  \bar{\alpha}^4  i \int \frac{d \omega}{2\pi} \int^{\Lambda} \frac{d^3 \veck}{(2\pi)^3}  
G_1^2(\omega,\veck) G_1^2(E_b/\hbar - \omega, - \veck) \\
& \propto  \frac{1}{r_*^2}   i  \int \frac{d \omega}{2\pi} \frac{1}{(\hbar \omega + i \delta)^2 } \frac{1}{(E_b - \hbar \omega + i \delta)^2}   \int^{\Lambda} \frac{d^3 \veck}{(2\pi)^3}    \label{eq:line2} \\
 & \propto - \frac{1}{k_b^6 r_*^2 \bar{a}^3}. \label{eq:powercount}
\end{align}
\end{subequations}
To obtain the last line we have used $ i  \int d \omega/2\pi  (\hbar\omega + i \delta)^{-2 } (E_b - \hbar\omega + i \delta)^{-2} 
= \hbar^{-1}\text{Res}[ (E_b - \hbar\omega + i \delta)^{-2},\hbar \omega=-i \delta] =-2/(\hbar E_b^3) \propto -1/k_b^6$. 

\subsubsection{ Leading order diagram for the $A_3$ term}

\begin{figure}[htp]
\begin{center}
\includegraphics[width=0.6\linewidth]
 {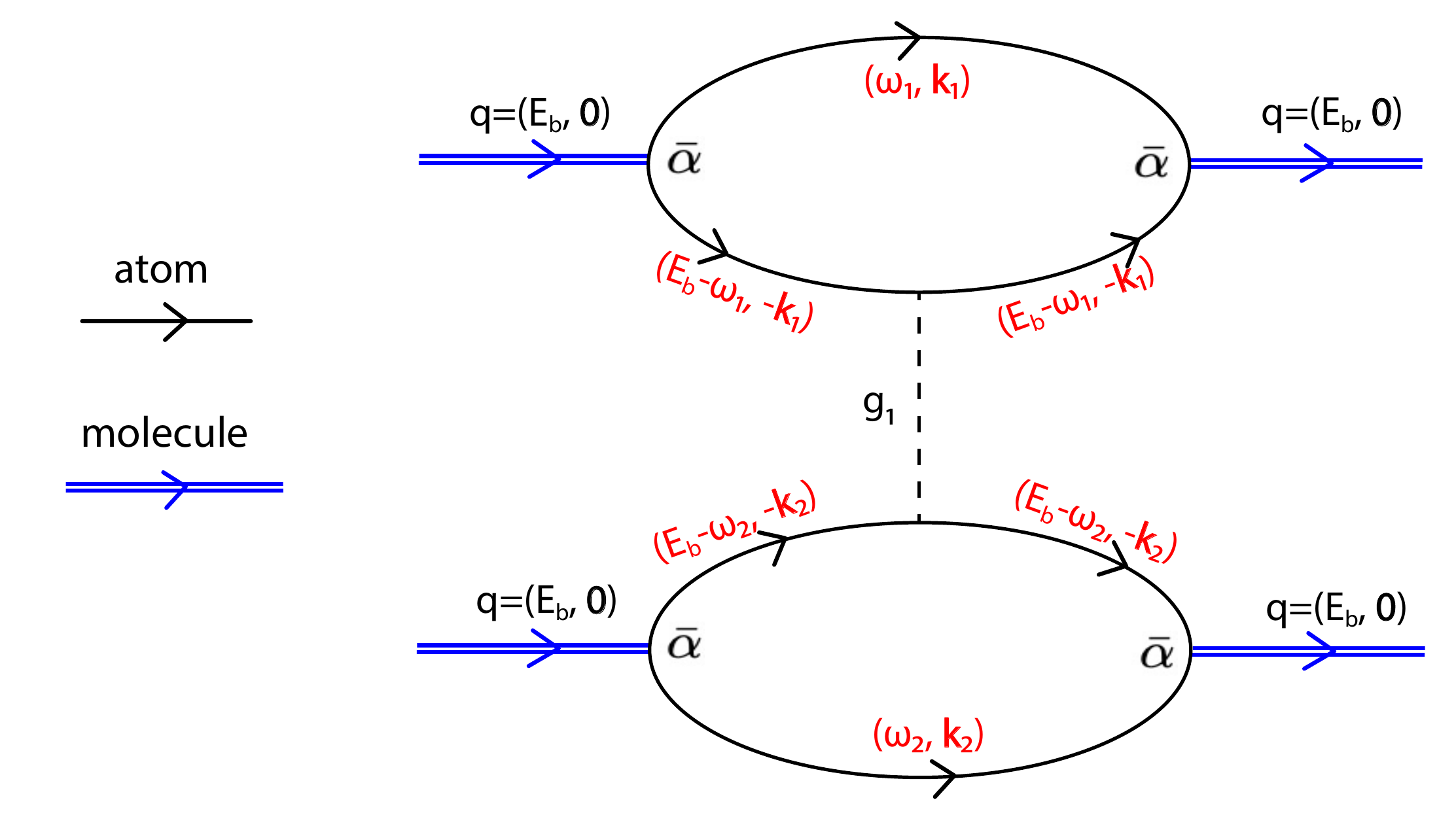}
\caption{Leading order diagram for the $A_3$ term in Eq.~\eqref{eq:ammunify}. 
In this diagram we have already considered the low energy limit for the incoming and outgoing external molecule legs,
by setting $q=(\Omega,\vecq)=(E_b/\hbar,\vec{0})$.
This diagram involves a combination of the atom background interaction $g_1$
and a product of 4 Feshbach coupling, $\bar{\alpha}^4$, resulting in the factor $\asbg/r_*^2$ in the $A_3$ term in Eq.~\eqref{eq:ammunify}.
}
\label{fig:diagramB}
\end{center}
\end{figure}

The leading order contribution from the $A_3$ term in Eq.~\eqref{eq:ammunify} can be represented by Fig.~\ref{fig:diagramB}. 
Following the calculation in Eq.~\eqref{eq:diagramA} we can estimate the contribution from Fig.~\ref{fig:diagramB}
to the scattering length $\amm$ as
\begin{subequations}
\begin{align}
\text{ Fig.~\ref{fig:diagramB} }  & \sim   g_1 \bar{\alpha}^4 \bigg[ i  \int \frac{d\omega_1}{2\pi} \int^\Lambda \frac{ d^3\veck_1}{(2\pi)^3}   
G_1(\omega_1,\veck_1)  \big[ G_1(E_b/\hbar -\omega_1,-\veck_1) \big]^2   \bigg]
\bigg[ i  \int \frac{d \omega_2 }{2\pi} \int^\Lambda \frac{d\veck_2}{(2\pi)^3}   G_1(\omega_2,\veck_2)  \big[ G_1(E_b/\hbar -\omega_2,-\veck_2) \big]^2 \bigg]\\
& \propto  \frac{\asbg}{ r_*^2}  \quad \bigg\{ i \int \frac{d\omega}{2\pi}  \frac{1}{\hbar \omega_1 + i \delta} \frac{1}{\big( E_b- \hbar\omega_1 + i \delta \big)^2 } \int^{\Lambda} \frac{ d^3\veck_1}{(2\pi)^3} \bigg\}^2   \\
& \propto   \frac{\asbg}{k_b^8 r_*^2 \bar{a}^6}. 
\end{align}
\end{subequations}

\section{Details of the variational analysis}
\label{sec:variational}

In this section, we provide details about our variational analysis, which we used to arrive at the stability phase diagrams in Fig. 2 of the main text
and the various formulae for the scattering lengths $\amm$ and $\ammeff$.

\subsection{Variational wavefunction and saddle point equations} 

We start with the many-body Hamiltonian $H$ given earlier in Eq.~\eqref{eq:Hamiltonian}. 
To approximate its true ground state at zero temperature $T=0$ we adopt the following many-body variational wavefunction, 
\begin{align} \label{eq:Psivar}
|\Psivar \rangle & =\frac{1}{\mathcal{N}}  e^{ \sum_\sigma \Psi_{\sigma 0} \sqrt{V}  a^\dagger_{\sigma 0}   + \sum_{\veck}^\prime \sum_{\sigma} \chi_{\sigma \veck}\; a^\dagger_{\sigma \veck} a^\dagger_{\sigma -\veck} } \vert 0 \rangle. 
\end{align}
In the exponent, the summation over the momentum $\veck$ is over half of the $\veck-$space. The prime in the $\veck$-summation implies that the point $\veck=0$ is excluded. 
The variational parameters $\Psi_{\sigma 0}$ and $\chi_{\sigma \veck}$, without loss of generality, can be taken to be real since all the interaction parameters in the Hamiltonian are real. 
The vacuum state $|0\rangle$ satisfies  $a_{\sigma\veck} |0\rangle =0$ for all annihilation operators $a_{\sigma\veck}$, 
and $\mathcal{N} = e^{\sum_\sigma V |\Psi_{\sigma0}|^2/2} \prod_{\sigma} \prod_{ \veck}^\prime (1- |\chi_{\sigma \veck}|^2)^{-1/2}$ is the normalization factor. 
Here, in the spirit of generalized Bogoliubov theory, only pair-wise correlations between atoms or between molecules are included in the exponent of the variational wavefunction, which is adequate for our focus of the detuning regime that is not too close to the resonance. 
In making this approximation we have also assumed that an inter-channel pair correlation does not play a key role. 

Then the trial ground state energy associated with $|\Psivar \rangle$ is
\begin{align}   \label{eq:Etri}
 &  \Otri [\Psi_{10}, \Psi_{20}, \chi_{1\veck},\chi_{2\veck}] =  \langle \Psivar | \hat{H} | \Psivar\rangle   = V \bigg\{ 
  \sum_{\sigma=1}^2  \bigg[ h_{\sigma \veck=0}   |\Psi_{\sigma 0}|^2 +  \frac{g_\sigma }{2}   |\Psi_{\sigma 0}|^4   +  \frac{g_\sigma}{2}  |x|^2 \bigg]
  - \alpha  ( (\Psi_{10}^*)^2 \Psi_{20} +c. c.)   \nonumber \\
&  +   \sum_{\sigma=1}^2 \bigg[  \frac{1}{V} \sum_{\veck \ne 0}  ( h_{\sigma \veck} + 2  g_\sigma |\Psi_{\sigma 0}|^2 + g_\sigma n_\sigma)  n_{\sigma \veck }  
    +    \frac{1}{V} \sum_{\veck \ne 0 }   \frac{g_\sigma}{2} \big( (\Psi_{\sigma 0}^*)^2   x_{\sigma \veck } + c.c.  \big) 
  \bigg]
 -  \frac{1}{V} \sum_{\veck \ne 0}  \alpha \big(  \Psi_{20} x^*_{1 \veck}  + c.c.   \big)
 \bigg\},
\end{align}
which is a functional of the parameters $\{\Psi_{10}, \Psi_{20}, \chi_{1\veck},\chi_{2\veck}\}$.
In writing down the above equation we have used result $\Psi_{\sigma 0}=\langle a_{\sigma 0} \rangle/ \sqrt{V}$ and 
also introduced $ x_{\sigma \veck}   \equiv  \langle a_{\sigma \veck}  a_{\sigma -\veck} \rangle = \chi_{\sigma \veck} / (1-|\chi_{\sigma \veck}|^2)$ ,  $ n_{\sigma \veck}   \equiv \langle a_{\sigma \veck}^\dagger a_{\sigma \veck} \rangle 
= |\chi_{\sigma \veck}|^2/(1-|\chi_{\sigma \veck}|^2)$, $ x _\sigma  =  V^{-1}\sum_{\veck \ne 0} x_{\sigma \veck}$, and $ n_\sigma  = V^{-1} \sum_{\veck \ne 0} n_{\sigma \veck}$.  Here and throughout the paper, $\langle \cdots \rangle \equiv \langle \Psivar | \cdots | \Psivar \rangle$. 
The variational parameter $\Psi_{10}$ ($\Psi_{20}$) represents the amplitude of the atomic (molecular) condensate. 
The expectation value $x_{\sigma \veck}$ is a Cooper-pair type correlation between two open-channel atoms or two closed-channel molecules, 
and, with our choice of the variational wavefunction, $n_\sigma/2$ correspondingly counts the number of these pairs. 

The correlation parameter $\langle a_{1 \veck}  a_{1, -\veck} \rangle$  or $x_{1\veck}$ plays an important role in undermining the molecular condensate stability. 
This parameter can, in various regions of the phase diagram,
either reflect a Cooper pairing, induced by an attraction between two atoms through the Feshbach coupling, or a quantum depletion effect, the latter induced by the presence of a nonzero order parameter $\Psi_{10}$ and by repulsive interaction $g_1$. 
Although at a general detuning $\bar{\nu}$, both mechanisms contribute to the pair correlation $x_{1\veck}$,
for our main focus of the negative detuning regime where the atomic BEC condensate vanishes $\Psi_{10}=0$,
the atom pair correlation $x_{1\veck}$ results completely from the Cooper pairing effect. 
On the other hand, the molecule pair correlation $x_{2\veck}$
is solely a quantum depletion effect as there is no attraction to bind two molecules in the Hamiltonian $H$.

Minimizing the trial ground-state energy $\Otri$ in Eq.~\eqref{eq:Etri} with respect to the variational parameters $\{ \Psi_{10}^*, \Psi_{20}^*, \chi_{1\veck}^*, \chi_{2\veck}^*\}$ leads to the following saddle-point equations~\cite{Wang2024}.
\begin{widetext}
\begin{subequations} \label{eq:saddle}
\begin{align}
0  & =  \big[  h_{1\veck=0}  + g_1 |\Psi_{10}|^2  + 2 g_1 n_1     \big] \Psi_{10} + g_1 \Psi_{10}^* x_1 - 2 \alpha  \Psi_{10}^* \Psi_{20},  \label{eq:Psi10} \\
0 & =\big[  h_{2 \veck=0}  + g_2 |\Psi_{20}|^2  + 2 g_2 n_2  \big] \Psi_{20} + g_2 \Psi_{20}^* x_2 - \alpha  ( x_1 + \Psi_{10}^2 ),  \label{eq:Psi20} \\
0 & = 2  \big[  h_{1\veck} + 2 g_1 (  |\Psi_{10}|^2 + n_1 )   \big] x_{1\veck}   +  \big[g_1 (x_1 + \Psi_{10}^2)  -2 \alpha \Psi_{20} \big] (2 n_{1\veck}+1) , \label{eq:x1k} \\
0 & = 2 \big[  h_{2\veck} + 2 g_2  (  |\Psi_{20}|^2 + n_2)     \big]  x_{2\veck} +    g_2 (x_{2}  +  \Psi_{20}^2) (2 n_{2\veck}+1).  \label{eq:x2k}
\end{align}
\end{subequations}
\end{widetext}
The last two sets of equations, derived from the derivative $\partial \Omega/\partial \chi_{\sigma \veck}^*$,  can be recast in a simpler form of a BCS-like gap equation
by introducing the pairing order parameter $\Delta_\sigma$,
\begin{align}
\Delta_\sigma = g_\sigma x_\sigma = g_\sigma \frac{1}{V}  \sum_{\veck} x_{\sigma \veck}.   \label{eq:DeltaDef}
\end{align}
To write the right hand side of this equation in terms of the order parameter $\Delta_\sigma$ itself so that the equation forms a self-consistent one,
we first substitute the results $x_{\sigma \veck} =\langle a_{\sigma \veck} a_{\sigma, -\veck} \rangle=\chi_{\sigma \veck}/(1-|\chi_{\sigma \veck}|^2)$
and $n_{\sigma \veck} = \langle a^\dagger_{\sigma \veck} a_{\sigma \veck} \rangle=|\chi_{\sigma \veck}|^2/(1-|\chi_{\sigma \veck}|^2)$
into Eqs.~\eqref{eq:x1k} and \eqref{eq:x2k} and rewrite the obtained two equations in the following common form,
\begin{align}
 \widetilde{\Delta}_{\sigma}^* \chi_{\sigma \veck}^2 + 2 \widetilde{\epsilon}_{\sigma \veck} \chi_{\sigma \veck}  +  \widetilde{\Delta}_{\sigma} =0, \label{eq:chisigmak}
\end{align}
where we have introduced 
\begin{subequations}
\begin{align}
 \widetilde{\epsilon}_{\sigma \veck}  & = h_{\sigma \veck} + 2 g_\sigma  (|\Psi_{\sigma 0}|^2 + n_\sigma), \\ 
\widetilde{\Delta}_1 & =\Delta_1 + g_1 \Psi_{10}^2 - 2\alpha \Psi_{20},   \label{eq:Delta1Tilde} \\
\widetilde{\Delta}_2& =  \Delta_2 + g_2 \Psi_{20}^2.   \label{eq:Delta2Tilde} 
\end{align}
\end{subequations}
Eq.~\eqref{eq:chisigmak} is quadratic in $\chi_{\sigma \veck}$ whose root can be easily written down as
\begin{align}
\chi_{\sigma \veck} = \frac{- \widetilde{\epsilon}_{\sigma \veck} + \sqrt{\widetilde{\epsilon}_{\sigma \veck}^2-  |\widetilde{\Delta}_{\sigma}|^2}}{ \widetilde{\Delta}_{\sigma}^*}
=  -  \frac{ \widetilde{\epsilon}_{\sigma \veck} - E_{\sigma \veck}}{ \widetilde{\Delta}_{\sigma}^*}, 
\label{eq:chiksolution}
\end{align}
where we have discarded the other unphysical solution.
Substituting Eq.~\eqref{eq:chiksolution} back into the right hand side of Eq.~\eqref{eq:DeltaDef} leads to
\begin{align}
\Delta_\sigma  & = -  g_\sigma \frac{1}{V}  \sum_{\veck} \frac{\widetilde{\Delta}_{\sigma}}{2 E_{\sigma \veck}}.  \label{eq:BCS}   
\end{align}
In this equation, $ E_{\sigma \veck}   = \sqrt{  \widetilde{\epsilon}_{\sigma \veck}^2  - |\widetilde{\Delta}_\sigma|^2 }$ is either the atomic (with $\sigma=1$) or the molecular  ($\sigma=2$) Bogoliubov excitation energy. 
Using Eq.~\eqref{eq:chiksolution} we can also re-express the trial ground-state energy $\Otri[\Psi_{10},\Psi_{20},\chi_{1\veck},\chi_{2\veck}]$ in Eq.~\eqref{eq:Etri} in terms of only four variables,
i. e.  $\Omega=\Omega[\Psi_{10},\Psi_{20},\Delta_1,\Delta_2]$.
The purpose of this rewriting is to effectively reduce the number of variational parameters, thereby making the numerical calculations of compressibility 
and also the derivations of
molecule scattering length later more tractable. 

After the rewriting, we need to solve four saddle-point equations for the four unknowns $\{\Psi_{10},\Psi_{20},\Delta_1,\Delta_2\}$, 
given by Eqs.~\eqref{eq:Psi10}, \eqref{eq:Psi20} and \eqref{eq:BCS}. Additionally, we impose
the particle number-density constraint, 
 $ n = (|\Psi_{10}|^2 + n_1) + 2 (|\Psi_{20}|^2 + n_2) $, which fixes the chemical potential $\mu$.
 Here,
\begin{align}
n_\sigma & = \frac{1}{V} \sum_\veck \frac{1}{2} \bigg(\frac{\widetilde{\epsilon}_{\sigma \veck}}{E_{\sigma \veck}} - 1 \bigg).      \label{eq:nsigma}
\end{align}

\begin{figure}[htp]
\begin{center}
\includegraphics[width=0.65\linewidth]{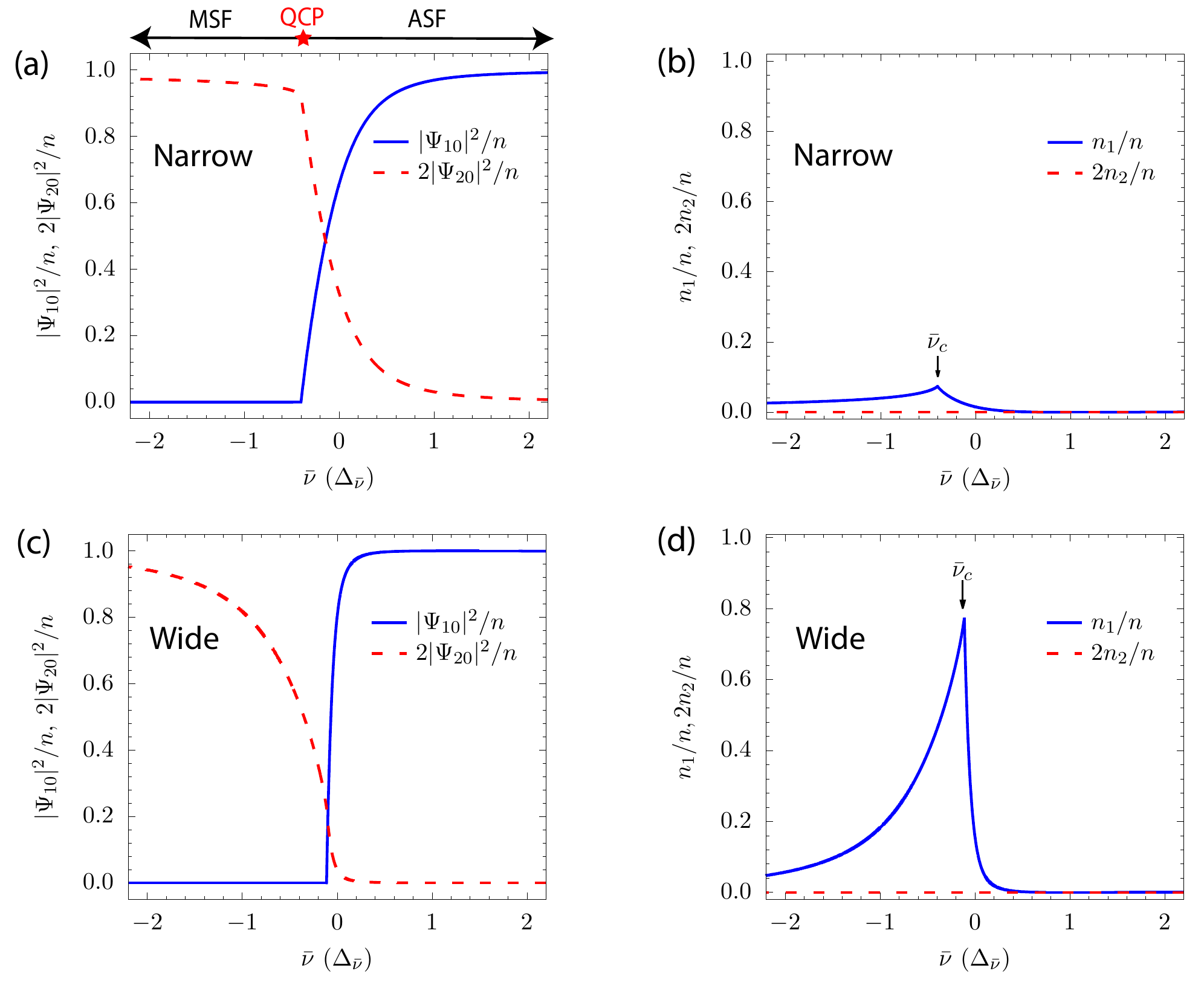}
\caption{Numerical results to the saddle-point equations for both a narrow (top row) and wide (bottom) resonance at a typical particle density $n \asbg^3=1.68\times 10^{-5}$. 
In the horizontal axes, the detuning $\bar{\nu}=\Delta \mu_m (B-B_0)$ is measured in terms of the resonance with 
$\Delta_{\bar{\nu}}=\Delta \mu_m \Delta B$. 
(a,c) Calculated condensate density $|\Psi_{10}|^2$ and  $|\Psi_{20}|^2$ as a function of the detuning $\bar{\nu}$,
(b,d) the corresponding number of atom pairs $n_1$. 
\textit{We note that these results are obtained without considering the stability issue. }
In (a,c), the parameters $\Psi_{10}$ and $\Psi_{20}$ describe the amplitudes of open-channel atom and closed-channel molecule condensates, respectively. 
In the MSF phase $\Psi_{10}$ is zero , while it is nonzero in the ASF phase. The two phases are separated by a QCP, marked by the red star in (a),
corresponding to the $\bar{\nu}_c$ line in the following Fig.~\ref{fig:Fig1}. 
The contrast in $n_1$ between (b) and (d) suggests that atom pairs play an important role in instabilities of molecular condensates in a wide resonance,
as we will see in Fig.~\ref{fig:Fig3}. 
Also shown in (b) and (d) is the negligible number of molecule pairs $n_2$, originating from the molecular quantum depletion effect. 
The parameters chosen for the narrow resonance in (a,b), as well as in the subsequent figures, are from the $g$-wave resonance of \Cs~at $B_0 = 19.849$ G used in Refs.~\cite{Zhang2021,Zhang2023,Wang2024} (see Table~\ref{tab:parameters}
for details);
for the wide resonance in (c,d) we increase the resonance width $\Delta_{\bar{\nu}}$ from its narrow-resonance value in (a,b) by $10^2$ times, while keeping all other parameters the same. 
 }
\label{fig:Fig1v2}
\end{center}
\end{figure}

\subsection{Numerical results for the saddle-point solutions}

Figure~\ref{fig:Fig1v2} presents our numerical solutions to the saddle-point equations for both narrow and wide resonances 
at a typical atomic particle density $n\asbg^3 =1.68 \times 10^{-5}$. The choice of interaction parameters for this calculation
was given earlier in Table~\ref{tab:parameters} and discussed in detail in Section~\ref{sec:parameters}.     

From Fig.~\ref{fig:Fig1v2}(a,c) we see that, as predicted in previous literature~\cite{Romans2004,Radzihovsky2004}, there is a quantum critical point at detuning $\nuc $, separating the atomic superfluid phase (ASF) at detuning larger than the QCP value $\bar{\nu} > \nu_c$, where both the atomic condensate amplitude $\Psi_{10}$ and the molecular condensate amplitude $\Psi_{20}$ are nonzero, 
and the molecular
superfluid phase (MSF) at detuning less than the QCP value $\bar{\nu} < \nuc$, where the molecular condensate amplitude is nonzero $\Psi_{20} \ne 0$ while 
the atomic condensate amplitude vanishes $\Psi_{10}=0$. 
A notable difference between the wide and narrow resonance cases
is the number of condensed closed-channel molecules, given by $|\Psi_{20}|^2$, near the QCP. 
While this number nearly saturates the total particle number density $n$ in the narrow case,
it corresponds to a small fraction of $n$ for the wide resonance considered. The remaining 
rather large fraction of the particles exist in the form of Cooper pairs of open-channel atoms.
This difference is a reflection of the fact that near the QCP detuning, a dressed molecule is almost completely closed-channel like for a narrow resonance,
while it is predominantly open-channel like for a sufficiently wide resonance.

In Figs~\ref{fig:Fig1v2}(b,d) we plot the corresponding number of particles that are present
in the form of Cooper (or Cooper-like) pairs of atoms (and molecules)~\footnote{Within our choice of the variational wavefunction, $n_{1}$ (and $n_2$) counts the number of pairs of atoms (and molecules).}, represented by $n_1$ (and $n_2$). 
In the narrow resonance case (Fig.~\ref{fig:Fig1v2}(b))
both $n_1$ and $n_2$ are quite small: the number of pairs $n_1$ is only a few percent of the total particle number even near the QCP. 
This should be contrasted with the wide resonance case in Fig.~\ref{fig:Fig1v2}(d),
where the fraction $n_1/n$ reaches about $80\%$ at the QCP. We will see that
the observed large fraction of pairs leads to a much stronger detrimental effect on the stability of the molecular condensates
for the wide resonance.

\subsection{Numerical results for the ground-state chemical potential}
\label{sec:chemicalpotential}

\begin{figure}[h]
\begin{center}
\includegraphics[width=\linewidth]{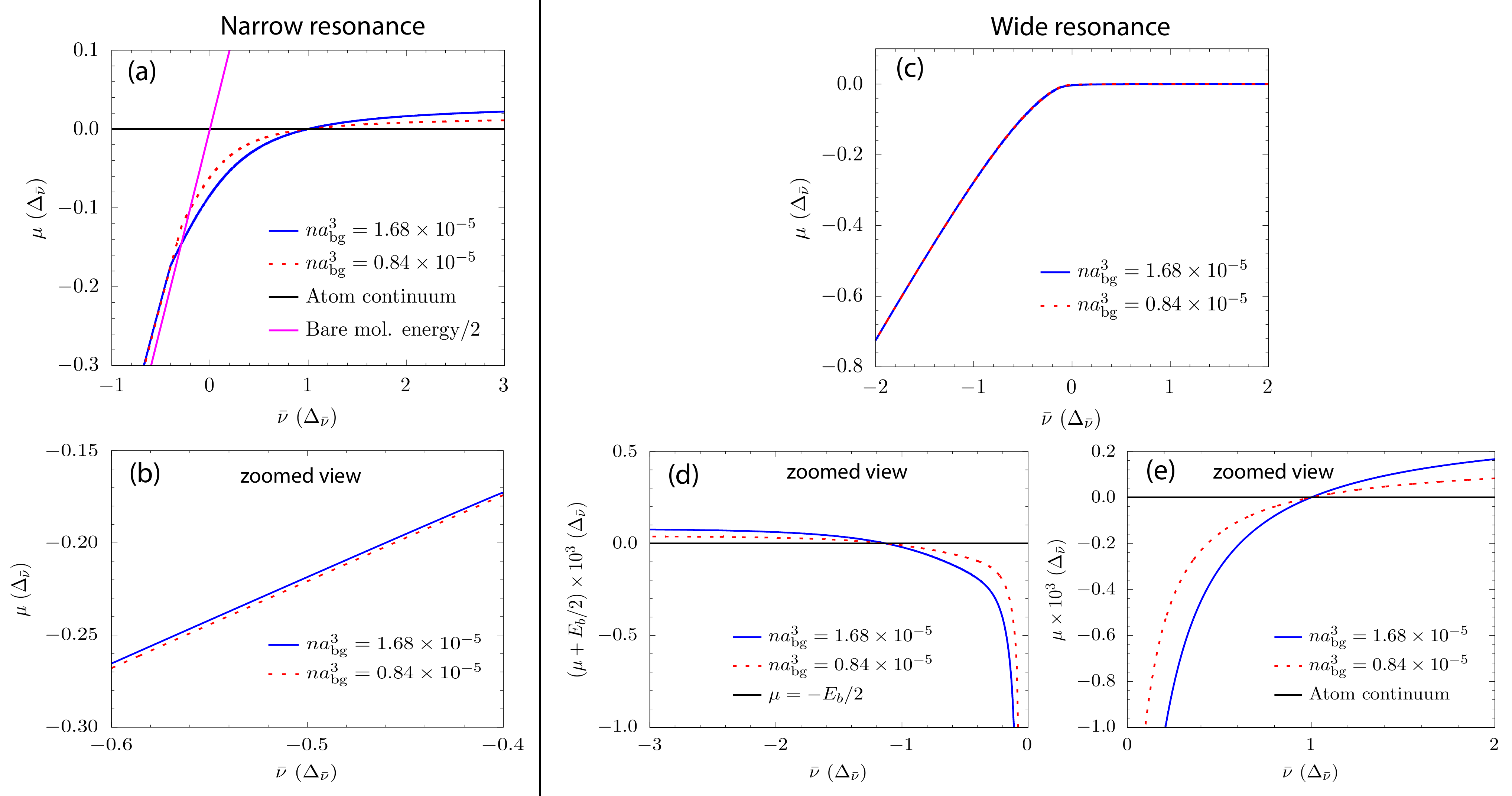}
\caption{Variational results of the ground-state atomic chemical potential $\mu$ (blue solid and red dashed lines) 
as a function of the detuning $\bar{\nu}=\Delta \mu_m (B-B_0)$ for two different representative densities.
Panels (a,b) and (c,d,e) correspond to the narrow and wide resonances studied in Fig. 2 of the main text (see also Fig.~\ref{fig:Fig1} below). 
In panel (a), the ``bare molecular energy/2" (magenta solid line) is equal to $\bar{\nu}/2$, and the ``atom continuum" (black solid line) corresponds to the threshold $\mu \equiv 0$.
The zoomed view in panel (b) shows that on the molecular side and away from the QCP, $\mu$ increases with density $n$, 
consistent with the compressibility being positive $\kappa>0$ in the corresponding regime of the phase diagram in Fig.~\ref{fig:Fig1}.
In panel (c) of the wide resonance case, the two chemical potential curves for different densities almost coincide when plotted on the scale of the resonance width $\Delta_{\bar{\nu}}=\Delta \mu_m \Delta B$.
This behavior on the atomic side comes from the fact that $\mu \sim (4\pi \hbar^2 \asbg/m_1) n \ll \Delta_{\bar{\nu}}$.
On the molecular side, the $n$-dependence of $\mu$ is hard to see in panel (c) because $\mu$ is dominated by
the large two-body molecular binding energy $E_b$, which is $n$-independent. 
Panel (d) presents the zoomed view of $\mu$ on the molecular side, after subtracting the dominant two-body contribution of the dressed molecular energy divided by two, i. e. $-E_b/2$.
Here, $E_b=\hbar^2 k_b^2/m_1$ is obtained by numerically solving the transcendental equation of the binding momentum $k_b$ in Eq.~\eqref{eq:root} (see below). 
Panel (e) shows the zoomed view of the chemical potential $\mu$ on the atomic side, from which one sees that the behavior of the chemical potential is similar to its narrow-resonance counterpart in panel (a).
}
\label{fig:muplot}
\end{center}
\end{figure}

The saddle point equations were solved together with the particle number density constraint, from which we can also obtain the chemical potential $\mu$. 
In Figs. 3(a,b) of the main text, we have already presented our numerical results for the chemical potential $\mu$ at two different representative densities, 
focusing primarily on the near-resonance detuning regimes. 
Here, we provide additional details about the behavior of the chemical potential across a broader range of detuning in Fig.~\ref{fig:muplot}. 

From the narrow-resonance results in Figs.~\ref{fig:muplot}(a,b) we first observe a sharp contrast in the behavior of the chemical potential 
$\mu$ at finite density between the far-off-resonance and the near-resonance
detuning regimes. In the former regime, the chemical potential $\mu$ exceeds two energy levels,  the non-interacting atomic continuum threshold $\mu=0$
on the atomic side and the ``bare molecular energy/2" on the molecular side. 
In contrast, near the resonance, the chemical potential falls below these two bare energy levels. 
As already discussed in the main text,
this observation has important implications that the inverse compressibility $\kappa^{-1}=d\mu/d n$ must be \textit{negative near and on both sides of the resonance}, 
while it is necessarily positive in the far-off-resonance regime (assuming both $\asbg$ and $\ammbg$ are positive).
In the near-resonance regime this is because the chemical potential $\mu$ increases with decreasing density $n$ (see Fig.~\ref{fig:muplot}(a))
and it approaches the two bare energy levels from below as the density $n$ decreases to zero.

Similar observations apply to the wide-resonance case depicted in Figs.~\ref{fig:muplot}(c,d,e), albeit with some differences. 
As shown in Fig.~\ref{fig:muplot}(e),
on the atomic side of the resonance the chemical potential $\mu$ behaves similarly to the narrow-resonance case.
However, differences arise on the molecular side.
In Fig.~\ref{fig:muplot}(c), we see that in the wide-resonance case, the chemical potential $\mu$ is almost completely dominated by the two-body molecular binding energy $-E_b$,
which differs from the bare molecular energy $\bar{\nu}$--- unlike in the narrow resonance scenario. 
Nevertheless, if we replace the ``bare molecular energy$/2$" from the discussions in the previous paragraph with $-E_b/2$, we can deduce the same
conclusion: the inverse compressibility $\kappa^{-1}$ must be negative near and on both sides of the resonance. 
The reasoning for the atomic side is the same as before.
On the molecular side, this conclusion comes from the chemical potential falling below the energy $-E_b/2$ near the resonance (see Fig.~\ref{fig:muplot}(d)).
Then, since the chemical potential $\mu$ should approach the energy level $-E_b/2$ from below as the density decreases to zero $n\rightarrow 0$, the derivative $d\mu/dn$ is necessarily negative, resulting in the unstable regime of the phase diagram in Fig.~\ref{fig:Fig1}(b).


\subsection{Numerical results for the ground-state compressibility $\kappa$}
\label{sec:phasediagram}

\begin{figure*}[htp]
\begin{center}
\includegraphics[width=0.75\linewidth,clip]{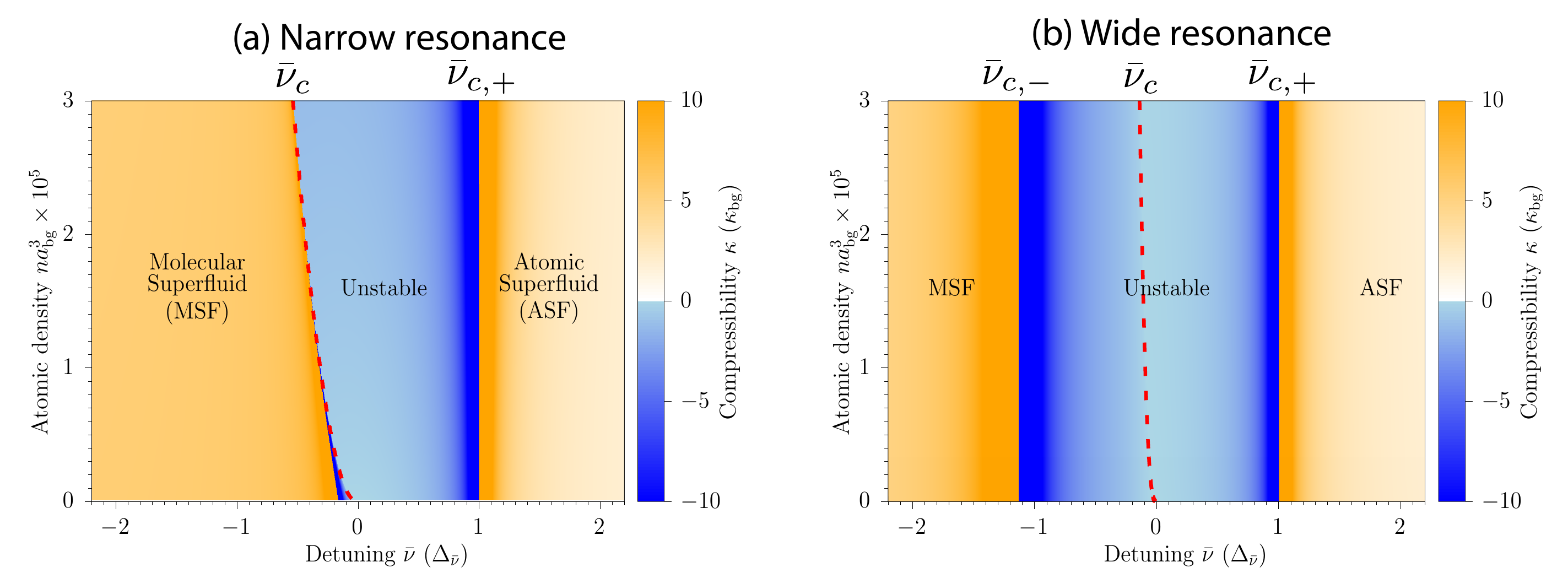}
\caption{Ground state phase diagram for a narrow  (a) and wide  (b) resonance.  
Plotted is a map of the compressibility  $\kappa= d n/ d \mu$ as a function of atom number density $n$ and detuning $\bar{\nu} =\Delta \mu_m (B-B_0)$. 
The compressibility $\kappa$ is normalized by $\kappa_{\mathrm{bg}}=m_1/(4\pi \hbar^2 a_{\mathrm{bg}})$ with $m_1$ the atomic mass and and $a_{\mathrm{bg}}$ the background scattering length of atoms. 
The ASF and MSF phases are stable in the orange region, and unstable in the blue.
The dashed line (red and labeled by $\bar{\nu}_c$), calculated in the current variational approach, is the QCP that separates the ASF from the MSF phase when the stability issue is not considered. 
The detuning $\bar{\nu}_{c,+}$ ($\bar{\nu}_{c,-}$) denotes the boundary that separates the unstable region from the stable ASF (MSF) phase region.
}
\label{fig:Fig1}
\end{center}
\end{figure*}

We now turn to the stability issue which we
address via the compressibility $\kappa \equiv d n/d\mu$.
We first compute $\kappa$ numerically from the calculated chemical potential $\mu$ as a function of particle density $n$.
A necessary condition for the calculated superfluid phase to be stable is a positive compressibility $\kappa >0$. Otherwise
the system will undergo a collapse~\cite{Gerton2000} in spatial space. 
Using the compressibility $\kappa$, 
we construct the ground-state stability phase diagrams for both the narrow and wide resonance cases,
as shown in Fig.~\ref{fig:Fig1}.  
Also shown in the phase diagrams is the QCP line, determined by the onset of the atomic condensate amplitude $\Psi_{10}$ before the stability issue is considered.
From these phase diagrams, we see that there is a relatively wide regime near the resonance, shaded in blue and labeled by ``Unstable", where the compressibility is negative $\kappa <0$.
Interestingly, for the narrow resonance in Fig.~\ref{fig:Fig1}(a), the phase boundary between this ``Unstable" and the stable MSF phase regime nearly coincides with the QCP line except at 
extremely low density $n$. 

Importantly, this narrow resonance case makes a study of 
quantum criticality (associated with the predicted QCP) possible from the MSF side 
of resonance.
This is a unique signature associated with a narrow resonance, which is
not possible in the wide-resonance case, as seen from Fig.~\ref{fig:Fig1}(b), 
where the Unstable-MSF phase boundary
is significantly separated from the presumed QCP line, so that the latter is
inaccessible. 
In Section~\ref{sec:accessibility} below, we will derive a generic condition for when such a near-coincidence occurs for a narrow resonance (see Eq.~\eqref{eq:criterion2} below).

Finally, we note that the other side of the unstable region which
interfaces the stable ASF phase
is nearly independent of the density $n$. In particular, the boundary detuning of this region, $\bar{\nu}_{c,+}$,
is right at $\bar{\nu}/\Delta_{\bar{\nu}}=+1$, which is where the two-body atomic scattering length $a_s$ changes its sign, for both the narrow and wide resonances.
This will become more clear in the following Fig.~\ref{fig:Fig4}. 


\subsection{Uncertainty estimation for phase boundaries in Fig.~\ref{fig:Fig1}}
\label{sec:uncertaintity}

Because our variational wavefunction ansatz, Eq.\eqref{eq:Psivar}, includes only Gaussian-level correlations between atoms or molecules, 
the impact of neglected higher-order correlations on the phase boundaries in Fig.\ref{fig:Fig1} and the molecular scattering length formulas (Eqs. (2)-(4) and (11) of the main text) 
warrants investigation. While the accuracy of our approach is not adequate precisely
at the Feshbach resonance, we demonstrate in the following
that it remains sufficient for determining the phase boundaries in Fig.~\ref{fig:Fig1}, as the boundary detunings are sufficiently far from resonance in 
both the wide and narrow-resonance cases. 

For the wide-resonance case in Fig.~\ref{fig:Fig1}(b), the MSF-Unstable phase boundary is located approximately one resonance width away from the resonance, 
where background interactions and leading-order Feshbach scattering dominate. These effects are already captured by our variational approach.

For the narrow-resonance case in Fig.\ref{fig:Fig1}(a), a more refined analysis is necessary because the MSF-unstable phase boundary is small compared to the resonance width. 
Although a precise evaluation would require incorporating three-atom correlations into Eq.\eqref{eq:Psivar} and recalculating the phase diagrams, a task beyond the 
scope of this paper, we provide two independent estimates demonstrating that the effect of these neglected higher-order correlations on the phase boundaries in Fig.~\ref{fig:Fig1} is negligible.

\subsubsection{ Estimation I}

As a first estimate, we compare the magnitude of the MSF-Unstable phase boundary detuning, $\bar{\nu}_{c,-}(n)$ (in Fig.~\ref{fig:Fig1}(a)), 
with the detuning of the first Efimov bound state near a narrow resonance. By neglecting three-atom correlations in our variational wavefunction, 
we omit correlations responsible for Efimov bound states, which become increasingly important near resonance. Thus, the position of the first Efimov 
bound state provides a strong indication of the significance of these higher-order effects.

According to Ref.\cite{Petrov2004a} by D. S. Petrov, the first Efimov bound state appears when the atomic scattering length, 
$a_s(\bar{\nu}) = \asbg (1-  \Delta \mu_m \Delta B / \bar{\nu}) $, satisfies $r_*/a_s(\bar{\nu}) = 2.2$, where $r_*=\hbar^2/(m_1 \asbg \Delta \mu_m \Delta B)$
(Eq.\eqref{eq:rstar}; denoted as $R^*$ in Ref.~\cite{Petrov2004a}). Therefore, the corresponding detuning $\bar{\nu}_{\text{1st, Efimov}}$ is

\begin{align}  \label{eq:Efimov}
\frac{\bar{\nu}_{\text{1st, Efimov}}}{ \Delta \mu_m \Delta B} =  \frac{1}{ 1- r_*/\asbg/2.2  } = \frac{1} {1- 1/(2.2 \times 0.006)} = - 0.013,
\end{align}
where we have used $\asbg/r_* =0.006$ for the narrow resonance in Fig.~\ref{fig:Fig1}(a). See Sec.~\ref{sec:parameters} for the choice of resonance parameters. 

From Fig.~\ref{fig:Fig1}(a) we see that the ``MSF-Unstable" phase boundary detuning $\bar{\nu}_{c,-}$ satisfies 
\begin{align} \label{eq:nucminus}
\frac{|\bar{\nu}_{c,-}(n)|}{ \Delta \mu_m \Delta B } \gtrsim 0.15
\end{align}
 for all densities, particularly for $n \asbg^3 \gtrsim 0.1 \times 10^{-5}$ where it coincides with the predicted QCP.
Combining Eqs.~\eqref{eq:Efimov} and \eqref{eq:nucminus} we conclude that 
\begin{align}
 \frac{|\bar{\nu}_{c,-}(n)|} { |\bar{\nu}_{\text{1st, Efimov}}| } \gtrsim 11 . 
\end{align} 
The large separation between $\bar{\nu}_{c,-}(n)$ and $ \bar{\nu}_{\text{1st, Efimov}}$ indicates that the effect of three-body (and higher-order) correlations on 
$\bar{\nu}_{c,-}$ is negligible, 
justifying the accuracy of our variational wavefunction for determining the phase boundaries in Fig.~\ref{fig:Fig1}.

\subsubsection{Estimation II}
\begin{figure}[htbp]
\begin{center}
\includegraphics[width=0.8\linewidth]{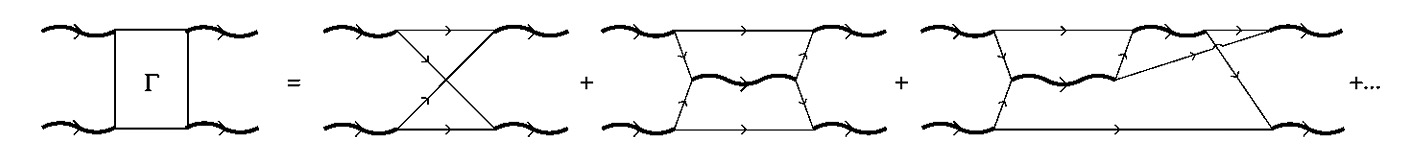}
\caption{Diagrammatic contributions to the two-molecule scattering amplitude, irreducible with respect to cutting two atom propagators. 
Thin straight lines represent atoms, vertices represent Feshbach coupling, and thick wavy lines denote molecules. The figure is adapted from Ref.~[\onlinecite{Levinsen2006}].}
\label{fig:Levinsen}
\end{center}
\end{figure}

As a second estimate we consider the ratio between a neglected leading higher-order contribution to $\amm$ and the leading one that we retained. 
The former arises from the second diagram 
on the right-hand side of the equal sign in Fig.~\ref{fig:Levinsen} (Fig. 11 of Ref.~\onlinecite{Levinsen2006}).
This diagram is 6th order in the Feshbach coupling strength $\bar{\alpha}$, while the term we retained (the first diagram on the right-hand side of Fig.~\ref{fig:Levinsen}) is 4th order.  
The ratio between the contributions of these two diagrams to the scattering length $\amm$ can be roughly estimated as follows,
\begin{align} \label{eq:estimate}
\frac{ \text{6th order term}}{ \text{4th order term}}   &\sim \bar{\alpha}^2 \int \frac{d\omega d\vec{k}}{(2\pi)^4}  \frac{1}{(\hbar \omega)^3} \bigg \vert_{\hbar \omega=|\bar{\nu}|, |\vec{k}|=\sqrt{m_1 |\bar{\nu}|}/\hbar} 
\sim \bar{\alpha}^2  \frac{4\pi}{16\pi^4}  \frac{\sqrt{m_1^3 /|\bar{\nu}|} }{ \hbar^3} \\
 & = \frac{4\pi \hbar^4}{ m_1^2 r_*} \frac{1}{4\pi^3} \frac{m_1^{3/2}}{ \hbar^3 \sqrt{|\bar{\nu}|}} = \frac{1}{\pi^2}  \frac{\hbar }{r_* \sqrt{m_1 |\bar{\nu}|}}.   \label{eq:estimate2}
\end{align}
The middle expression in the first line arises from the following considerations:
\begin{itemize}
\item The 6th order diagram in Fig.~\ref{fig:Levinsen} contains two additional vertices compared to the 4th order diagram, contributing a factor of $\bar{\alpha}^2$.
\item The 6th order diagram involves one more loop integral over an intermediate frequency $\omega$ and momentum $\vec{k}$, resulting in the factor $\int d\omega d\vec{k} / (2\pi)^4$.
\item The 6th order diagram contains three more propagators (one molecular and two atomic), each roughly contributing a factor of $1/(\hbar\omega)$. 
\item For a rough estimate one can set the frequency and momentum at their characteristic values determined by the molecular binding energy,
 i. e. $\hbar \omega=|\bar{\nu}|$ and $|\vec{k}|=\sqrt{m_1 |\bar{\nu}|}/\hbar$.  
\end{itemize}
To obtain Eq.~\eqref{eq:estimate2} we have used $r_*=4\pi \hbar^4/(m_1^2 \bar{\alpha}^2) =\hbar^2/(m_1 \asbg \Delta \mu_m \Delta B)$ 
(see Eq.~\eqref{eq:rstar}). 
Equation~\eqref{eq:estimate2} shows that the importance of the higher order terms depends on the detuning, as expected.    \\

From Fig.~\ref{fig:Fig1}(a) we see that the magnitude of the MSF-Unstable phase boundary detuning $\bar{\nu}_{c,-}(n)$ increases with density $n$,
satisfying
\begin{align} \label{eq:nucminus2}
|\bar{\nu}_{c,-}(n)| \ge  |\bar{\nu}_{c,-}(n=0)| & = \frac{\hbar^2}{m_1 r_*^2} (\frac{r_*}{ \ammbg })^{2/3} \approx 0.15 \; \Delta \mu_m \Delta B. 
\end{align}
Here, we have used the zero-density expression for $\bar{\nu}_{c,-}(n=0)$, derived by solving
\begin{align}
\frac{4\pi \hbar^2 \ammbg}{m_2 }=  \bar{\alpha}^4 \frac{1}{V} \sum_\veck \frac{1} {2 (\hbar^2\veck^2/2m_1 -\bar{\nu}/2)^3 }
\end{align}
for $\bar{\nu}$
and replacing $\bar{\alpha}^2$ in the solution with $r_*$ using Eq.~\eqref{eq:rstar}. 
The result $0.15 \Delta \mu_m \Delta B$ in Eq.~\eqref{eq:nucminus2} can be obtained either by directly evaluating $ \frac{\hbar^2}{m_1 r_*^2} (\frac{r_*}{ \ammbg })^{2/3}$
using the narrow-resonance parameters given in Section~\ref{sec:parameters}, or by reading it directly from Fig.~\ref{fig:Fig1}(a). 

Combining Eqs.~\eqref{eq:estimate2} and \eqref{eq:nucminus2} we find
\begin{align}
\frac{ \text{6th order term}}{ \text{4th order term}}   & \lesssim \frac{ 1 }{\pi^2}  \frac{\hbar}{r_* \sqrt{m_1 |\bar{\nu}_{c,-}(n=0)|}} 
= \frac{1}{\pi^2} (\frac{ \ammbg }{r_*})^{1/3} \approx \frac{1}{\pi^2} 0.008^{1/3} 
=2\%.
\end{align}
Alternatively, the ratio can be expressed as
\begin{align}
\frac{ \text{6th order term}}{ \text{4th order term}}   & \lesssim \frac{ 1 }{\pi^2}  \frac{\hbar}{r_* \sqrt{m_1 |\bar{\nu}_{c,-}(n=0)|}} 
=  \frac{1}{\pi^2} \frac{ \sqrt{ \hbar^2 / ( m_1 a_{\text{bg}}^2 \Delta_{\bar{\nu}}} ) } { \sqrt{ |\bar{\nu}_{c,-}(n=0)|/\Delta_{\bar{\nu}} } }
\approx  \frac{0.008}{\sqrt{0.15} }  = 2\%,
\end{align}
where we have used Eq.~\eqref{eq:nucminus2} and the definition $\Delta_{\bar{\nu}}= \Delta \mu_m \Delta B$. 
Despite the approximate nature of the estimate, 
the small ratio of $2\%$ strongly supports our conclusion that the neglected higher-order correlations have a negligible impact on the phase boundaries in Fig.~\ref{fig:Fig1}.

\section{Effective molecular scattering length $\ammeff(n)$}
\label{sec:insight}

To understand the stark contrast in the compressibility between the narrow and wide resonances, we introduce the following effective molecule-molecule interaction parameter $g_{2}^{\mathrm{eff}}$
and the corresponding effective molecular scattering length $\ammeff$,
\begin{align}
\gtwoeff
& \equiv \frac{ (\mathcal{Z}^{\mathrm{eff}})^2 }{2}
\frac{\partial^4 \Otri[\Psi_{10}^s, \Psi_{20}, \Delta_1^s, \Delta_2^s,\mu]/V }{\partial \Psi_{20}^2 \partial (\Psi_{20}^*)^2 } \bigg \vert_{\mu},   \label{eq:g2effL1} \\
\ammeff & = \frac{m_2}{4\pi \hbar^2} \gtwoeff.   \label{eq:ammeffdef}
\end{align}
In Eq.~\eqref{eq:g2effL1}, the ground-state energy $\Otri$ was given in Eq.~\eqref{eq:Etri} and we have shown its dependence on the chemical potential explicitly.
The superscript ``$s$" in the pairing order parameter $\Delta_1^s=\Delta_1^s[\mu,\Psi_{20}]$ and others indicate that these quantities are set to their saddle-point values 
and are viewed as functions of the closed-channel molecular condensate amplitude $\Psi_{20}$ and the chemical potential $\mu$ only. 
To define the interaction $\gtwoeff$ for dressed molecules, instead of the bare closed-channel molecules, we have introduced the closed-channel fraction 
of dressed molecular condensate $\Zeff$ in Eq.~\eqref{eq:g2effL1}, which will be discussed in detail below.

The idea behind the definition of $\gtwoeff$ in Eq.~\eqref{eq:g2effL1} is to view both the closed-channel molecule condensate $\Psi_{20}$
and the atom-pair condensate $\Delta_1$ as part of a single dressed-molecule condensate $\Psi_{20}^{\mathrm{dressed}}$.
Then on the molecular side of the QCP, where $\Psi_{10}=0$, there is only one BEC condensate $\Psi_{20}^{\mathrm{dressed}}$,
for which we can follow the usual mean-field theory of a simple one-component BEC to define 
\begin{align}
\gtwoeff & =\frac{1}{2}  \frac{ \partial^4 \Otri }{ \partial (\Psi_{20}^{\mathrm{dressed}})^2
\partial (\Psi_{20}^{*, \mathrm{dressed}} )^2} \bigg \vert_{\mu}.  \label{eq:g2dress}
\end{align}
One might be concerned about the beyond-mean-field effects on $\gtwoeff$ due to the molecular depletion effect, represented by $\Delta_2$
in our $\Omega$. However, this latter effect is negligible, as evidenced by the magnitude of $n_2$ in Fig.~\ref{fig:Fig1v2}.
Now we can understand why the pre-factor $(\mathcal{Z}^{\mathrm{eff}})^2$ is needed in the definition of $\gtwoeff$ in Eq.~\eqref{eq:g2effL1}:
this factor is included to ensure that Eq.~\eqref{eq:g2effL1} is equivalent to \eqref{eq:g2dress},
since $\Zeff $ is defined as
\begin{align}
\Zeff  & \equiv  \frac{|\Psi_{20}|^2}{|\Psi_{20}|^2 + n_1/2} = \frac{|\Psi_{20}|^2}{|\Psi_{20}^{\mathrm{dressed}}|^2}.  \label{eq:Zeff}
\end{align}
In other words, $\Zeff$ describes the closed-channel fraction of the dressed molecule condensate. It can be viewed as the many-body analog of the two-body closed-channel fraction $\mathcal{Z}$ for a dressed molecule. 
The dressed molecule is an admixture of bare closed-channel molecules and open-channel pairs, which can be defined as~\cite{Romans2004,Duine2004} 
\begin{align}  \label{eq:dressedmol}
 |\Psi_{\mathrm{2\veck=0}}^{\mathrm{dressed}} \rangle & \equiv  \sqrt{\mathcal{Z}} a_{2\veck=0}^\dagger |0\rangle + V^{-1}\sum_{\veck^\prime} P_{\veck^\prime} a_{1\veck^\prime}^\dagger a_{1,-\veck^\prime}^\dagger | 0\rangle,
\end{align}
where $P_\veck$ is the $\veck$-dependent part of the dressed-molecule wavefunction that derives from the Cooper pairing of open-channel atoms.
In the zero density limit, $P_{\veck}$ is directly related to the $\chi_{1\veck}$ parameter in our variational wavefunction $|\Psivar \rangle$ in Eq.~\eqref{eq:Psivar}.
Correspondingly, $\Zeff \rightarrow \Z$ and $\ammeff \rightarrow \amm$ where $\amm$ is the two-body scattering length. 
We note that in this limit, the fact that the $\Z$ factor plays an important role in the calculation of the two-body molecular scattering length $\amm$ is consistent with previous diagrammatic calculations~\cite{Levinsen2006,Levinsen2011} of $\amm$, albeit in the context of fermionic BCS-BEC crossover. 

We now carry out the quartic derivative in Eq.~\eqref{eq:g2effL1} and arrive at (for details, see the next Section~\ref{sec:g2effZ2})
\begin{align}
\frac{\gtwoeff}{(\Zeff)^2}
&  \approx \widetilde{ \bar{g}}_{2} -  \frac{1}{2} \alpha \frac{\partial^3 x_1}{\partial \Psi_{20}^2 \partial \Psi_{20}^* }   \label{eq:g2effL2} \\
& \approx \widetilde{\bar{g}}_2  - \widetilde{\bar{\alpha}}^4 \bigg\{ 
\frac{1}{V}  \sum_\veck \frac{1}{2 E_{1\veck}^3} - 2 g_1  \bigg[ \frac{1}{V} \sum_\veck  \frac{ \widetilde{\epsilon}_{1\veck}}{ 2 E_{1\veck}^3 } \bigg]^2 \bigg\},   \label{eq:g2effL3}
\end{align}
where we have introduced
\begin{subequations} \label{eq:alphabartilde}
\begin{align}
\widetilde{\bar{g}}_2 & =g_2/(1+g_2 \frac{1}{V}\sum_\veck \frac{1}{2 E_{2\veck}}),    \label{eq:g2bartildedef}  \\
\widetilde{\bar{\alpha}} & =\sqrt{2}\alpha/(1+g_1 \frac{1}{V}\sum_\veck  \frac{1}{2 E_{1\veck}}), \label{eq:alphabartildedef} 
\end{align}
\end{subequations}
which are the many-body counterparts of the corresponding two-body interaction parameters
$\bar{g}_2$ and $\bar{\alpha}$.
Recall that 
$\bar{g}_2$ and $\bar{\alpha}$ are related to the bare parameter $g_2$ and $\alpha$ 
by $\bar{g}_2=g_2/(1+g_2 V^{-1}\sum_\veck 1/2 \epsilon_{2\veck}) $, which is equivalent to Eq.~\eqref{eq:g2}, and
$\bar{\alpha}=\sqrt{2} \alpha/(1+ g_1 V^{-1}\sum_\veck 1/2 \epsilon_{1\veck})$, which is equivalent to Eq.~\eqref{eq:alpharelation}. 
Here, $\epsilon_{1\veck}=\hbar^2 \veck^2/2m_1$ and $\epsilon_{2\veck}=\hbar^2 \veck^2/2m_2$. 

Some remarks are in order regarding Eq.~\eqref{eq:g2effL3}. We note that the interaction parameter appearing in the second term inside the curly brace is the bare $g_1$, not the renormalized
one $\bar{g}_1=4\pi \hbar^2 a_{\mathrm{bg}}/m_1=g_1/(1+ g_1 V^{-1}\sum_\veck 1/2 \epsilon_{1\veck})$.
This discrepancy arises because this term is derived from the Hartree-Fock self-energy $2 g_1 n_1$
that contributes to the expression for $E_{1\veck}$. 
It is well known that the current variational wavefunction ansatz in Eq.~\eqref{eq:Psivar} 
does not treat this self-energy in a fully self-consistent manner, leaving the $g_1$ parameter in this self-energy unrenormalized. 

In an ideal theory, one would expect the $g_1$ in Eq.~\eqref{eq:g2effL3} to be replaced by $\bar{g}_1$~\cite{Leggett2003}, 
which is what we do in this supplement material as well as in the main text when we derive formulas for $\ammeff$ and $\amm$.
However, for the numerical illustrations in all our figures that compare the inverse compressibility $\kappa^{-1}$ with the effective molecular scattering length $\ammeff$, we use the bare $g_1$ in Eq.~\eqref{eq:alphabartildedef},
to ensure that the Hartee-Fock self-energy effects are treated consistently between the two.  

\subsection{Detailed derivations of Eq.~\eqref{eq:g2effL3}}
\label{sec:g2effZ2}
In this subsection, we provide detailed steps for how we arrive at Eq.~\eqref{eq:g2effL3}. 
Our derivation starts with the trial ground-state energy $\Otri$ in Eq.~\eqref{eq:Etri},
from which we obtain its first-order derivative with respect to the bare molecule condensate amplitude $\Psi_{20}^*$ as
\begin{subequations}
\begin{align}
\frac{\partial \Otri/V}{\partial \Psi_{20}^*} & \equiv  \frac{\partial ( \Otri[\Psi_{10}^s, \Psi_{20}, \Delta_1^s, \Delta_2^s, \mu]/V)}{\partial \Psi_{20}^*} \bigg\vert_\mu  \label{eq:1stderivative1}\\
  & =  [h_{20} + g_2 |\Psi_{20}|^2 + 2 g_2 n_2] \Psi_{20} + g_2 \Psi_{20}^* x_2 - \alpha (x_1 + \Psi_{10}^2).    \label{eq:1stderivative2}
\end{align}
\end{subequations}
As mentioned earlier, the superscript ``$s$" in $\Psi_{10}^s, \Delta_1^s$, etc. means that these quantities are set to their saddle point values,
meaning they are viewed as functions of $\{\Psi_{20},\mu \}$ and are defined implicitly through the saddle-point Eqs.~\eqref{eq:Psi10} and \eqref{eq:BCS}. 
From Eq.~\eqref{eq:1stderivative1} to  Eq.~\eqref{eq:1stderivative2} we have used the fact that indirect dependence on $\Psi_{20}$ through
$\Delta_1^s$ (as well as through $\Psi_{10}^s$ and others) does not contribute since the associated terms are proportional to the derivative $\partial (\Omega/V)/\partial \Delta_1$,
which vanishes when evaluated at the saddle point $\Delta_1 =\Delta_1^s$. \textit{For brevity, the superscript ``$s$" of $\{\Delta_1^s,\Psi_{10}^s, \Delta_2^s\}$ has been omitted in Eq.~\eqref{eq:1stderivative2} and will be omitted in the subsequent expressions of this section as well. }

Next, we calculate the second-order derivative $\partial^2 (\Otri/V) / (\partial \Psi_{20} \partial \Psi_{20}^*)$ from Eq.~\eqref{eq:1stderivative2} and obtain 
\begin{align}
\frac{\partial^2 (\Otri/V)}{\partial \Psi_{20} \partial \Psi_{20}^*}  & = 2 g_2 |\Psi_{20}|^2 + 2 g_2 n_2 +2 g_2 \Psi_{20} \frac{\partial n_2}{\partial \Psi_{20}}
+ g_2 \Psi_{20}^* \frac{\partial x_2}{\partial \Psi_{20}} -\alpha \frac{\partial x_1}{\partial \Psi_{20}}.  \label{eq:2ndderivative}
\end{align}
Here, we have dropped the $\Psi_{10}^2$ term from Eq.~\eqref{eq:1stderivative2} because we focus on the molecular side of the QCP, where $\Psi_{10}^s = 0$. 
From Eq.~\eqref{eq:2ndderivative} we then derive the third-order derivative $\partial^3 (\Otri/V) /(\partial \Psi_{20} \partial (\Psi_{20}^*)^2 )$,
\begin{align}
\frac{\partial^3 (\Otri/V)}{\partial \Psi_{20} \partial (\Psi_{20}^*)^2 }  & = 2 g_2 \Psi_{20}  + 2 g_2  \frac{\partial n_2}{\partial \Psi_{20}^*} +  2 g_2 \Psi_{20}  \frac{\partial^2 n_2}{ \partial \Psi_{20} \partial \Psi_{20}^*} 
 + g_2  \frac{\partial x_2}{\partial \Psi_{20}}  + g_2 \Psi_{20}^* \frac{\partial^2 x_2}{\partial \Psi_{20} \partial \Psi_{20}^*} - \alpha \frac{\partial^2 x_1}{\partial \Psi_{20} \partial \Psi_{20}^* }.
\end{align}
Finally, we obtain the quartic derivative
\begin{widetext}
\begin{align}
\frac{\partial^4 (\Otri/V)}{\partial \Psi_{20}^2 \partial (\Psi_{20}^*)^2 }  & = 2 g_2 + 4 g_2  \frac{\partial^2 n_2}{\partial \Psi_{20} \partial \Psi_{20}^*} + 2 g_2 \Psi_{20} \frac{\partial^3 n_2}{\partial \Psi_{20}^2 \partial \Psi_{20}^* }   + g_2  \frac{\partial^2 x_2}{\partial \Psi_{20} \partial \Psi_{20}} 
 + g_2 \Psi_{20}^* \frac{\partial^3 x_2}{\partial \Psi_{20}^2 \partial \Psi_{20}^*} - \alpha \frac{\partial^3 x_1}{\partial \Psi_{20}^2 \partial \Psi_{20}^* }.
\end{align}
\end{widetext}
On the right-hand side, the 2nd to 5th terms depend on either $n_2$ or $x_2$, both of which are very small. We only incorporate their effects
to the extent that they renormalize the first term, the bare interaction parameter $g_2$, to the renormalized one
$\widetilde{\bar{g}}_2=g_2/(1+ g_2 V^{-1}\sum_\veck 1/2 E_{2\veck})$. 
With this approximation, we then derive from Eq.~\eqref{eq:g2effL1}
\begin{align}
\frac{\gtwoeff }{ (\Zeff)^2 } & \approx  \widetilde{\bar{g}}_{2}  -  \frac{1}{2} \alpha \frac{\partial^3 x_1}{\partial \Psi_{20}^2 \partial \Psi_{20}^* },  \label{eq:appg2effx1}
\end{align}
which corresponds to the previous Eq.~\eqref{eq:g2effL2}.

To further carry out the derivative of $x_1$ in Eq.~\eqref{eq:g2effL2} we go back to the corresponding saddle point Eq.~\eqref{eq:BCS},
which contains the implicit dependence of $x_1$ on $\Psi_{20}$. 
By using $\widetilde{\Delta}_1=\Delta_1+g_1\Psi_{10}^2-2\alpha \Psi_{20} = g_1 x_1 - 2\alpha \Psi_{20} $, we can formally solve Eq.~\eqref{eq:BCS}
and obtain
\begin{align} 
x_1 & = \frac{2\alpha  F}{1+ g_1 F} \Psi_{20},  \quad  \text{with} \quad F \equiv   \frac{1}{V}\sum_\veck \frac{1}{ 2 E_{1\veck}}.  \label{eq:x1solution}
\end{align}
Then
\begin{subequations}
\begin{align}
\frac{\partial x_1}{\partial \Psi_{20}}   & =   \frac{2\alpha }{1+ g_1 F} F + \frac{2\alpha \Psi_{20} }{(1+g_1 F)^2} \frac{\partial F}{\partial \Psi_{20}}, \\
\frac{\partial^2  x_1}{\partial \Psi_{20} \partial \Psi_{20}^*}  & =  \frac{2\alpha }{(1+ g_1 F)^2 } \frac{\partial F}{\partial \Psi_{20}^*}  +  \frac{2\alpha \Psi_{20}}{(1+ g_1 F)^2} \frac{\partial^2 F}{\partial \Psi_{20} \partial \Psi_{20}^* }   -   \frac{4\alpha  g_1 \Psi_{20}}{(1+ g_1 F)^2} \frac{\partial F}{\partial \Psi_{20}}  \frac{\partial F}{\partial \Psi_{20}^*} ,   \\
\frac{\partial^3  x_1}{\partial \Psi_{20}^2 \partial \Psi_{20}^*}  & = \frac{4\alpha }{(1+ g_1 F)^2 } \frac{\partial^2 F}{\partial \Psi_{20} \partial \Psi_{20}^*} - \frac{8\alpha g_1}{(1+ g_1 F)^2} \frac{\partial F}{\partial \Psi_{20}}  \frac{\partial F}{\partial \Psi_{20}^*}   +   \frac{2\alpha \Psi_{20}}{(1+ g_1 F)^2} \frac{\partial^3 F}{\partial \Psi_{20}^2 \partial \Psi_{20}^* }    \nonumber \\ 
&  -  \frac{8\alpha g_1  \Psi_{20} }{(1+ g_1 F)^3} \frac{\partial^2 F}{\partial \Psi_{20} \partial \Psi_{20}^* } \frac{\partial F}{\partial \Psi_{20}} 
-  \frac{4\alpha  g_1 \Psi_{20} }{(1+ g_1 F)^3} \frac{\partial^2 F}{\partial \Psi_{20}^2 } \frac{\partial F}{\partial \Psi_{20}^*} 
 + \frac{12\alpha  g_1^2 \Psi_{20}}{(1+ g_1 F)^4} ( \frac{\partial F}{\partial \Psi_{20} })^2 \frac{\partial F}{\partial \Psi_{20}^*}. 
\end{align}
\end{subequations}
On the right-hand side of the equation of $\partial^3  x_1/(\partial \Psi_{20}^2 \partial \Psi_{20}^*)$ above, the last five terms
vanish in the zero-density limit.
This is because: the last four terms are explicitly $\propto \Psi_{20}$, and the second term involves $\partial F/\partial \Psi_{20} \propto \Psi_{20}^*$ (because of gauge invariance).
At finite density $n$, these terms are smaller than the $\partial^2 F/(\partial \Psi_{20} \partial \Psi_{20}^*)$ term,
because of their higher-order dependencies in one of the two dimensionless small parameters, 
$g_1 n/E_{1\veck=0}$ and $\alpha \sqrt{n}/E_{1\veck=0}$.
Therefore, as long as $\bar{\nu}$ is not too close to the QCP so that $E_{1\veck=0}\gtrsim \{ g_1 n, \alpha \sqrt{n}\}$,
one can safely drop these terms. 
This approximation is further validated by our numerical results in Fig.~\ref{fig:Fig4},
where we compare the numerical $\kappa^{-1}$ with our $\ammeff$
and find a good agreement. 
With this approximation we have
\begin{align}   \label{eq:x1derivativeL3}
\frac{\partial^3  x_1}{\partial \Psi_{20}^2 \partial \Psi_{20}^*}  & \approx \frac{4\alpha }{(1+ g_1 F)^2 } \frac{\partial^2 F}{\partial \Psi_{20} \partial \Psi_{20}^*}. 
\end{align}

We still need to compute the derivative $\partial^2 F / \partial \Psi_{20} \partial \Psi_{20}^* $ in Eq.~\eqref{eq:x1derivativeL3}. 
From the definition of $F$ in Eq.~\eqref{eq:x1solution} and the expression of $E_{1\veck}$ below Eq.~\eqref{eq:BCS} we obtain
\begin{subequations}
\begin{align}
\frac{\partial F}{\partial \Psi_{20}}  & =  \frac{1}{V} \sum_{\veck} \frac{1}{2} \frac{-1}{2 E_{1\veck}^3} \bigg[- \widetilde{\Delta}_1^* \frac{\partial \widetilde{\Delta}_1}{\partial \Psi_{20}}  + 2 \widetilde{\epsilon}_{1\veck} 2 g_1 \frac{\partial n_1}{\partial \Psi_{20}} \bigg] \\
\frac{\partial^2 F}{\partial \Psi_{20} \partial \Psi_{20}^* }  & = \frac{1}{V} \sum_{\veck} \frac{1}{2} \frac{-1}{2 E_{1\veck}^3} 
\bigg[  - \frac{\partial \widetilde{\Delta}_1^*}{\partial \Psi_{20}^* } \frac{\partial \widetilde{\Delta}_1}{\partial \Psi_{20}}  
+ 2 \widetilde{\epsilon}_{1\veck} 2 g_1 \frac{\partial^2 n_1}{\partial \Psi_{20} \partial \Psi_{20}^* } \bigg]     \nonumber \\
&  \approx \frac{2\alpha^2}{(1+ g_1 F )^2} \bigg\{ 
\frac{1}{V} \sum_{\veck}  \frac{1}{2 E_{1\veck}^3} - 2 g_1  \bigg[ \frac{1}{V}  \sum_{\veck}  \frac{ \widetilde{\epsilon}_{1\veck}}{ 2 E_{1\veck}^3 } \bigg]^2  
 \bigg\}.     \label{eq:F3derivative}
\end{align}
\end{subequations}
To obtain the last line we have used 
\begin{subequations}
\begin{align}
\frac{ \partial \widetilde{\Delta}_1 }{ \partial \Psi_{20} } & = \frac{\partial }{\partial \Psi_{20}} \frac{ - 2\alpha \Psi_{20}}{1+ g_1 F} \approx \frac{ - 2\alpha }{1+ g_1 F} \\
\frac{\partial n_1}{\partial \Psi_{20}}  & = \frac{1}{V}\sum_{\veck} \bigg\{ 
\frac{1}{2} \bigg[ \frac{1}{E_{1\veck}}  - \frac{\widetilde{\epsilon}_{1\veck}^2}{ E_{1\veck}^3} \bigg] (2 g_1 \frac{\partial n_1}{\partial \Psi_{20}})  +  \frac{1}{2} \frac{\widetilde{\epsilon}_{1\veck}}{(-2) E_{1\veck}^3} (- \widetilde{\Delta}_1^*) \frac{\partial \widetilde{\Delta}_1}{\partial \Psi_{20}}
\bigg\}
 \approx  \frac{1}{V} \sum_{\veck} \frac{\widetilde{\epsilon}_{1\veck}}{4 E_{1\veck}^3} \widetilde{\Delta}_1^* \frac{\partial \widetilde{\Delta}_1}{\partial \Psi_{20}}, \\
\frac{\partial^2 n_1}{\partial \Psi_{20} \partial \Psi_{20}^* }  & \approx \frac{1}{V}\sum_{\veck} \frac{\widetilde{\epsilon}_{1\veck}}{4 E_{1\veck}^3} \frac{\widetilde{\Delta}_1^* }{\partial \Psi_{20}^*}\frac{\partial \widetilde{\Delta}_1}{\partial \Psi_{20}} 
 \approx \frac{1}{V} \sum_{\veck} \frac{\widetilde{\epsilon}_{1\veck}}{4 E_{1\veck}^3} (\frac{2\alpha}{1+ g_1 F} )^2. 
\end{align}
\end{subequations}
Again, we have dropped subleading terms. We have also used $\widetilde{\Delta}_1=g_1 x_1 - 2\alpha \Psi_{20}$ from Eq.~\eqref{eq:Delta1Tilde}, 
with $x_1$ given in Eq.~\eqref{eq:x1solution},
and the expression of $n_1=\frac{1}{V}\sum_\veck (1/2) (\widetilde{\epsilon}_{1\veck}/E_{1\veck} -1)$ from Eq.~\eqref{eq:nsigma}. 

Substituting the result of $\partial^2 F / \partial \Psi_{20} \partial \Psi_{20}^* $ in Eq.~\eqref{eq:F3derivative} into Eq.~\eqref{eq:x1derivativeL3} 
and then inserting the resulting expression into Eq.~\eqref{eq:appg2effx1}, we arrive at the final expression of $\gtwoeff/(\Zeff)^2$ in Eq.\eqref{eq:g2effL3}.

\subsection{Comparing the effective molecular scattering length $\ammeff$ with the inverse compressibility $\kappa^{-1}$}
\label{sec:ammmanybody}

Using the expression for $\gtwoeff/(\Zeff)^2$ in Eq.~\eqref{eq:g2effL3} and that for $\Zeff$ in Eq.~\eqref{eq:Zeff} 
we can now evaluate the $\ammeff$ in Eq.~\eqref{eq:ammeffdef}, and compare $\ammeff$ to the numerical results of the inverse compressibility $\kappa^{-1}$
to see how well $\kappa^{-1}$ is described by the $\ammeff$ defined. 
Before making this comparison, we first discuss the implications of the formula of $\gtwoeff$ in Eq.~\eqref{eq:g2effL3}.

We first note that for a positive molecular background interaction $g_2$, the term that drives $\gtwoeff$ negative is
$- \widetilde{\bar{\alpha}}^4 V^{-1} \sum_\veck 1/ (2 E_{1\veck}^3)$. 
The contribution of this term can still be schematically represented by the Feynman diagram in Fig.~\ref{fig:DiagramC}(a),
which we previously presented for the two-body scattering length $\amm$. 
The main differences in the current context are (1) the two-body Feshbach coupling constant $\bar{\alpha}$ in that figure is now replaced by
its many-body counterpart $ \widetilde{\bar{\alpha}}$, and (2) the intermediate atomic propagator lines in Fig.~\ref{fig:DiagramC}(a) are now replaced by those of
the corresponding Bogoliubov quasiparticles, due to the presence of molecular condensates. 
This latter difference is the origin of the $1/(2E_{1\veck}^3)$ factor.

The overall minus sign in the negative term, $- \widetilde{\bar{\alpha}}^4 V^{-1} \sum_\veck 1/ (2 E_{1\veck}^3)$, contrasts with 
the situation when the open-channel atoms are \textit{fermionic}, as we will see from Eq.~\eqref{eq:g2effF} in the following Section~\ref{sec:contrast}, where we derive the fermionic counterpart of Eq.~\eqref{eq:g2effL3}. 
This contrast highlights the important role 
played by the statistics of the underlying atoms in contributing to the effective molecular scattering length $\ammeff$.
The origin of this sign difference stems from the fact that the scattering process in Fig.~\ref{fig:DiagramC}(a)
involves a net exchange of two intermediate atomic excitations---whether they are the bare atoms in vacuum or the Bogoliubov quasiparticles
in the presence of condensates---whose contribution to the molecular scattering amplitude differ by a minus sign between
the bosonic and fermionic atom scenarios.

\begin{figure}[htp]
\begin{center}
\includegraphics[width=0.75\linewidth,trim=0 0 0 0,clip]{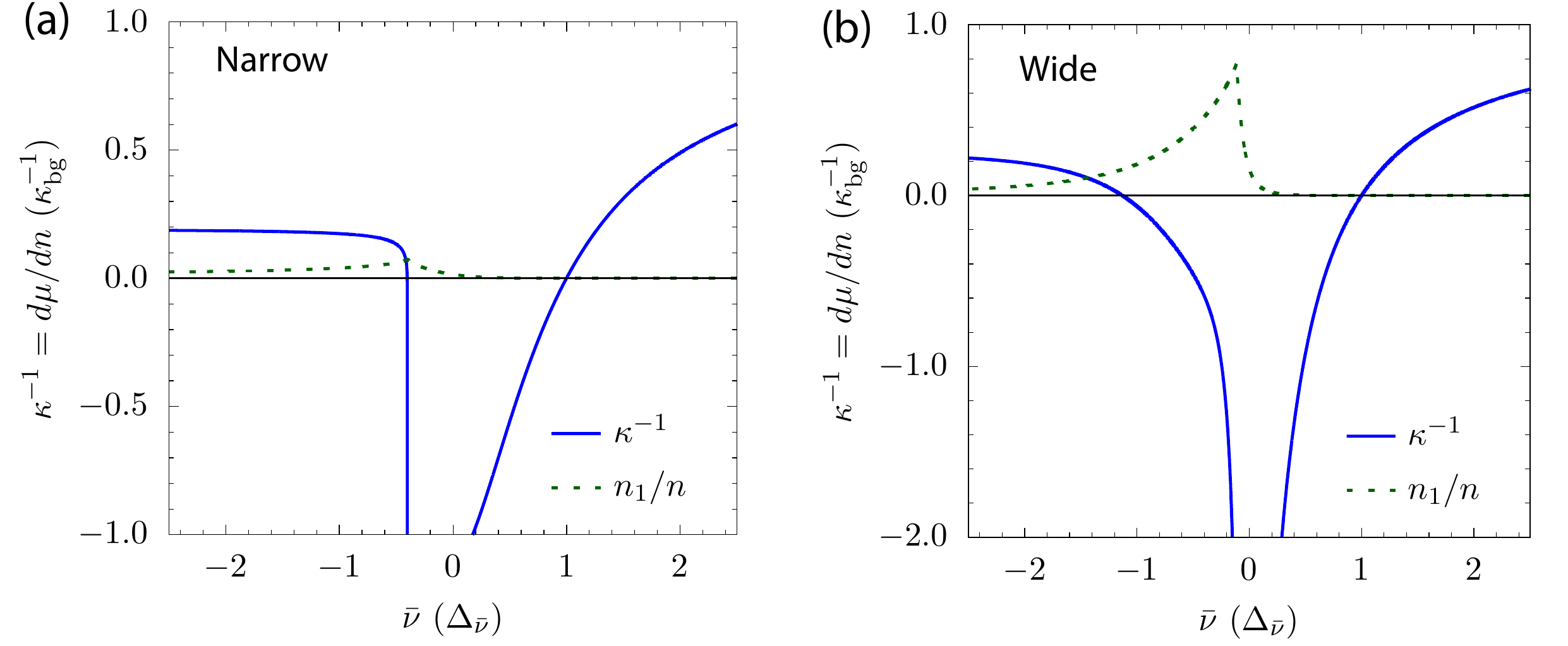}
\caption{Correlation between the number of atom pairs ($n_1/n$ in dark-green dashed lines) and the appearance of a wider unstable regime on the molecular side of the QCP (the cusps in the $n_1/n$ line). The results are plotted for a fixed particle density $n \asbg^3=1.68\times 10^{-5}$.
The blue solid lines are a constant-density line cut of the results in Fig.~\ref{fig:Fig1}, except that in Fig.~\ref{fig:Fig1}, 
we plotted the compressibility $\kappa$, instead of its inverse $\kappa^{-1}$, for a better contrast at the boundary between the stable and unstable regimes. In both the narrow and wide resonance cases, the inverse compressibility $\kappa^{-1}$ remains finite at the resonance point and it shows a weak kink singularity (not shown) right at the QCP detuning $\bar{\nu}_c$. 
}
\label{fig:Fig3}
\end{center}
\end{figure}

From Eq.~\eqref{eq:g2effL2} we also observe that the $\widetilde{\bar{\alpha}}^4$ term in Eq.~\eqref{eq:g2effL3} 
can be traced back to the term $\partial^3 x_1/(\partial \Psi_{20}^2 \partial \Psi_{20}^* ) $ in Eq.~\eqref{eq:g2effL2}, which is a derivative of the atom pairing correlation $x_1$ with respect to the closed-channel molecular condensate amplitude $\Psi_{20}$ (as well as its complex conjugate).
The form of this derivative indicates that the underlying physics originates from a fluctuating inter-conversion between molecules $\Psi_{20}$ and atom pairs $x_1\propto \Delta_1$, showing that the atom pairs play an important role in driving $\ammeff$ negative, thereby undermining the molecular condensate stability.
This relation is clearly illustrated in Fig.~\ref{fig:Fig3}, where a direct correlation is observed between
the emergence of a more extensive unstable detuning regime in the wide-resonance case
and the presence of a substantial number of atom Cooper pairs.

\begin{figure}[htp]
\begin{center}
\includegraphics[width=0.75\linewidth,trim=0 0 0 0,clip]{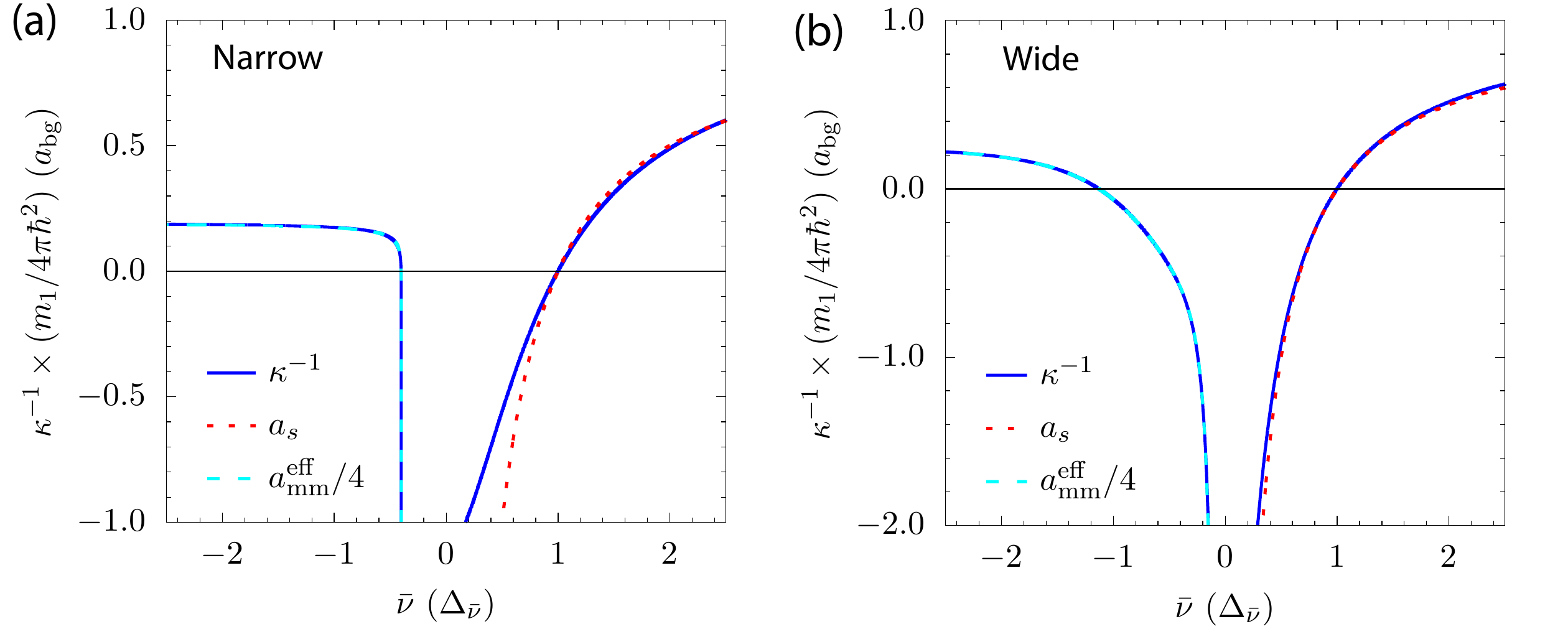}
\caption{ Comparison of the inverse compressibility $\kappa^{-1}$ (in blue solid lines) to the atomic scattering length $a_s$ (in red dashed lines) and the effective molecule-molecule scattering length $\ammeff$ (in cyan dashed lines) for both the narrow (a) and wide (b) resonances. 
 We only show results of $a_s$ and $\ammeff$ for positive and negative detunings, respectively. The effective molecular scattering length $\ammeff$ is defined from the effective
molecule-molecule interaction $g_{2}^{\mathrm{eff}}$ in Eqs.~\eqref{eq:g2effL1}. In the asymptotic limit of large negative and positive detuning, $\ammeff$ and $a_s$ approach the background value $\ammbg$ and $\asbg$, respectively.
}
\label{fig:Fig4}
\end{center}
\end{figure}

We now numerically evaluate $\ammeff$ using Eqs.~\eqref{eq:ammeffdef}, \eqref{eq:g2effL3}, and \eqref{eq:Zeff},
and compare it with the (rescaled) $\kappa^{-1}$ in Fig.~\ref{fig:Fig4}. 
Surprisingly, the $\ammeff$ we defined agrees almost exactly with the numerical $\kappa^{-1}$ for all negative detunings up to the QCP,
for both the narrow and wide resonances. The difference is not discernable to the eyes.
What this means is that the terms that we have neglected in obtaining Eq.~\eqref{eq:g2effL3} are indeed negligible. 

Since $\kappa^{-1}$ is almost equal to $\ammeff$, we can now use the formula of $\ammeff$, or more precisely that of $\gtwoeff$ in Eq.~\eqref{eq:g2effL3}
to understand the contrast between the wide and narrow resonance cases in Figs.~\ref{fig:Fig1} and \ref{fig:Fig4}.
For a wide resonance, the large $\widetilde{\bar{\alpha}}^4$ factor in Eq.~\eqref{eq:g2effL3}
quickly drives $\gtwoeff$ negative at a detuning far from the QCP as the detuning approaches the QCP from the molecular side.
Recall that the QCP is the point where $E_{1\veck}$ becomes gapless. 
In contrast, for a very narrow resonance, the very small $\widetilde{\bar{\alpha}}^4$ in Eq.~\eqref{eq:g2effL3} means that
$\gtwoeff$ becomes negative only when $E_{1\veck=0}$ is sufficiently small, i. e. when $\bar{\nu}$ is sufficiently close to $\bar{\nu}_c$. 

In the above discussions, we have neglected the effect of $(\Zeff)^2$. However, this does not mean it plays no role. 
For a wide resonance at a large negative detuning, $\Zeff$ decreases as the detuning increases towards the resonance.
As a result, the effective scattering length $\ammeff$ decreases from its background value $\ammbg$.
This tends to drive $\ammeff$ to zero, which in the zero density limit results in the $A_1$ term in Eq.~\eqref{eq:ammunify} of $\amm$,
as we will see in the following Section~\ref{sec:zeronlimit}. 
On the other hand, for the narrow resonance, $(\Zeff)^2\approx 1$ at all detuning up to the QCP, so it does not play an important role. 
Considering the effect of $(\Zeff)^2$ sharpens the contrast between narrow and wide resonances.
We emphasize that our formula for the effective scattering length $\ammeff$ in Eq.~\eqref{eq:ammeffdef} has already incorporated this $(\Zeff)^2$ effect. 

\section{Two-body molecular scattering length $\amm$}
\label{sec:zeronlimit}

In this section, we will
\begin{enumerate}
\item Provide details on how we arrive at the expression for the two-body molecular scattering length $\amm$ presented earlier in Eq.~\eqref{eq:ammunify} by considering the
zero-density limit of the many-body effective scattering length $\ammeff$, which we defined in the previous section.
\item Show analytically that, in the zero density limit, the compressibility inverse $\kappa^{-1}$ calculated from $d\mu/dn$ is \textit{exactly} equal
to the effective scattering length $\ammeff$, which is defined from the quartic derivative of the ground state energy in Eq.~\eqref{eq:g2effL1}.
This exact relation, though obtained for zero density, partially explains the excellent agreement between the calculated $\kappa^{-1}$ and $\ammeff$ in Fig.~\ref{fig:Fig4} for finite density.  
\end{enumerate}

\subsection{Zero-density limit of $\ammeff(n) \rightarrow \amm$}
\label{sec:ammderive}

In the zero density limit, the interaction $\widetilde{\bar{g}}_2$ in Eq.~\eqref{eq:g2effL3} approaches its two-body counterpart, given by
$ \bar{g}_2=4\pi \hbar^2 \ammbg /m_2$
\footnote{Ideally, in the zero density limit, $\protect \widetilde{\bar{g}}_2$ should coincide exactly with $\bar{g}_2$.
However, this is not the case in our theory, as the molecule Bogoliubov quasiparticle energy $E_{2\veck}$ is not exactly
gapless, despite the molecules being condensed. This discrepancy should be viewed as an artifact arising from the approximate nature of the variational wavefunction we have employed. Physically, we expect that $E_{2\veck} \rightarrow \epsilon_{2\veck}$
in the zero density limit, so we simply replace $\protect \widetilde{\bar{g}}_2$ with $\bar{g}_2$
when deriving formulas for $\amm$. However, for numerical comparisons between $\kappa^{-1}$ and $\amm$ (or $\ammeff$),
we do not make such replacements.}.
The atomic Bogoliubov quasiparticle energy $E_{1\veck}$ in Eq.~\eqref{eq:g2effL3} becomes the bare atomic energy dispersion 
$h_{1\veck}=\epsilon_{1\veck} -\mu= \hbar^2 \veck^2/2m_1+E_b/2$, where $E_b$ the molecule binding energy. 
Substituting both $E_{1\veck}$ and $\widetilde{\epsilon}_{1\veck}$ in Eq.~\eqref{eq:g2effL3}  with $h_{1\veck}$, completing the integrals over the momentum $\veck$,
and using the definition of the length scale $r_*$ from Eq.~\eqref{eq:rstar},
we obtain
\begin{align} \label{eq:g2effzeron}
\frac{ \gtwoeff }{\Z^2 } & = \frac{4\pi \hbar^2 \ammbg }{m_2} - \frac{2\pi \hbar^2}{m_1} \frac{1}{ k_b^4 r_*^2 \bar{a}}  \frac{C_3 -4 C_2^2 \asbg / \bar{a} }{(1- C_1 \asbg / \bar{a} )^4},
\end{align}
where on the left hand side we have already replaced the factor $(\Zeff)^2$ with its two-body counterpart $\Z^2$.
In the above equation, $C_1,C_2$ and $C_3$ are three functions of the dimensionless product $k_b \bar{a}$, and their expressions were given earlier in Eqs.~\eqref{eq:C1}, \eqref{eq:C2}, and \eqref{eq:C3},
respectively. 
These functions come from the following integrals over $\veck$, 
\begin{subequations} \label{eq:C1C2C3}
\begin{align}
 \int^\Lambda \frac{d\veck } { (2\pi)^3} \frac{1}{ 2 h_{1\veck} }  &  = \int^\Lambda \frac{d \veck}{(2\pi)^3}  \frac{1}{2 (\hbar^2 \veck^2/2m_1 +E_b/2)}  
 =   \frac{m_1 \Lambda}{2\pi^2 \hbar^2 } \bigg\{ 1- \frac{k_b}{\Lambda}  \tan^{-1} \frac{\Lambda}{k_b}  \bigg\}   \equiv \frac{m_1}{4\pi \hbar^2 \bar{a} } \bigg(1- C_1 \bigg) , \\
 \int^\Lambda \frac{d\veck } { (2\pi)^3} \frac{1}{ 2 h_{1\veck}^2 }  & = \frac{m_1^2 \Lambda}{2\pi^2 \hbar^4 k_b^2 }  \bigg\{  \frac{k_b}{\Lambda} \tan^{-1} \frac{\Lambda}{k_b}  -\frac{  1 } { \Lambda^2 / k_b^2 +1 }\bigg\} \equiv   \frac{m_1^2}{4\pi \hbar^4 k_b^2 \bar{a} } C_2, \\
 \int^\Lambda \frac{d\veck } { (2\pi)^3} \frac{1}{ 2 h_{1\veck}^3 }  &  =  \frac{m_1^3 \Lambda}{4\pi^2 \hbar^6  k_b^4 }  \bigg\{  \frac{k_b}{\Lambda} \tan^{-1} \frac{\Lambda}{k_b}  +   \frac{  \Lambda^2 / k_b^2  -1 } {( \Lambda^2 / k_b^2 +1)^2}\bigg\}  \equiv \frac{m_1^3 }{8\pi \hbar^6 k_b^4 \bar{a}  } C_3,
\end{align}
\end{subequations}
 where we have used the relation $\Lambda \equiv \pi/(2 \bar{a})$.  
In writing down Eq.~\eqref{eq:g2effzeron} we have also made the replacement $g_1\rightarrow \bar{g}_1=4\pi \hbar^2 \asbg/m_1$. See the discussions below Eq.~\eqref{eq:g2effL3}
for why we do this replacement. 

To compute the two-body scattering length $\amm$ from Eq.~\eqref{eq:g2effzeron} we still need the expression of the closed-channel fraction $\Z$, which can be obtained from the zero-density limit of the closed-channel fraction $\Zeff$ of the dressed molecular condensates  in Eq.~\eqref{eq:Zeff}, i. e.
\begin{align} 
\mathcal{Z}  & = \lim_{n\rightarrow 0 } \Zeff  =   \lim_{n\rightarrow 0 } \frac{1}{1+ n_1/(2|\Psi_{20}|^2)} 
 =  \frac{2 k_b r_* (1- C_1 \asbg/\bar{a})^2}{C_2/(k_b \bar{a})+2 k_b r_* (1- C_1 \asbg/\bar{a} )^2},   \label{eq:zeronZ}
\end{align}
where we have used 
\begin{align}
\lim_{n\rightarrow 0} \frac{ n_1 }{2 |\Psi_{20}|^2} = \lim_{n\rightarrow 0}  \frac{d n_1}{2 d |\Psi_{20}|^2} = \frac{1}{2 k_b r_* } \frac{C_2/(k_b \bar{a})}{ (1- C_1 \asbg/\bar{a})^2 },  \label{eq:dn1dPsi20v2}
\end{align}
whose derivation will be given later in Section~\ref{sec:zeronexact} (see Eq.~\eqref{eq:dn1dPsi20} below). 
One can also derive the same expression for the two-body closed-channel fraction $\mathcal{Z}$ by calculating the closed-channel molecule propagator 
in the context of a two-atom scattering problem in vacuum. 
As a check of the correctness of  Eq.~\eqref{eq:zeronZ}, we can compare its limit at $\asbg=0$
with the expression of $\Z$ found in the literature (see the appendix of Ref.~\cite{Levinsen2011} for example),
and we find they agree. 
 
Now, multiplying Eq.~\eqref{eq:g2effzeron} by $\Z^2$ with $\mathcal{Z}$ from Eq.~\eqref{eq:zeronZ} we obtain
\begin{align}  \label{eq:ammuniv1}
\amm & \equiv \lim_{n\rightarrow 0} \frac{m_2}{4\pi \hbar^2} \gtwoeff  
= \bigg\{ \ammbg - \frac{1}{ k_b^4 r_*^2 \bar{a}}  \frac{C_3 - 4 (C_2)^2 \asbg/\bar{a} }{(1- C_1 \asbg/\bar{a})^4} \bigg\} 
\bigg[ \frac{2 k_b r_* (1- C_1 \asbg/\bar{a})^2}{C_2/(k_b \bar{a})+2 k_b r_* (1- C_1 \asbg/\bar{a})^2} \bigg]^2. 
\end{align}
In Section~\ref{sec:contrast}, we will present the fermionic counterpart formula of Eq.~\eqref{eq:ammuniv1}.
Next, we rewrite Eq.~\eqref{eq:ammuniv1} in the form of Eq.~\eqref{eq:ammunify} as follows,
\begin{subequations} \label{eq:ammABC2}
\begin{align}
\amm & = \bigg\{ \ammbg - \frac{1}{ k_b^4 r_*^2 \bar{a}}  \frac{C_3 - 4 C_2^2 \asbg/\bar{a} }{(1- C_1 \asbg/\bar{a})^4} \bigg\}  \Z^2 \\
&= \ammbg - \ammbg (1- \Z^2)  - \frac{1}{k_b^4 r_*^2 \bar{a}}  \frac{C_3 \Z^2} {(1- C_1 \asbg/\bar{a})^4 } + \frac{\asbg}{k_b^4 r_*^2 \bar{a}^2} \frac{4 C_2^2 \Z^2 }{(1- C_1 \asbg/\bar{a})^4 }  \\
& \equiv \ammbg - \frac{\ammbg}{k_b^4 r_* \bar{a}^3} A_1   - \frac{1}{k_b^6 r_*^2 \bar{a}^3} A_2 + \frac{\asbg}{k_b^8 r_*^2 \bar{a}^6 } A_3.  \label{eq:ammABC}
\end{align}
\end{subequations}
One can verify that the expressions for $\{A_1, A_2, A_3\}$ are given by those in Eqs.~\eqref{eq:A}, \eqref{eq:B}, and \eqref{eq:C}. 
This rewriting clearly shows that, indeed, the $A_1$ term originates from a molecular self-energy effect; more specifically, it results from
the reduction of the closed-channel fraction $\Z$ due to Feshbach coupling between the two channels. 
Additionally, we also observe that the $A_1$ term by itself can only reduce the scattering length $\amm$ from its background value $\ammbg$ to zero
but can never make it negative, since the factor $\Z^2$ is positive semi-definite. 

\subsection{Analytical derivation of the zero-density limit of the inverse compressibility $\kappa^{-1}$}
\label{sec:zeronexact}

In this subsection, we derive the zero-density limit of the inverse compressibility $\kappa^{-1}=d\mu/dn$ analytically within our variational approach and show
that, after being rescaled to match the physical dimensions, it is exactly equal to the two-body scattering length $\amm$ we derived in Eq.~\eqref{eq:ammuniv1}, which
was defined from the quartic derivative of the ground-state energy in Eq.~\eqref{eq:g2effL1}. 
While this exact equivalence is expected, it is not guaranteed a priori given the approximate nature of our variational
wavefunction treatment of the problem. Therefore, it is reassuring that we can show the two are indeed in exact agreement. 
Additionally, in the following derivation, we will also show how Eq.~\eqref{eq:zeronZ} is obtained.

Our derivation starts with the total particle number density constraint,
\begin{subequations}
\begin{align}
 n  & = 2  |\Psi_{20}|^2 + n_1 + 2 n_2, \\
\Rightarrow dn & = 4 \Psi_{20} d\Psi_{20}  + d n_1.  \label{eq:dneq1}
\end{align}
\end{subequations}
In the first line, we have used the fact that for the considered negative detuning regime, $\Psi_{10}=0$.
From the first line to the second we have dropped the molecular depletion term $dn_2$, since $d n_2$ is negligible compared to $dn$, i. e. $dn_2/dn \propto dn_2/d|\Psi_{20}|^2 \rightarrow 0$.
For simplicity, we have also treated $\Psi_{20}$ as a real parameter in this subsection to make the derivation easier to follow. 

The next step is to express the two terms on the right-hand side of Eq.~\eqref{eq:dneq1} in terms of $d\mu$ and $d n$
by using the saddle-point equations for $\{\Psi_{10}, \Psi_{20}, \Delta_\sigma \}$ in Eqs.~\eqref{eq:saddle} and \eqref{eq:BCS}. 
We first take the $\Psi_{20}$ saddle-point equation~\eqref{eq:Psi20}, 
divide its both hand sides by $\Psi_{20}$ which is nonzero (we will send $\Psi_{20}\rightarrow 0$ at the end of the calculation, but not in the intermediate step),
and then differentiate the resulting equation to obtain
\begin{subequations} 
\begin{align}
2 d\mu & =  2 \frac{g_2}{1+g_2 K} \Psi_{20} d\Psi_{20} -  \frac{2\alpha^2}{(1+g_1 F)^2} d F,  \label{eq:dmudA} \\
\Rightarrow 2 d \mu & = 2 \frac{g_2}{1+g_2 K} \Psi_{20} d\Psi_{20} -  \frac{2\alpha^2}{(1+g_1 F)^2} \bigg\{ -I  (- d\mu+ 2 g_1 d n_1)  + J  \big( \frac{ 2\alpha}{1+ g_1 F} \big)^2 \Psi_{20}  d\Psi_{20}\bigg\} .       \label{eq:dmudA2}
\end{align}
\end{subequations}
To arrive at the first line, we have dropped terms that are higher order than linear in the density $n$. 
The term that is $\propto \Psi_{20}d\Psi_{20}$ in Eq.~\eqref{eq:dmudA} is derived from $d \big( g_2 |\Psi_{20}|^2 + g_2 x_2 \big)$, for which we have used the following expression for $x_2$,
\begin{align}
x_2 &  =  - \frac{ g_2 K }{1+g_2 K}  \Psi_{20}^2  \quad \text{with} \quad  K  =  \frac{1}{V}  \sum_\veck \frac{1}{2 E_{2\veck}},  \label{eq:x2solution}
\end{align}
which itself is derived from Eq.~\eqref{eq:BCS}.
The $dF$ term in Eq.~\eqref{eq:dmudA} is obtained from $-\alpha d x_1$ with the expression of $x_1$ taken from Eq.~\eqref{eq:x1solution},
which involves $F=(1/V)\sum_\veck 1/(2 E_{1\veck})$.
From Eq.~\eqref{eq:dmudA} to \eqref{eq:dmudA2}, we have used
$ d F  = -I  (- d\mu+ 2 g_1 d n_1)  +  J \big( \frac{ 2\alpha}{1+ g_1 F} \big)^2 \Psi_{20}  d\Psi_{20}$,
with $I$ and $J$ two integrals over $\veck$, given by
\begin{subequations} \label{eq:IJexp}
\begin{align}
I & \equiv  \lim_{n\rightarrow 0} \frac{1}{V}\sum_\veck  \frac{\widetilde{\epsilon}_{1\veck}}{2 E_{1\veck}^3} =  \frac{1}{V}\sum_\veck  \frac{1}{2 h_{1\veck}^2} =\frac{m_1^2}{4\pi \hbar^4 k_b^2 \bar{a} } C_2,  \label{eq:Iexp}\\
J & \equiv  \lim_{n\rightarrow 0}  \frac{1}{V}\sum_\veck \frac{1}{2 E_{1\veck}^3} =   \frac{1}{V}\sum_\veck \frac{1}{2 h_{1\veck}^3} =\frac{m_1^3 }{8\pi \hbar^6 k_b^4 \bar{a}  } C_3.  \label{eq:Jexp}
\end{align}
\end{subequations}
Here, we have used Eq.~\eqref{eq:C1C2C3}. 

Then from the expression of $n_1=1/(2 V) \sum_\veck (\widetilde{\epsilon}_{1\veck}/E_{1\veck} -1)$ and $\widetilde{\Delta}_1= g_1 x_1 - 2\alpha \Psi_{20}=-2\alpha \Psi_{20}/(1+g_1 F)$ (see Eq.~\eqref{eq:x1solution}) we obtain
\begin{align}
d n_1 & =  \frac{1}{V}\sum_\veck  \frac{\widetilde{\epsilon}_{1\veck}}{4 E_{1\veck}^3} d |
\widetilde{\Delta}_1|^2   =  I  ( \frac{ 2\alpha }{1+ g_1 F})^2 \Psi_{20}  d\Psi_{20}.   \label{eq:dn1dPsi20}
\end{align}
Again, to arrive at this equation we have dropped terms that are higher order in the density $n$.
We note that this equation immediately leads to Eq.~\eqref{eq:dn1dPsi20v2} used in the previous Section~\ref{sec:ammderive}, if we use Eq.~\eqref{eq:Iexp} for the expression of $I$, along with the following Eq.~\eqref{eq:g2alphazeron} for $\alpha/(1+ g_1 F)$ and the definition of $\bar{\alpha}$ in terms of $r_*$ provided in Eq.~\eqref{eq:rstar}.

Eqs.~\eqref{eq:dneq1}, \eqref{eq:dmudA2}, and \eqref{eq:dn1dPsi20} form a set of three coupled but independent equations for the differentials $\{d n_1, d\Psi_{20}, d\mu, d n\}$.
By using linear combinations of the three equations, we can eliminate two differentials $\{ dn_1, d\Psi_{20}\}$ to obtain a relation between $d\mu$ and $d n$,
from which we deduce
\begin{align}
\kappa^{-1} = \frac{d\mu}{dn}  & = \frac{1}{4} \frac{ \widetilde {\bar{g}}_2 - \widetilde{\bar{\alpha}}^4 ( J -  2 g_1 I^2)}{(1 + \widetilde{\bar{\alpha}}^2 I/2)^2 }.  \label{eq:kappainvexact}
\end{align}
In writing down this expression, we have used Eq.~\eqref{eq:alphabartilde} for both $\widetilde{\bar{g}}_2$ and $\widetilde{\bar{\alpha}}$, which in the zero
density limit becomes
\begin{subequations} \label{eq:g2alphazeron}
\begin{align}
\widetilde{\bar{g}}_2 & =\frac{g_2} {1+g_2 \frac{1}{V}\sum_\veck 1/(2 E_{2\veck} ) } \overset{n \rightarrow 0}{=} \frac{g_2} {1+g_2 \frac{1}{V}\sum_\veck 1/(2\epsilon_{2\veck}) } 
=\bar{g}_2= \frac{4\pi \hbar^2 \ammbg}{m_2},   \\
\widetilde{\bar{\alpha}} & =\frac{\sqrt{2}\alpha}{1+g_1 F} = \frac{\sqrt{2}\alpha} {1+g_1 \frac{1}{V}\sum_\veck  1/(2 E_{1\veck}) } \overset{n \rightarrow 0}{=}  
\frac{\sqrt{2}\alpha} {1+g_1 \frac{1}{V}\sum_\veck  1/(2 h_{1\veck })} = \frac{\sqrt{2}\alpha} {1+g_1 (m_1/(4\pi \hbar^2 \bar{a} )) (1- C_1 )}
= \frac{\bar{\alpha}}{1- C_1 \asbg/\bar{a} }. 
\end{align} 
\end{subequations}
Substituting Eq.~\eqref{eq:g2alphazeron} for $\{\widetilde{\bar{g}}_2,\widetilde{\bar{\alpha}} \}$ and Eq.~\eqref{eq:IJexp} for $\{I,J\}$ into Eq.~\eqref{eq:kappainvexact},
replacing the bare $g_1$ in Eq.~\eqref{eq:kappainvexact} with $\bar{g}_1=4\pi \hbar^2 \asbg/m_1$ as we have done for the scattering length $\amm$,
and then using Eq.~\eqref{eq:rstar} to substitute $\bar{\alpha}^2$ with $1/r_*$,
we see that the right hand side of Eq.~\eqref{eq:kappainvexact} is identical to Eq.~\eqref{eq:ammuniv1} of $\amm$,
i. e. 
\begin{align}
\kappa^{-1}=\frac{1}{4}  \bigg(\frac{4\pi \hbar^2}{m_2} \amm \bigg),
\end{align}
On the right hand side above, the prefactor $1/4$ appears because in the definition of the inverse compressibility $\kappa^{-1}$, the differentials $d \mu$ and $d n$ are defined for atoms, not for molecules.


\section{Accessibility of the QCP in a narrow resonance and derivations of Eq. (4) of the main text for the many-body effective scattering length $\ammeff$}
\label{sec:accessibility}

In Fig.~\ref{fig:Fig1}, as well as in Fig. 2 of the main text, we have seen that
for a narrow resonance the boundary detuning $\bar{\nu}_{c,-}$ between the stable MSF phase and the unstable regime
nearly coincides with the QCP $\bar{\nu}_c$, except when the density $n$ is extremely low. 
In the following, we describe quantitatively when such a near-coincidence occurs. 
At the end, we also show how we arrive at Eq. (4) of the effective scattering length $\ammeff$ in the main text. 

We first determine the QCP detuning $\bar{\nu }_c$ from the onset condition for the atomic condensate amplitude $\Psi_{10}$. 
From the saddle-point Eq.~\eqref{eq:Psi10} of $\Psi_{10}$ right at the QCP we obtain
\begin{align}
 \mu_1 & \equiv \mu =  2 g_1 n_1 - \frac{2 \alpha}{1+ g_1 F} \Psi_{20}.
\end{align}
Recall that $F \equiv V^{-1}\sum_\veck 1/2E_{1\veck}$.
From Eq.~\eqref{eq:Psi20} of $\Psi_{20}$ we have 
\begin{align}
\mu_2 \equiv 2\mu_1 -\nu_c \approx  \frac{g_2}{1+g_2 K} |\Psi_{20}|^2 - \alpha \frac{2\alpha F}{1+g_1 F}
\end{align}
where $K =V^{-1}\sum_\veck 1/2E_{2\veck}$ and we have neglected the small molecule depletion contributions from the term that is $\propto g_2 n_2$ in Eq.~\eqref{eq:Psi20}.
Combining the two equations of $\mu_1$ and $\mu_2$ above leads to
\begin{subequations}
\begin{align}
\bar{\nu }_c   & = \nu_c -  \frac{m_1}{4\pi \hbar^2 \bar{a}} \frac{\bar{\alpha}^2}{1-\asbg/\bar{a}}= 2\mu_1 - \mu_2  -    \frac{m_1}{4\pi \hbar^2 \bar{a}} \frac{\bar{\alpha}^2}{1-\asbg/\bar{a}}  \label{eq:nuceq1}\\
&  =  4 g_1 n_1 - \frac{4\alpha}{1+ g_1 F} \Psi_{20} - \frac{g_2}{1+g_2 K} |\Psi_{20}|^2 +  \frac{2\alpha^2 F}{1+g_1 F} -  \frac{m_1}{4\pi \hbar^2 \bar{a}} \frac{\bar{\alpha}^2}{1-\asbg/\bar{a}} \label{eq:nuceq2} \\
&  \approx - \frac{g_2}{1+g_2 K} |\Psi_{20}|^2 - \frac{4\alpha}{1+ g_1 F} \Psi_{20}   \label{eq:nuceq3} \\
& \approx   - 2 \sqrt{2} \bar{\alpha} \Psi_{20} 
\approx  - 2 \bar{\alpha} \sqrt{n},  \label{eq:nuceq4}
\end{align}
\end{subequations}
In the first line, we have used the relation between the bare $\nu$ and the renormalized $\bar{\nu}=\Delta \mu_m (B-B_0)$ as given in Eq.~\eqref{eq:nurelation}. 
To go from Eq.~\eqref{eq:nuceq2} to Eq.~\eqref{eq:nuceq3} we have (1) dropped the small $4g_1 n_1$ term associated with the atom pairing
and (2) also omitted the fourth and fifth terms in Eq.~\eqref{eq:nuceq2} as they are of higher order in the Feshbach coupling strength $\alpha$.
To arrive at Eq.~\eqref{eq:nuceq4} from Eq.~\eqref{eq:nuceq3}, we have assumed that the term that is $\propto g_2$ is also smaller than the $\alpha$ term, because of their different dependences on the density $n$~\footnote{This is the case in our numerics for the phase diagrams shown in Fig.~\ref{fig:Fig1}(a). When it is not the case, one needs to keep both terms in Eq.~\eqref{eq:nuceq3}. }.
Additionally, we have used the approximations $F=V^{-1}\sum_\veck 1/(2E_{1\veck}) \approx V^{-1}\sum_\veck 1/(2 \epsilon_{1\veck})$ such that $\alpha/(1+ g_1 F) \approx \bar{\alpha}/\sqrt{2}$.
The terms neglected are either higher order in the Feshbach coupling constant or higher order in density. 
To obtain the final result on the right hand side of the last equality in Eq.~\eqref{eq:nuceq4}, we have used $|\Psi_{20}|^2 \approx n/2$.

Now, if we assume that the lower boundary detuning $\nubdry$, which separates the unstable regime from the stable MSF phase regime in the phase diagram in Fig.~\ref{fig:Fig1},
is very close to the QCP detuning $\bar{\nu}_c$, i. e. $\delta \equiv |\nubdry - \bar{\nu}_c|/|\bar{\nu}_{c,-}| \ll 1$, we can then determine the boundary detuning $\nubdry$ from Eq.~\eqref{eq:g2effL3} analytically as follows.
Since the inverse compressibility $\kappa^{-1}$ is almost equal to the effective molecular scattering length $\ammeff$ (see Fig.~\ref{fig:Fig4} and discussions in Section~\ref{sec:ammmanybody})
and $\bar{\nu}_{c,-}$ is by definition the point where $\kappa^{-1}$ changes sign, the boundary detuning $\bar{\nu}_{c,-}$ is therefore also where $\ammeff$, or equivalently $\gtwoeff$, changes sign.
Hence, right at $\bar{\nu}=\bar{\nu}_{c,-}$ we have
\begin{subequations} \label{eq:nunm}
\begin{align}
0& = \gtwoeff  \nonumber \\
 & \approx  \bar{g}_2 -  \bar{\alpha}^4 \frac{1}{V} \sum_\veck \frac{1}{2 E_{1\veck}^3}     \label{eq:nunmL1}  \\
& \approx \bar{g}_2 -  \bar{\alpha}^4  \intdk \frac{ 1 }{2 \sqrt{(\epsilon_{1\veck} +|\nubdry|/2)^2 -(\bar{\nu}_c/2)^2 }^3}  \label{eq:nunmL2}  \\
& \approx \bar{g}_2 -  \bar{\alpha}^4  \frac{ \sqrt{2m_1}^3}{4\pi^2 \hbar^3 |\bar{\nu}_{c,-}|^{3/2}}  \int_0^\infty \frac{x^2 dx}{\sqrt{x^4 + x^2 + \delta/2}^3}   \label{eq:nunmL3}  \\
& \approx \bar{g}_2 -  \bar{\alpha}^4  \frac{\sqrt{2m_1}^3}{4\pi^2 \hbar^3 |\bar{\nu}_c|^{3/2}}  \big[ \frac{1}{2} \ln \frac{2}{\delta}  - 2(1- \ln 2)+ \mathcal{O}( \delta) \big]. \label{eq:nunmL4} 
\end{align}
\end{subequations}
In writing down Eq.~\eqref{eq:nunmL1} we have approximated the many-body interaction parameters $\widetilde{\bar{g}}_2$ and $\widetilde{\bar{\alpha}}$ in Eq.~\eqref{eq:alphabartilde} by their
two-body counterpart, $\bar{g}_2$ and $\bar{\alpha}$. 
The difference between $\widetilde{\bar{g}}_2$ and $\bar{g}_2$ is $\sim \mathcal{O}(n_2)$ and can be neglected;
the one between $\widetilde{\bar{\alpha}}$ and $\bar{\alpha}$ is higher order in the Feshbach coupling strength $\alpha$, and therefore can be also neglected for a narrow resonance.
From Eq.~\eqref{eq:nunmL1} to Eq.~\eqref{eq:nunmL2} we have used $|\widetilde{\Delta}_1| =|g_1 x_1 - 2 \alpha \Psi_{20}| \approx  2 |\alpha \Psi_{20}|
\approx \sqrt{2} |\bar{\alpha} \Psi_{20} | \approx |\bar{\nu}_c|/2$ and $\widetilde{\epsilon}_{1\veck}\approx  \epsilon_{1\veck} +\bar{\nu}/2$ so that
$E_{1\veck}=\sqrt{\widetilde{\epsilon}_{1\veck}^2 - |\widetilde{\Delta}_1|^2} = \sqrt{(\epsilon_{1\veck} +|\nubdry|/2)^2 -(\bar{\nu}_c/2)^2 }$ for $\bar{\nu} =\nubdry$.
To obtain the last line in Eq.~\eqref{eq:nunmL4} we have used
$ \int_0^\infty \frac{x^2 dx}{\sqrt{x^4 + x^2 + \delta/2}^3} = \int_0^\infty \frac{x^2 dx}{\sqrt{x^4 + x^2 + \delta/2}^3} -  \int_0^1 \frac{x^2dx }{\sqrt{x^2+\delta/2}^3 } 
+ \int_0^1 \frac{x^2dx }{\sqrt{x^2+\delta/2}^3 } = - (1-\ln 2)  +\mathcal{O}(\delta) + \int_0^1 \frac{x^2dx }{\sqrt{x^2+\delta/2}^3 } = - 2 (1-\ln 2)  + \frac{1}{2} \ln \frac{2}{\delta} +\mathcal{O}(\delta)$. 

Dropping the $\mathcal{O}(\delta)$ term in Eq.~\eqref{eq:nunmL4} and solving this equation for $\delta$ we obtain
\begin{align}
\delta  & =\frac{|\nubdry - \bar{\nu}_c|}{|\bar{\nu}_{c,-}|} \approx 0.6 \exp\bigg\{- 2\sqrt{2} \pi^{\frac{7}{4}}  (n (\ammbg)^3)^{\frac{3}{4}}  (\frac{r_*}{\ammbg})^{\frac{5}{4}}   \bigg\}.   \label{eq:criterion}
\end{align}
The exponent comes from
\begin{align}
 \bar{g}_2 \frac{4\pi^2 \hbar^3 |\bar{\nu}_c|^{3/2}}{(2m_1)^{3/2} \bar{\alpha}^4} & \approx \frac{4\pi \hbar^2 \ammbg }{m_2} \frac{4\pi^2 \hbar^3 (2 \bar{\alpha} \sqrt{n})^{3/2} }{ (2m_1)^{3/2} \bar{\alpha}^4} 
 =\sqrt{2} \pi^{\frac{7}{4}}  n^{3/4} \ammbg (\frac{4\pi \hbar^4}{m_1^2 \bar{\alpha}^2})^{\frac{5}{4}} 
 = \sqrt{2} \pi^{\frac{7}{4}}  (n (\ammbg)^3)^{\frac{3}{4}}  (\frac{r_*}{\ammbg})^{\frac{5}{4}},
\end{align}
where we have used Eqs.~\eqref{eq:nuceq4} and \eqref{eq:rstar}. 
Equation~\eqref{eq:criterion} leads to the following generic criterion that needs to be satisfied in order for $\delta \ll 1$,
\begin{align}
(n (\ammbg)^3)^{\frac{3}{4}}  (\frac{r_*}{ \ammbg })^{\frac{5}{4}} \gtrsim \frac{1}{2\sqrt{2} \pi^{ \frac{7}{4}}}.  \label{eq:criterion2}
\end{align}

Using Eqs.~\eqref{eq:criterion} and \eqref{eq:criterion2} we can now quantitatively understand why the MSF-Unstable phase boundary in Fig.~\ref{fig:Fig1}(a)
nearly coincide with the QCP at moderate densities. For example, if we consider the density used in Figs.~\ref{fig:Fig3} and \ref{fig:Fig4}, 
i. e. $n \asbg^3  = 1.68 \times 10^{-5}$, then we find $n (\ammbg)^3 = 4.2 \times 10^{-5}$ and $r_*/ \ammbg  \approx 124$.
This leads to $ (n (\ammbg)^3)^{\frac{3}{4}}  ( r_* / \ammbg)^{\frac{5}{4}} \approx 0.2 \gg 1/(2\sqrt{2}\pi^{7/4})$
and $\delta \sim 10^{-2}$, explaining why $\bar{\nu}_{c,-}$ and $\bar{\nu}_c$ almost coincide with each other at such density in  Fig.~\ref{fig:Fig1}(a). 
For this estimate we have used the narrow-resonance parameters from Table~\ref{tab:parameters}.
On the other hand, Eq.~\eqref{eq:criterion} also implies that when $n$ is low enough, $\delta$ eventually becomes large, i. e.
$\nubdry$ and $\nuc$ are well separated. This is also observed in Fig.~\ref{fig:Fig1}(a). 
In other words, the behavior of $\delta$ eventually resembles that of the wide resonance case in Fig.~\ref{fig:Fig1}(b), when $n$ is sufficiently small. 
We note that a similar point regarding the density effect on the distinctions between narrow and wide resonances has been made previously
in Ref.~\cite{Gurarie2007} in the context of fermionic BCS-BEC crossover. 

\subsection{Derivation of Eq. (4) of the main text}
Following the derivation of the lower boundary detuning $\nubdry$ in Eq.~\eqref{eq:nunm} we can also derive an expression for $\ammeff$, for the detuning $\bar{\nu}$ that satisfies $|\bar{\nu} -\bar{\nu}_c| \ll |\bar{\nu}|$,
\begin{subequations}
\begin{align}
\ammeff &= \frac{m_2}{4\pi \hbar^2} \gtwoeff \approx \ammbg - \asbg \frac{(\Delta \mu_m \Delta B)^2}{\sqrt{\hbar^2/(m_1 \asbg^2)}} \frac{1}{|\bar{\nu}|^{3/2}} \frac{2\sqrt{2}}{\pi} \ln \frac{\bar{\nu}^2}{\bar{\nu}^2 -\bar{\nu}_c^2}   \label{eq:ammeffnarrow} \\
& = \ammbg\bigg[ 1- \big( \frac{\Delta_B^{\text{narrow}}}{B_0-B} \big)^{\frac{3}{2}}  \; \frac{2\sqrt{2}}{\pi} \ln \frac{\bar{\nu}^2}{\bar{\nu}^2 -\bar{\nu}_c^2} \bigg], \label{eq:ammeffnarrow2}
\end{align}
\end{subequations}
where in the first line we have only kept the leading order contribution for the second term (, corresponding to the $\ln(2/\delta)$ term in Eq.~\eqref{eq:nunmL4}). 
To obtain the second line, we have used Eq.~\eqref{eq:DeltaBn} for the definition of $\Delta_B^{\text{narrow}}$.
Clearly, Eq.~\eqref{eq:ammeffnarrow} can not be applied to
zero density since in this limit $\bar{\nu}_c=0$ and the condition $|\bar{\nu} -\bar{\nu}_c| \ll |\bar{\nu}|$ is never satisfied. 
Instead, for the zero density $n=0$ we use $\ammeff(n=0)=\amm$ with the expression of $\amm$ given earlier in Eq.~\eqref{eq:ammnarrow}. 
Therefore, as an interpolation between Eq.~\eqref{eq:ammeffnarrow2} and Eq.~\eqref{eq:ammnarrow}, we can use the following formula for both finite and zero-density cases:
\begin{align}
\ammeff(n) & =  \ammbg\bigg[ 1- \big( \frac{\Delta_B^{\text{narrow}}}{B_0-B} \big)^{\frac{3}{2}}  \; \bigg( 1+ \frac{2\sqrt{2}}{\pi} \ln \frac{\bar{\nu}^2}{\bar{\nu}^2 -\bar{\nu}_c^2(n)}  \bigg) \bigg],
\end{align}
which is nothing but Eq. (4) of the main text. In this equation, $\bar{\nu}_c(n) \approx -2\bar{\alpha} \sqrt{n}$ (See Eq.~\eqref{eq:nuceq4}). 

\section{Contrast with \textit{fermionic} superfluids in a Feshbach resonance}
\label{sec:contrast}

In the main text, we emphasized that the overall minus sign associated with the $-\widetilde{\bar{\alpha}}^4$ term in Eq.~\eqref{eq:g2effL3} of
the many-body effective molecular scattering length $\ammeff$
or equivalently with the $-C/ r_*^2$ term in Eq.~\eqref{eq:ammABC} of the two-body molecular scattering length $\amm$ is fundamentally related to
the bosonic statistics of the atoms. 
In this section, we demonstrate this explicitly by carrying out similar calculations for the scattering lengths $\amm$ and $\ammeff$ in the context of
a Feshbach resonance of \textit{fermionic} atoms. 
Since most of the derivations will parallel those presented in the previous Sections~\ref{sec:insight} and \ref{sec:zeronlimit}, 
we will outline only the main steps and highlight some of the key differences.  

We start with the following two-channel Hamiltonian for a fermionic atom Feshbach resonance~\cite{Stajic2004,Ohashi2002}
\begin{align} \label{eq:HamiltonianF}
\hat{H} & =  \sum_{\veck} \sum_{\sigma=\uparrow, \downarrow} h_{1\veck} a^\dagger_{\veck \sigma} a_{\veck \sigma}  + \sum_{\vecq} h_{2\vecq} b^\dagger_\vecq b_\vecq  
 -\alpha  \frac{1}{\sqrt{V}} \sum_{\veck, \vecq} ( b^\dagger_{\vecq} a_{-\veck \downarrow} a_{ \vecq +\veck \uparrow} + h.c.  )
 \nonumber \\
& 
+  g_1 \frac{1}{V} \sum_{\veck_1,\veck_2,\veck_3} a^\dagger_{\veck_1 \uparrow }a^\dagger_{\veck_2, \downarrow} a_{\veck_3 \downarrow} a_{\veck_1+\veck_2-\veck_3, \uparrow}  
+  \frac{g_2}{2} \frac{1}{V} \sum_{\vecq_1,\vecq_2,\vecq_3} b^\dagger_{\vecq_1 }b^\dagger_{\vecq_2} b_{\vecq_3} b_{\vecq_1+\vecq_2-\vecq_3},
\end{align}
where $a_{\veck \sigma}$ is now the annihilation operator for a \textit{fermionic} open-channel atom with spin $\sigma$ 
while $b_\vecq$ is that for a (bosonic) closed-channel molecule. 
Note that for the fermionic problem, we use subscripts of $\{1,2\}$ to denote the two channels, with $1$ and $2$ representing the open and closed channels, respectively,
which is consistent with previous conventions.
However, $\sigma=\{\uparrow,\downarrow\}$ now stands for fermionic spins. 
The fermionic Hamiltonian $\hat{H}$ does not support a quantum phase transition as one changes the detuning between the non-interacting energy band bottom of the two channels; instead, at zero temperature it describes only a crossover between a fermionic superfluid that is more BCS like and one that is more BEC like.
This is of course because the fermionic atoms themselves can not BEC condense without forming pairs. 

The regularization conditions for the bare parameters in $\hat{H}$ are almost identical to their bosonic counterpart given earlier in Eqs.~\eqref{eq:regularization} and \eqref{eq:renormquant}
of Section~\ref{sec:parameters}, except that now 
\begin{align}
\alpha  & = \bar{\alpha} /(1-\asbg/\bar{a} ),    \label{eq:alpharelationF}
\end{align}
which differs from Eq.~\eqref{eq:alpharelation} by a $1/\sqrt{2}$ on the right hand side. 
The definitions of $\bar{\alpha}$ remain the same as in Eq.~\eqref{eq:renormquant}. 

To describe the ground-state BCS-BEC crossover we adopt the following variational wavefunction,
\begin{align} \label{eq:PsivarF}
| \Psivar  \rangle & =\mathcal{N}^{-1} e^{\Psi_{20} \sqrt{V} b_{\vecq=0}^\dagger 
+ \sum_\veck \chi_{1\veck} a^\dagger_{\veck \uparrow} a_{-\veck \downarrow}^\dagger 
+ \sum_\vecq \chi_{2\vecq} b^\dagger_{\vecq } b_{-\vecq}^\dagger 
} | 0 \rangle.
\end{align}
The main difference between this wavefunction and that in Eq.~\eqref{eq:Psivar} is that, for the fermion problem, we do not have a $\Psi_{10}$ condensate.
As before, $\chi_{1\veck}$ in the exponent characterizes Cooper pairing between atoms
while $\chi_{2\vecq}$ describes a pairing correlation effect resulting from the quantum depletion of closed-channel molecules. 

Following the procedure in Sections~\ref{sec:variational} and \ref{sec:insight}, we
can derive the counterpart of Eq.~\eqref{eq:g2effL3} for the fermion case as
\begin{subequations}
\begin{align}
\frac{\gtwoeff}{(\Zeff)^2} & \equiv \frac{1}{2V} \frac{\partial^4 \Otri[\Psi_{20},\Delta_1^s,\Delta_2^s, \mu]}{\partial \Psi_{20}^2 \partial (\Psi_{20}^*)^2 } \bigg\vert_{\mu}
\approx \widetilde{\bar{g}}_2  -  \frac{\alpha}{2}  \frac{\partial^3 x_1}{\partial \Psi_{20}^2 \partial \Psi_{20}^* }    \\
&  \approx \widetilde{\bar{g}}_2  + \frac{1}{2} \widetilde{\bar{\alpha}}^4 \bigg\{  
\frac{1}{V}  \sum_\veck \frac{1}{2 E_{1\veck}^3} + g_1  \bigg[ \frac{1}{V} \sum_\veck  \frac{ \widetilde{\epsilon}_{1\veck}}{ 2 E_{1\veck}^3 } \bigg]^2 \bigg\}. 
  \label{eq:g2effF}
\end{align}
\end{subequations}
In this equation, the definitions of $\widetilde{\bar{g}}_2$, $x_1$, $\Zeff$, and so on are almost the same as in Sections~\ref{sec:variational} and \ref{sec:insight}
with the following exceptions: (1) for the current fermionic problem the expression of $\widetilde{\bar{\alpha}}$ is given by
\begin{align}
\widetilde{\bar{\alpha}} & = \frac{ \alpha } {1+g_1 \frac{1}{V}\sum_\veck  \frac{1}{2 E_{1\veck}} },  \label{eq:alphabartildeF}
\end{align}
which differs from its bosonic version in Eq.~\eqref{eq:alphabartildedef}  by a factor of $1/\sqrt{2}$ on the right hand side. 
(2) Another difference lies in the form of the atom Bogoliubov quasiparticle energy, which now takes the form
of $E_{1\veck}=\sqrt{\widetilde{\epsilon}_{1\veck}^2+|\widetilde{\Delta}|^2}$.
This differs from its bosonic counterpart by a minus sign under the square root. 
Here, $\widetilde{\epsilon}_{1\veck} = h_{1\veck} + g_1 n_{1\sigma} $ (where the spin $\sigma$ is not summed)
and $\widetilde{\Delta}_1 =  \Delta_1 - \alpha \Psi_{20} =g_1 x_1 - \alpha \Psi_{20}$,
with $n_{1\sigma}=V^{-1} \sum_\veck (1/2) [1- \widetilde{\epsilon}_{1\veck}/ E_{1\veck} ]$ the spin-resolved density of fermions. 
These expressions are slightly different from their bosonic counterpart in Section~\ref{sec:variational}.
(3) In the expression of many-body closed-channel molecular fraction $\mathcal{Z}^{\mathrm{eff}}$, although it is still defined by Eq.~\eqref{eq:Zeff},
the density $n_1$ in the denominator needs to be summed over spin, i. e. $n_1 = \sum_{\sigma=\uparrow, \downarrow} n_{1\sigma}$. 
 
What is most striking in Eq.~\eqref{eq:g2effF} is the \textit{positive} sign in front of the $\widetilde{\bar{\alpha}}^4$ term, which is opposite to its bosonic counterpart in Eq.~\eqref{eq:g2effL3}.
As a result, Cooper pairing of fermions due to Feshbach coupling does not drive $\gtwoeff$ negative and therefore does not cause instability of molecular condensates. 
Algebraically, this sign difference arises because $E_{1\veck}$ differs by a minus sign under its square root between the fermion and boson cases.
This results in a sign change in the derivative $\partial^3 x_1/\partial \Psi_{20}^2 \partial \Psi_{20}^* $ since $x_1=V^{-1}\sum_\veck (-\widetilde{\Delta}_1)/2 E_{1\veck}$ depends on $E_{1\veck}$. Ultimately, all these sign differences reflect the different statistics of the particles being paired.

\subsection{Zero density limit}
Using Eq.~\eqref{eq:g2effF} for $\gtwoeff/(\Zeff)^2$ and Eq.~\eqref{eq:Zeff} for $\Zeff$ we can then arrive at the expression of $\gtwoeff$ and $\ammeff$ (see Eq.~\eqref{eq:ammeffdef} for its definition). 
After that, we again consider the zero density limit of the many-body molecular scattering length $\ammeff$ to derive its two-body counterpart $\amm$. 
The final result is
\begin{align}  \label{eq:ammuniv1F}
\amm & \equiv  \lim_{n\rightarrow 0} \frac{m_2}{4\pi \hbar^2} \gtwoeff  
= \bigg\{ \ammbg  + \frac{1}{2} \frac{1}{ k_b^4 r_*^2 \bar{a}}  \frac{C_3 + 2 C_2^2 \asbg/\bar{a} }{(1- C_1 \asbg/\bar{a})^4} \bigg\} 
\bigg[ \frac{2 k_b r_* (1- C_1 \asbg/\bar{a})^2}{C_2/(k_b \bar{a})+2 k_b r_* (1- C_1 \asbg/\bar{a})^2} \bigg]^2.
\end{align}
The definitions of the parameters $\{r_*, C_1, C_2, C_3\}$ in this equation are the same as those given previously in Eqs.~\eqref{eq:rstar} and \eqref{eq:C123} (or \eqref{eq:C1C2C3}) for the boson problem. The term in the square bracket of Eq.~\eqref{eq:ammuniv1F} represents the two-body closed-channel molecular fraction $\Z$ of a dressed molecule.
This expression of $\Z$ is identical to Eq.~\eqref{eq:zeronZ}
that we derived for the boson problem,
which should not be surprising, as the two-atom scattering problem does not depend on the statistics of the atoms being scattered.

The main difference between Eq.~\eqref{eq:ammuniv1F} and its bosonic counterpart in Eq.~\eqref{eq:ammuniv1}
is that the minus sign in front of the $1/( k_b^4 r_*^2 \bar{a})$ term in Eq.~\eqref{eq:ammuniv1} is now replaced by $+1/2$.
This change arises from the difference in sign in front of the $\widetilde{\bar{\alpha}}^4$ term between Eqs.~\eqref{eq:g2effF} and \eqref{eq:g2effL3}. 
Due to this sign difference, the molecular scattering length $\amm$ in Eq.~\eqref{eq:ammuniv1F} is always positive if both $\ammbg$ and $\asbg$ are positive.
Even if $\ammbg$ and/or $\asbg$ are negative, near the resonance---where $k_b$ is small---the $C_3$ term in Eq.~\eqref{eq:ammuniv1F} will
dominate, ensuring that $\amm$ remains positive so that the fermion pair superfluid is always stable near the resonance. 
This sharp contrast with the bosonic case highlights the importance of atomic statistics on the
stability of molecular condensates in a Feshbach resonance.

For the $\amm$ in Eq.~\eqref{eq:ammuniv1F} we can also consider its asymptotic limiting form in the three detuning regimes that we have
considered in Section~\ref{sec:amm123} for the bosonic case. 
\begin{enumerate}
\item {\bf Large negative detuning regime I}: i. e.  $(-\bar{\nu})  \gg \hbar^2/ m_1\bar{a}^2 \quad  \& \&  \quad   (-\bar{\nu})  \gg \hbar^2 / (m_1 r_* \bar{a})$.
For this regime,
\begin{subequations}
\begin{align} 
a_{\mathrm{mm}}  &  \approx   \ammbg \bigg[ 1 - \bigg( \frac{\Delta_B^{\text{wide}}}{B_0-B} \bigg)^2  \bigg],
 \label{eq:ammwideF} \\
\text{with }\quad
\Delta_B^{\text{wide}}   & =  \frac{\pi}{\sqrt{6}}  \sqrt{ \frac{\hbar^2}{m_1 (\bar{a} - \asbg)^2\Delta \mu_m} }  \sqrt{\Delta B}. 
\end{align}
\end{subequations}
This expression is identical to Eq.~\eqref{eq:ammwide} for the boson case. This is not surprising since the two terms in Eq.~\eqref{eq:ammwideF}
come from a leading expansion of the term $\ammbg \Z^2$ in the inverse binding momentum $1/k_b$, as can be seen from Eq.~\eqref{eq:ammABC2},
and the closed-channel molecular fraction $\Z$ does not depend on whether the atoms are bosons or fermions. 
It is important to note that Eq.~\eqref{eq:ammwideF} does not imply that $\amm$ changes sign at detuning $B-B_0=-\Delta_B^{\text{wide}}$,
since the above asymptotic formula breaks down before this detuning point is reached. 
\item {\bf Small negative detuning regime II for a wide resonance}: i. e. $E_b \ll  (-\bar{\nu}) \ll \frac{\hbar^2}{m_1 \bar{a}^2}$.
This is the universal regime discussed in the literature on fermionic BCS-BEC crossover. 
 For this regime, Eq.~\eqref{eq:ammuniv1F} reduces to
\begin{align} 
a_{\mathrm{mm}}  & \approx  4 \asbg + \frac{2}{k_b} = 4  \asbg \bigg[ 1 + \frac{ \Delta B/2}{B_0-B} \bigg],
\end{align}
which is the fermionic counterpart of Eq.~\eqref{eq:ammwide2}. 
Note the difference in the sign of the resonant term inside the square bracket between the two cases. 
If we drop the background contribution from the $4\asbg$ term, then 
\begin{align}
\amm  & \approx  +  \frac{2}{k_b} \approx +2 a_s.  \label{eq:BornF}
\end{align}
This is the familiar Born result~\cite{Brodsky2006,Levinsen2006,Pieri2000} for the two-body molecular scattering length $\amm$, as discussed in the literature on the fermionic BCS-BEC crossover. 
Both exact solutions of the Schrodinger equation for the four-body scattering problem~\cite{Petrov2004} and diagrammatic calculations~\cite{Brodsky2006,Levinsen2006} have shown
that, \textit{in the resonance regime}, the exact answer 
is $\amm=0.6 a_s$, instead of $2a_s$. 
The discrepancy arises because the zero density limit of our variational wavefunction $\Psivar$
in Eq.~\eqref{eq:PsivarF} does not account for all scattering processes that contribute to $\amm$.
To go beyond the Born result in Eq.~\eqref{eq:BornF}, it is necessary to include higher-order correlations among atoms
in the exponent of $\Psivar$~\cite{Tan2006}, beyond just 
the Cooper pairing term $a_{\veck,\uparrow}^\dagger a_{-\veck,\downarrow}^\dagger$.  
We emphasize, however, that these neglected higher-order contributions are not particularly consequential for our bosonic problem, since our focus is away from the near-resonance regime. \\

For the bosonic problem considered in the previous sections, including these higher-order correlations presents a significant challenge as they will introduce
effects associated with the formation of Efimov and tetramer bound states~\cite{Gogolin2008,DIncao2009}, which can considerably increase the complexity of calculation of $\amm$.
Fermions do not suffer from this issue and Ref.~\cite{Tan2006} suggests that progress can be made in improving the value of $\amm$  for this universal regime. \\

Finally, we comment on the applicability of the diagrammatic-integral-equation method from Ref.~[\onlinecite{Levinsen2006}] to the bosonic systems discussed in previous sections. 
This approach is inapplicable because Ref.~[\onlinecite{Levinsen2006}] neglects crucial background interactions between atoms and molecules, which are essential for the stability of molecular 
condensates in Fig.\ref{fig:Fig1}. The success of Ref.~[\onlinecite{Levinsen2006}] rests on the assumption that these background interactions are negligible; incorporating them alongside Feshbach 
coupling introduces significant, potentially insurmountable, complexities. Moreover, even without considering background interactions, applying Ref.~[\onlinecite{Levinsen2006}] to our bosonic system
is unfeasible, because the presence of Efimov and tetramer bound states precludes a direct solution of their integral equations for the molecular scattering amplitudes.

\item {\bf Intermediate negative detuning regime for a narrow resonance} i. e. the detuning regime such that $\{1/\bar{a} \sim 1/|\asbg| \} \gg \sqrt{m_1 (-\bar{\nu})}/\hbar \gg 1/\sqrt{r_* \ammbg} \gg 1/r_*$. 
For this regime,
\begin{subequations} \label{eq:ammnarrowF}
\begin{align}
\amm  & \approx \ammbg \bigg[ 1 + \bigg( \frac{ \Delta_B^{\text{narrow}} }{B_0-B} \bigg)^{ \frac{3}{2}}   \bigg] , \\
\text{with } \quad
 \Delta_B^{\text{narrow}}    & = \frac{1}{2}  \bigg( \frac{\asbg}{\ammbg} \bigg)^{ \frac{2}{3}}  \bigg( \frac{ m_1 \asbg^2 \Delta \mu_m }{\hbar^2}   \bigg)^{\frac{1}{3}}  \;  (\Delta B)^{\frac{4}{3}},
\end{align}
\end{subequations}
which is the fermionic counterpart of Eq.~\eqref{eq:ammnarrow}. Writing the molecular scattering length $\amm$ in the above form, we have already assumed the molecular background scattering length to be positive, $\ammbg>0$
~\footnote{If $\ammbg<0$, which is more relevant for a fermionic BCS-BEC crossover, one needs to replace the $\amm$ formula in Eq.~\eqref{eq:ammnarrowF} with $\amm=\ammbg + \asbg (\Delta_B^{\text{narrow}}/|B-B_0|)^{3/2}$ and redefine $\Delta_B^{\text{narrow}}$ as $\Delta_B^{\text{narrow}} =(1/2) \big( m_1 \asbg^2 \Delta \mu_m / \hbar^2 \big)^{\frac{1}{3}}  \;  (\Delta B)^{\frac{4}{3}}$.
This form of $\amm$ shows more clearly that the resonant term in Eq.~\eqref{eq:ammnarrowF} actually does not depend on $\ammbg$. }.
If we drop the non-resonant term $\ammbg$ in Eq.~\eqref{eq:ammnarrowF}, as well as the background term $\asbg$ in Eq.~\eqref{eq:asdef} for the atomic scattering length $a_s$,
we can then rewrite Eq.~\eqref{eq:ammnarrowF} as
\begin{align} 
\amm
& \approx   \frac{1}{2} \sqrt{\frac{a_s}{r_*}} a_s,   \label{eq:ammuniv3F}
\end{align}
where we have used the definition of the length scale $r_*$ in Eq.~\eqref{eq:rstar}. 
This result, which is valid when all the background interaction effects can be ignored, has also been previously obtained by the authors of Refs.~\cite{Gurarie2007,Levinsen2011}. 
\end{enumerate}

 \bibliography{References2}
 